\documentclass[12pt]{article}

\usepackage{setspaceff}
\usepackage{amssymb}
\usepackage{enumerate}
\usepackage{amsmath}
\usepackage{bm}
\usepackage{bbm}
\usepackage[dvips]{graphicx}
\usepackage{pstricks}
\usepackage{pstricks-add}
\usepackage{pst-plot}
\usepackage{tikz}
\usepackage{scalefnt}

\usepackage{pdfpages}
\usepackage{pgfplots}
\pgfplotsset{compat=newest}
\usepackage[utf8]{inputenc}

\usetikzlibrary{arrows}
\usetikzlibrary{arrows.meta}

\usepackage{tikz-3dplot}
\usetikzlibrary{patterns}

\advance\textwidth      by  51pt
\advance\oddsidemargin  by -25.5pt
\advance\evensidemargin by -25.5pt
\advance\textheight     by  8\baselineskip
\advance\topmargin      by -4\baselineskip

\def\abststretch{1.00}
\def\bodystretch{1.20}
\def\refstretch {1.10}
\setfootnotestretch{1.05}
\setfloatstretch{1}

\let\bodyenvironmentsize=\normalsize
\let\bodyproofsize=\normalsize  
\let\appenvironmentsize=\small  
\let\appproofsize=\small        

\newcommand{\astrut}[1]{\rule{0em}{#1ex}}
\newcommand{\eref}[1]{{\rm (\ref{#1})}} 
\newcommand{\divn}[2]{\mbox{$\frac{#1}{#2}$}}

\def\KRC{Keller et al.\ (2005)}
\def\KR{Keller and Rady (2010)}

\def\Dt         {\tilde{\Delta}_\varepsilon}
\def\E          {{\cal E}^\Delta}
\def\Exp {\mathbb{E}\!}

\def\Pr {\mbox{Pr}}

\def\d  {\delta}
\def\l  {\Lambda}

\def\phat  {\hat{p}}
\def\plow  {\tilde{p}}
\def\pp    {\bar{p}}
\def\pr    {\underline{p}}
\def\ptilde {\tilde{p}}

\def\N{I\! \! N}

\def\wbar {\overline{w}}
\def\wlow {\underline{w}}

\def\kp   {\underline{\kappa}}  
\def\kr   {\overline{\kappa}}   
\def\K          {{\cal K}}

\def\Kp         {{\cal K}(p;\wbar,\wlow)}

\def\Tone       {T_1^\Delta}
\def\Tstar      {T_*^\Delta}
\def\Tp         {\underline{T}^\Delta}
\def\Tr         {\overline{T}^\Delta}

\def\VIpp       {V_{1,\pp}}
\def\VNpeps     {V_{N,p_\varepsilon}}

\def\VNpb       {V_{N,\breve{p}}}
\def\VNphat     {V_{N,\phat}}
\def\VNplow     {V_{N,\plow}}
\def\VNpr       {V_{N,\pr}}
\def\VNulp      {V_{N,\underline{p}}}
\def\wp         {\underline{w}^\Delta}  
\def\wr         {\overline{w}^\Delta}   
\def\Wsup       {\overline{W}^\Delta}
\def\Winf       {\underline{W}^\Delta}
\def\Wt         {\widetilde{W}^{\Delta,\varepsilon}}

\def\WsupSSE    {\Wsup_{\rm SSE}}
\def\WinfSSE    {\Winf_{\rm \, SSE}}
\def\WsupPBE    {\Wsup_{\rm PBE}}
\def\WinfPBE    {\Winf_{\rm \, PBE}}

\newtheorem{thm}  {Theorem}
\newtheorem{prop} {Proposition}
\newtheorem{lem}  {Lemma}[section] 

\def\QED{\hfill \rule{3mm}{3mm}}
\newcommand{\proof}      [1]{\proofsize
        {\sc Proof:}\ #1 \ \QED
        \par \environmentsize \vskip8truept}

\newcommand{\proofof}    [3]{\noindent \proofsize
        {\sc Proof of #1 \ref{#2}:}\ #3 \ \QED
        \par \environmentsize \vskip8truept}
\newcommand{\qed}{\nobreak \ifvmode \relax \else
\ifdim\lastskip<1.5em \hskip-\lastskip
\hskip1.5em plus0em minus0.5em \fi \nobreak
\vrule height0.75em width0.5em depth0.25em\fi}

\newcommand\AppendixIn{         
    \newpage
    \section*{Appendix}

    \parskip0.0ex
    \let\environmentsize=\appenvironmentsize
    \let\proofsize=\appproofsize
    \environmentsize

    \renewcommand{\thesection}{\Alph{section}}  
    \renewcommand{\theequation}{\thesection.\arabic{equation}}  

    \setcounter{section}{0}
}
\newcommand\AppendixOut{        
    \renewcommand{\thesection}{\arabic{section}}    
    \renewcommand{\theequation}{\arabic{equation}}  

    \parskip1.5ex
    \let\environmentsize=\bodyenvironmentsize
    \let\proofsize=\bodyproofsize
    \environmentsize
}

\date{
This version: \today \\
}

\title{
\sc Overcoming Free-Riding in Bandit Games\thanks{
This paper supersedes our earlier paper ``Strongly Symmetric Equilibria in Bandit Games''
(circulated in 2014 as Cowles Discussion Paper No.~1956
and SFB/TR 15 Discussion Paper No.~469)
which considered pure Poisson learning only.
Thanks for comments and suggestions are owed to
the editor,
three anonymous referees,
and seminar participants at
Aalto University,
Austin,
Berlin,
Bonn,
City University of Hong Kong,
Collegio Carlo Alberto,
Duisburg-Essen,
Edinburgh,
Exeter,
Frankfurt
(Goethe University, Frankfurt School of Finance and Management),
London
(Queen Mary, LSE),
Lund,
Maastricht,
Mannheim,
McMaster University,
Microsoft Research New England,
Montreal,
Oxford,
Paris
(S\'{e}minaire Roy,
S\'{e}minaire Parisien de Th\'{e}orie des Jeux,
Dauphine),
Princeton,
Southampton,
St.\ Andrews,
Sydney,
Toronto,
Toulouse,
University of Western Ontario,
Warwick,
Zurich,
the 2012 International Conference on Game Theory at Stony Brook,
the 2013 North American Summer Meeting of the Econometric Society,
the 2013 Annual Meeting of the Society for Economic Dynamics,
the 2013 European Meeting of the Econometric Society,
the 4th Workshop on Stochastic Methods in Game Theory at Erice,
the 2013 Workshop on Advances in Experimentation at Paris II,
the 2014 Canadian Economic Theory Conference,
the 8th International Conference on Game Theory and Management in St.\ Petersburg,
the SING 10 Conference in Krakow,
the 2015 Workshop on Stochastic Methods in Game Theory in Singapore,
the 2017 Annual Meeting of the Society for the Advancement of Economic Theory in Faro,
the 2019 Annual Conference of the Royal Economic Society,
and
the 2020 International Conference on Game Theory at Stony Brook.
Part of this paper was written during a visit to the Hausdorff Research Institute for Mathematics
at the University of Bonn under the auspices of the Trimester Program
``Stochastic Dynamics in Economics and Finance''.
Johannes H\"{o}rner acknowledges funding from the Cowles Foundation and the Agence nationale de la recherche under grant ANR-17-EURE-0010 (Investissements d'Avenir program).
Nicolas Klein acknowledges financial support from the Fonds de Recherche du Qu\'{e}bec Soci\'{e}t\'{e} et Culture
and the Social Sciences and Humanities Research Council of Canada.
Sven Rady acknowledges financial support from Deutsche Forschungsgemeinschaft through SFB/TR 15 (project A08) and SFB/TR 224 (project B04).
}}
\author{
Johannes H\"{o}rner\thanks{Yale University, 30 Hillhouse Ave., New Haven, CT 06520, USA, and TSE (CNRS), and CEPR, {\tt johannes.horner@yale.edu}.}
\and
Nicolas Klein\thanks{Universit\'{e} de Montr\'{e}al, D\'{e}partement de Sciences \'{E}conomiques,
C.P.\ 6128 succursale Centre-ville; Montr\'{e}al, H3C 3J7, Canada, and CIREQ, {\tt kleinnic@yahoo.com}.}
\and
Sven Rady\thanks{University of Bonn, Adenauerallee 24-42, D-53113 Bonn, Germany, and CEPR, {\tt rady@hcm.uni-bonn.de}.}
}

\begin{document}
\maketitle
\setcounter{page}{0}
\thispagestyle{empty}
\setstretch{\abststretch}

\newpage
\vspace{4ex}
\begin{abstract}
\noindent
This paper considers a class of experimentation games with L\'{e}vy bandits encompassing those of Bolton and Harris (1999) and Keller, Rady and Cripps (2005). Its main result is that efficient (perfect Bayesian) equilibria exist whenever players' payoffs have a diffusion component. Hence, the trade-offs emphasized in the literature do not rely on the intrinsic nature of bandit models but on the commonly adopted solution concept (Markov perfect equilibrium). This is not an artifact of continuous time: we prove that efficient equilibria arise as limits of equilibria in the discrete-time game. Furthermore, it suffices to relax the solution concept to strongly symmetric equilibrium.

\vspace{1ex}

\vspace{1ex}
\noindent
{\em JEL} {\sc Classification Numbers:}
C73, 
D83. 
\end{abstract}
\newpage

\AppendixOut                        
\setstretch{\bodystretch}

\section{Introduction
\label{sec:intro}}

The goal of this paper is to evaluate the role of the Markov assumption in strategic bandit models. Our main finding is that it is the driving force behind the celebrated trade-off between the free-riding and encouragement effects (Bolton and Harris, 1999). More precisely, we show that free-riding does not prevent efficiency from being achievable in equilibrium when learning involves a Brownian component, as in the Bolton-Harris model.
In the pure Poisson case, free-riding can be overcome entirely if payoff arrivals are not very informative, and partially if they are.

Our framework follows Bolton and Harris (1999), \KRC\ and \KR. Impatient players repeatedly choose between a risky and a safe arm. They share a common prior about the risky arm, which can be of one of two types.
Learning occurs via the players' payoffs, which are publicly observable and, in the case of the risky arm, type-dependent.
In terms of expected payoffs, a risky arm of the good type dominates the safe arm, which in turn dominates a risky arm of the bad type.
The payoff process which we assume for the risky arm is the simplest that encompasses both the Brownian motion in Bolton and Harris (1999) and the Poisson process in \KRC\ and \KR. Unlike ours, these three papers focus on Markov perfect equilibria of the game in continuous time with the posterior probability of a good risky arm as the state variable.

Understanding the role of the Markov refinement requires discretizing the timeline of the game because defining standard game-theoretic notions, such as perfect Bayesian equilibrium, raises conceptual problems in continuous-time games with observable actions (Simon and Stinchcombe, 1989). We then let the time interval between successive opportunities to revise actions vanish in order to get a clean characterization and a meaningful comparison with the literature.
The efficient behavior in this frequent-action limit is for all players to use the risky arm when they are sufficiently confident that it is of the good type, and for all of them to use the safe arm otherwise.

To give some very rough intuition for our results, the best equilibria have a flavor of ``grim trigger.''
Efficiency obtains if, and only if, the following holds when the common belief is right above the threshold where a social planner would stop all experimentation: a player that deviates from the risky to the safe arm would find it best to always use the safe arm thereafter---independent of the outcome of all other players' choices at that instant, and \textit{assuming} that all other players react to the deviation by using the safe arm exclusively thereafter.
Intuitively, having all other players stop experimenting forever is the worst punishment a defecting player can face.
If the best response to this punishment is to also play the safe arm, a unilateral deviation from the risky to the safe arm stops \emph{all} experimentation.
Then, each player effectively faces the same trade-off as the social planner, weighing the informational benefits of all experiments against the cost (in terms of current expected payoffs) of a single one.

A complicating factor, however, is that bandit models are stochastic games: the common belief about the risky arm evolves.
Will the threat be carried out in the case ``good news'' obtains, as the posterior belief might be so optimistic that all players find it preferable to adopt the risky arm, even in case of a single deviation?
Also, if a player expects experimentation to stop in the next instant
absent any good news about the risky arm, it had better be that the good news event be sufficiently likely; otherwise, what good is the threat that other players stop experimenting?

Hence, the question is whether the threat of the punishment in case of good news is both credible and the corresponding event sufficiently likely.
When there is a diffusion component, it is the leading force in players' belief updating, and good news is likely: the Brownian path is just as likely to go up as it is to go down (in contrast, the Poisson component is much less likely to yield good news).
The threat is credible, moreover: the belief does not jump up, but rather ticks up slightly, so that,
in the next period, it will almost certainly stay in a tight neighborhood of the current belief.
As the efficient belief threshold is strictly smaller than that of a single player experimenting in isolation,
the belief will thus remain in a region where playing safe is a best response to everyone else doing so.
Our criterion is then satisfied.

The situation is less clear in the pure Poisson case.
If good news arrives there, the belief jumps up and may reach a region in which the deviating player would find it optimal to use the risky arm.
This may increase the incentive to deviate and reduce the other player's ability to punish.
If good news are conclusive, for example, they lead all players to play risky forever, so there is no scope for any punishment whatsoever.
For inconclusive news, ``large'' jumps in beliefs allow for some punishment, but not enough for efficiency.
``Small'' jumps, however, are compatible with efficiency, and for the same reason as above: good news generated by other players keeps the posterior belief in a region where each player finds it optimal to use the safe arm if everybody else does so.

At a more technical level, the difference between our findings for payoff processes with and without a Brownian component has to do with the standard deviation of the ``noise'' in observed payoffs, which determines how fast the informational benefit of an experiment with the risky arm vanishes as the discretization period shrinks.
In view of the details needed to formulate an intuition along these lines, we postpone this until after the statement of our main result in Section \ref{sec:result}.

Irrespective of whether efficiency can be achieved or not, we show that both the highest and lowest average equilibrium payoff is attainable with strongly symmetric equilibria (SSEs),
that is, equilibria in which all players use the same continuation strategy for any given history, independent of their identity (\textit{e.g.}, regardless of whether they had been the sole deviator).
Moreover, both the highest and lowest payoff are obtained by an alternation between the same two Markov strategies: one that yields the highest payoff for any belief, and which governs play as long as no player deviates, and one that yields the lowest payoff, given that play reverts to the other Markov strategy at some random time.
Both these Markov strategies are cutoff strategies that have all players use the risky arm if and only if the current belief exceeds some threshold.
There is no need to resort to more complicated perfect Bayesian equilibria (PBEs). Of course, PBE need not involve symmetric payoffs, but we show that in terms of total payoffs across players, there is no difference between SSE and PBE: the best and worst total (and so average) equilibrium payoffs coincide.\footnote{One appealing property of SSEs is that payoffs can be studied via a coupled pair of functional equations that extends the functional equation characterizing MPE payoffs  (see Proposition \ref{prop:chara-discrete}).}
We further show that the worst average equilibrium payoff equals the optimal payoff of a single player experimenting in isolation.

Two caveats are in order.
First, we have pointed out the importance of studying the discrete-time game to make sense of PBE (as well as to factor out equilibria that are continuous-time quirks, such as the ``infinite-switching equilibria'' of \KRC, which have no equivalent in discrete time). However, our results are asymptotic to the extent that they only hold when the time interval between rounds is small enough. There is no qualitative difference between an arbitrarily small uptick vs.\ a discrete jump when the interval length is bounded away from zero. Our results rely heavily on what is known about the continuous-time limits, and especially on the analyses in Bolton and Harris (1999), \KR, and Cohen and Solan (2013).
To the extent that some of our proofs are involved, it is because they require careful comparison and convergence arguments. Because we rely on discrete time, we must settle on a particular discretization. We believe that our choice is natural:
players may revise their action choices at equally spaced time
opportunities, while payoffs and information accrue in continuous
time, independent of the duration of the intervals.\footnote{That is, ours is
the simplest version of inertia strategies as introduced by Bergin
and MacLeod (1993).} Nonetheless, other discretizations might conceivably yield different
predictions.

Second, our results do not cover all bandit games. Indeed, an explicit characterization of the single-agent and planner solutions, on which we build, requires some restrictions on the payoff process.
Here, we follow Cohen and Solan (2013) in ruling out bad-news jumps.\footnote{
Unlike Cohen and Solan (2013), we further rule out learning from the size of a lump-sum payoff.
We believe that this restriction is inconsequential; see the concluding comments for further details.}
Moreover, our framework does not subsume the ``breakdowns'' model of Keller and Rady (2015).\footnote{
The technical difficulty there is that the value functions cannot be solved in closed form. They are defined recursively, with the functional form depending on the number of
breakdowns triggering an end to all experimentation. We return to this scenario in the concluding comments.}

Our paper belongs to the growing literature on strategic bandits.
We have already mentioned the standard references in this literature.
Studying the undiscounted limit of the experimentation game, Bolton and Harris (2000) consider the Brownian motion case, while Keller and Rady (2020) allow for L\'{e}vy processes.
A number of authors have extended the exponential bandit framework of \KRC.
Klein and Rady (2011) and Das, Klein and Schmid (2020) investigate games in which the quality of the risky arm is heterogeneous across players.
Dong (2018) endows one player with superior information regarding the state of the world,
Marlats and M\'{e}nager (2021) examine strategic monitoring,
and Thomas (2021) analyzes congestion on the safe arm.
All these papers work in continuous time and rely on MPE as the solution concept.\footnote{
To the best of our knowledge, the MPE concept is adopted by all papers in the literature on strategic bandits unless they consider agency models (the principal having commitment, one solves for a constrained optimum rather than an equilibrium) or drop the assumption of perfect monitoring of actions and payoffs. Examples of the latter include Bonatti and H\"{o}rner (2011), Heidhues et al.\ (2015), and Rosenberg et al.\ (2007). As there is no common belief that could serve as a state variable, these authors use Nash equilibrium or one of its refinements (such as perfect Bayesian equilibrium).}
Here, we focus on the canonical bandit game with discounting and homogenous risky arms,
but relax the solution concept by considering a sequence of discrete-time games.\footnote{
Hoelzemann and Klein (2021) suggest that MPE may be a decent predictor of subjects' behavior in a laboratory experiment of the \KRC\ setting. They reject the hypothesis that subjects played according to the welfare-maximizing PBE constructed here. Rather, subjects adopted non-cutoff and turn-taking behaviors, which are quite reminiscent of \KRC's simple MPEs. The paper leaves open the question of what deterred subjects from the simple on-path cutoff behavior of the best equilibrium.}

Our results show that the conclusions drawn from strategic-experimentation models may crucially depend on the equilibrium concept being used.
When strategic experimentation is embedded in a richer environment, however, the robustness of MPE-based findings depends on the fine details of the game.
Two papers from the industrial organization literature that build on \KRC\ may serve to illustrate this.
Besanko and Wu (2013) study how learning and product-market externalities affect incentives to cooperate in R\&D.
Our results apply to the case that the overall externality is positive (so there is an incentive to free-ride on other firms' R\&D efforts); the best SSE then involves experimentation at full intensity down to the same cutoff as in the symmetric MPE.
The comparison between research competition and research cooperation thus becomes simpler, but the main insights remain unchanged.
In Besanko, Tong and Wu's (2018) analysis of research subsidies, our results would again reduce firms' free-riding (see their footnote 19), and hence increase the incentives to invest under the different subsidy types that they consider.
If there is no shadow cost of public funds, the funding agency can overcome free-riding through the design of its subsidy program, so MPE is not restrictive in this case.
With such a shadow cost, by contrast, this is no longer true: even if the agency chooses the subsidy that minimizes the risk of underinvestment in R\&D, this investment is not ``flat-out'' above the resulting cutoff, so the best SSE would again improve matters here.

Our paper also contributes to the literature on SSE.
Equilibria of this kind have been studied in repeated games since Abreu (1986). They are known to be
restrictive. First, they make no sense if the model itself fails to be
symmetric. However, as Abreu (1986) notes for repeated games, they are
(i) easily calculated, being completely characterized by two
simultaneous scalar equations; (ii) more general than static Nash, or even
Nash reversion; and even (iii) without loss in terms of total welfare,
at least in some cases, as in ours. See also Abreu, Pearce and
Stacchetti (1986) for the optimality of symmetric equilibria within a
standard oligopoly framework and Abreu, Pearce and Stacchetti (1993)
for a motivation of the solution concept based on a notion of equal
bargaining power. Cronshaw and Luenberger (1994) conduct a more
general analysis for repeated games with perfect monitoring, showing
how the set of SSE payoffs can be obtained by solving for the largest
scalar solving a certain equation. Hence, our paper shows that
properties (i)--(iii) extend to bandit games, with ``Markov perfect''
replacing ``Nash'' in statement (ii) and ``functional'' replacing ``scalar'' in (i): as mentioned above, a pair of functional equations
replaces the usual Hamilton-Jacobi-Bellman (HJB) (or Isaacs) equation from optimal control.

The paper is organized as follows.
Section \ref{sec:model} introduces the model.
Section \ref{sec:continuous-time} characterizes the efficient solution when actions can be chosen in continuous time and shows that MPEs cannot achieve efficiency.
Section \ref{sec:discrete-time} presents the game in which actions can
only be adjusted at regularly spaced points in time, the discrete-time game or discrete game for short.
Section \ref{sec:result} contains the main results regarding the set of equilibrium payoffs in the discrete game as the time between consecutive choices tends to zero.
Section \ref{sec:construction} is devoted to the construction of SSE in the discrete game.
Section \ref{sec:functional-equations} studies functional equations
that characterize SSE payoffs in both the discrete game and the continuous-time limit.
Section \ref{sec:conclu} concludes the paper.
Appendix \ref{app:auxiliary} presents auxiliary results on the evolution of beliefs and on various payoff functions.
The proofs of all other results are relegated to Appendix \ref{app:proofs}.

\section{The Model
\label{sec:model}}
Time $t \in [0,\infty)$ is continuous.
There are $N \geq 2$ players,
each facing the same two-armed bandit problem with one safe and one risky arm.

The safe arm generates a known constant payoff $s > 0$ per unit of time.
The distribution of the payoffs generated by the risky arm depends on the state of the world, $\theta \in \{0,1\}$, which nature draws at the outset with $\mathbb{P}\left[ \theta = 1 \right] = p$. Players do not observe $\theta$, but they know $p$.
They also understand that the evolution of the risky payoffs depends on $\theta$.
Specifically, the payoff process $X^n$ associated with player $n$'s risky arm evolves according to
$$
dX^n_t = \alpha_\theta\,dt+\sigma\,dZ^n_t + h\,dN^n_t,
$$
where
$Z^n$ is a standard Wiener process,
$N^n$ is a Poisson process with intensity $\lambda_\theta$,
and the scalar parameters $\alpha_0, \alpha_1, \sigma, h, \lambda_0, \lambda_1$ are known to all players.
Conditional on $\theta$, the processes $Z^1,\ldots,Z^N,N^1,\ldots,N^N$ are independent.
As $Z^n$ and $N^n - \lambda_\theta t$ are martingales, the expected payoff increment from using the risky arm over an interval of time $[t, t + dt)$ is
$m_\theta \, dt$ with $m_\theta = \alpha_\theta + \lambda_\theta h$.

Players share a common discount rate $r>0$.
We write $k_{n,t} = 0$ if player $n$ uses the safe arm at time $t$ and $k_{n,t} = 1$ if the player uses the risky arm at time $t$.
Given actions $(k_{n,t})_{t \geq 0}$ such that $k_{n,t} \in \{0,1\}$ is measurable with respect to the information available at time $t$,
player $n$'s total expected discounted payoff, expressed in per-period units, is
$$
\Exp \left[ \int_0^\infty r e^{-rt} \left[(1-k_{n,t}) s + k_{n,t} m_\theta\right] \, dt \right],
$$
where the expectation is over both the random variable $\theta$ and the stochastic process $(k_{n,t})$.\footnote{
Note that we have not yet defined the set of strategies available to each player and hence are silent at this point on how the players' strategy profile actually induces a stochastic process of actions
$(k_{n,t})_{t \geq 0}$ for each of them.
We will close this gap in two different ways in Sections \ref{sec:continuous-time} and \ref{sec:discrete-time}: by imposing Markov perfection in the former and a discrete time grid of revision opportunities in the latter.}

We make the following assumptions:
(i) $m_0 < s < m_1$, so each player prefers the risky arm to the safe
arm in state $\theta=1$ and prefers the safe arm to the risky arm in state $\theta=0$.
(ii) $\sigma > 0$ and $h > 0$, so the Brownian payoff component is always present and jumps of the Poisson component entail positive lump-sum payoffs;
(iii) $\lambda_1 \geq \lambda_0 \geq 0$, so jumps are at least as frequent in state $\theta=1$ as in state $\theta=0$.

Players begin with a common prior belief about $\theta$, given by the probability $p$ with which nature draws state $\theta = 1$.
Thereafter, they learn about this state in a Bayesian fashion by observing one another's actions and payoffs; in particular, they hold common posterior beliefs throughout time.
A detailed description of the evolution of beliefs is presented in Appendix \ref{app:beliefs}.
When $\lambda_1=\lambda_0$ (and hence $\alpha_1 > \alpha_0$), the
arrival of a lump-sum payoff contains no information about the state
of the world, and our setup is equivalent to that in Bolton and Harris
(1999), with the learning being driven entirely by the Brownian payoff component.
When $\alpha_1=\alpha_0$ (and hence $\lambda_1 > \lambda_0$), the
Brownian payoff component contains no information, and our setup is
equivalent to that in \KRC\ or \KR, depending on whether $\lambda_0 =
0$ or $\lambda_0 > 0$, with the learning being driven entirely by the arrival of lump-sum payoffs.\footnote{
\KRC\ and \KR\ consider compound Poisson processes where the distribution of lump-sum payoffs (and their mean $h$) at the time of a Poisson jump is independent of, and hence uninformative about, the state of the world.
By contrast, Cohen and Solan (2013) allow for L\'{e}vy processes where
the size of lump-sum payoffs contains information about the state, but
a lump sum of any given size arrives weakly more frequently in state $\theta=1$.}

\section{Efficiency and Markov Perfect Equilibria in Continuous Time
\label{sec:continuous-time}}

The authors cited in the previous paragraph assume that players use pure
Markov strategies in continuous time with the posterior belief as the state variable,
so that $k_{n,t}$ is a time-invariant deterministic function of the probability $p_t$ assigned to state $\theta = 1$ at time $t$.\footnote{
In the presence of discrete payoff increments, one actually has to
take the left limit $p_{t-}$ as the state variable, owing to the
informational constraint that the action chosen at time $t$ cannot
depend on the arrival of a lump sum at $t$.
In the following, we simply write $p_t$ with the understanding that the left limit is meant whenever this distinction is relevant.
Note that $p_{0-}=p_0$ by convention.}
In this section, we show how some of their
insights generalize to the present setting.
First, we present the efficient benchmark.
Second, we show that efficient behavior cannot be sustained as an MPE.

Consider a planner who maximizes the \emph{average} of the players' expected payoffs in continuous time by selecting an entire action profile $(k_{1,t},\ldots,k_{N,t})$ at each time $t$.
The corresponding average expected payoff increment is
$$
\left[ \left( 1 - \frac{K_t}{N} \right) s + \frac{K_t}{N} m_\theta \right] dt \qquad \text{with} \qquad K_t = \sum_{n=1}^{N} k_{n,t}.
$$
A straightforward extension of the main results of Cohen and Solan (2013) shows that the evolution of beliefs also depends on $K_t$ only\footnote{Cf.\ Appendix \ref{app:beliefs}.}
and that the planner's value function, denoted by $V_N^*$, has the following properties.

First, $V_N^*$ is the unique once-continuously differentiable solution of the HJB equation
$$
v(p) = s + \max_{K \in \{0,1, \ldots, N\}} K \left[b(p,v) - \frac{c(p)}{N} \right]
$$
on the open unit interval subject to the boundary conditions $v(0) = m_0$ and $v(1) = m_1$.
Here,
\begin{equation} \label{eq:b}
b(p,v)
= \frac{\rho}{2r}p^2(1-p)^2 v''(p)
- \frac{\lambda_1-\lambda_0}{r} \, p(1-p) \, v'(p)
+ \frac{\lambda(p)}{r} \, \left[ v(j(p)) - v(p) \right]
\end{equation}
can be interpreted as the expected informational benefit of using the risky arm when continuation payoffs are given by a (sufficiently regular) function $v$.\footnote{
Up to division by $r$, this is the infinitesimal generator of the process of posterior beliefs for $K = 1$, applied to the function $v$; cf.\ Appendix \ref{app:beliefs} for details.}
The first term on the right-hand side of \eref{eq:b} reflects Brownian learning, with
$$
\rho = \frac{(\alpha_1-\alpha_0)^2}{\sigma^2}
$$
representing the signal-to-noise ratio for the continuous payoff component.
The second term captures the downward drift in the belief when no
Poisson lump sum arrives.
The third term expresses the discrete change in the overall payoff
once such a lump sum arrives, with the belief jumping up from $p$ to
$$
j(p) = \frac{\lambda_1 p}{\lambda(p)};
$$
this occurs at the expected rate
$$
\lambda(p) = p \lambda_1 + (1-p) \lambda_0.
$$
The function
$$
c(p) = s - m(p)
$$
captures the opportunity cost of playing the risky arm in terms of expected current payoff forgone;
here,
$$
m(p) = p m_1 + (1-p) m_0
$$
denotes the risky arm's expected flow payoff given the belief $p$.
Thus, the planner weighs the shared opportunity cost of each experiment on the risky arm against the learning benefit, which accrues fully to each agent because of the perfect informational spill-over.

Second, there exists a cutoff $p_N^*$ such that all agents using the safe arm $(K = 0$) is optimal for the planner when $p \leq p_N^*$, and all agents using the risky arm ($K = N$) is optimal when $p > p_N^*$.
This cutoff is given by
$$
p_N^* = \frac{\mu_N (s - m_0)}{(\mu_N + 1) (m_1 - s) + \mu_N (s - m_0)}\,,
$$
where $\mu_N$ is the unique positive solution of the equation
$$
\frac{\rho}{2}\mu(\mu+1)+(\lambda_1-\lambda_0)\mu+\lambda_0\left(\frac{\lambda_0}{\lambda_1}\right)^\mu-\lambda_0-\frac{r}{N} = 0.
$$
Both $\mu_N$ and $p_N^*$ increase in $r/N$.
Thus, the interval of beliefs for which all agents using the risky arm is efficient widens with the number of agents and their patience.

Third, the value function satisfies $V_N^*(p)=s$ for $p \leq p_N^*$, and
\begin{equation}\label{eq:coopval}
V_N^*(p)=m(p)+\frac{c(p_N^*)}{u(p_N^*;\mu_N)}\ u(p;\mu_N) > s
\end{equation}
for $p > p_N^*$, where
$$
u(p;\mu) = (1-p) \left(\frac{1-p}{p}\right)^\mu
$$
is strictly decreasing and strictly convex for $\mu > 0$. The function
$V_N^*$ is strictly increasing and strictly convex on $[p_N^*,1]$.

By setting $N=1$, one obtains the single-agent value function $V_1^*$ and corresponding cutoff $p_1^* > p_N^*$.

Now consider $N \geq 2$ players acting noncooperatively.
Suppose that each of them uses a Markov strategy with the common belief as the state variable.
As in Bolton and Harris (1999), \KRC\ and \KR, the HJB equation for
player $n$ when he faces opponents who use Markov strategies is given by
$$
v_n(p) = s + K_{\neg n}(p) b(p,v_n)+ \max_{k_n \in \{0,1\}} k_n \left[b(p,v_n) - c(p) \right],
$$
where $K_{\neg n}(p)$ is the number of $n$'s opponents that use the risky arm.
That is, when playing a best response, each player weighs the
opportunity cost of playing risky against his own informational benefit only.
Consequently, $V_N^*$ does not solve the above HJB equation when player $n$'s opponents use the efficient strategy.
Efficient behavior therefore cannot be sustained in MPE.

To obtain existence of a symmetric MPE, the above authors actually allow the players to allocate one unit of a perfectly divisible resource across the two arms at each point in time, so the fraction allocated to the risky arm can be $k_{n,t} \in [0,1]$.
The symmetric MPE is unique and has all players play safe on an interval $[0,\ptilde_N]$ with $p_N^* < \ptilde_N < p_1^*$, play risky on an interval $[p_N^\dagger,1]$ with $p_N^\dagger > \ptilde_N$, and use an interior allocation on $(\ptilde_N,p_N^\dagger)$; see Keller and Rady (2010, Proposition 4), for example.
An adaptation of the proof of that proposition yields the same result for the payoff processes that we consider here.\footnote{Details are available from the authors on request.}

Figure 1 illustrates the payoff function $\tilde{V}_N$ of the symmetric MPE together with the cooperative value function $V_N^*$ and the single-agent value function $V_1^*$ for the parameters $(r,s,\sigma,\alpha_1,\alpha_0,h,\lambda_1,\lambda_0,N)=(1,1,1,0.1,0,1.5,1,0.2,5)$,
implying $\rho = 0.01$, $m_1 = 1.6$, $m_0 = 0.3$ and $(p_N^*,\ptilde_N,p_1^*,p_N^\dagger) \simeq (0.27,0.40,0.45,0.53)$.
The comparatively large gap between $\tilde{V}_N$ and $V_N^*$ reflects the double inefficiency of the MPE:
it not only involves a higher cutoff (hence, an earlier stop to all use of the risky arms) but also entails too low an intensity of experimentation on an intermediate range of beliefs.

\begin{figure}[h]
\centering
\begin{picture}(175.00,100.00)(0,0)
\put(10,00){\scalebox{1.45}{\includegraphics{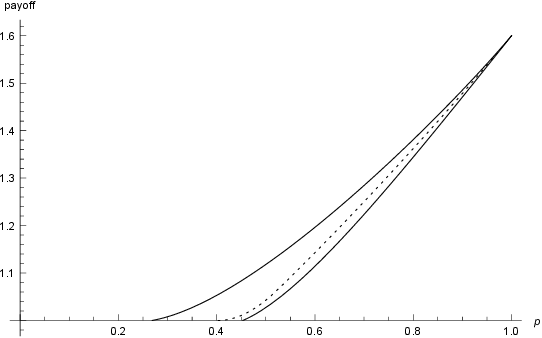}}}
\end{picture}
\begin{quote}
\caption{Payoffs $V_N^*$ (upper solid curve), $\tilde{V}_N$ (dotted) and $V^*_1$ (lower solid curve) for
$(r,s,\sigma,\alpha_1,\alpha_0,h,\lambda_1,\lambda_0,N)=(1,1,1,0.1,0,1.5,1,0.2,5)$.
}
\end{quote}
\end{figure}

\section{The Discrete Game
\label{sec:discrete-time}}

Henceforth, we restrict players to changing their actions $k_{n,t} \in \{0,1\}$
only at the times $t=0, \Delta, 2 \Delta, \ldots$ for some fixed $\Delta > 0$.
This yields a discrete-time game evolving in a continuous-time framework; in particular, the payoff processes are observed continuously.\footnote{ \label{fn:discretization}
While arguably natural, our discretization remains nonetheless \textit{ad hoc}, and other discretizations might yield other results.
Not only is it well known that the limits of the solutions of the discrete-time models
might differ from the continuous-time solutions, but the particular
discrete structure might also matter; see, among others, M\"{u}ller (2000), Fudenberg and Levine (2009), H\"{o}rner and Samuelson (2013), and Sadzik and Stacchetti (2015).
In H\"{o}rner and Samuelson (2013), for instance, there are multiple solutions to the optimality equations, corresponding to different boundary conditions, and to select among them, it is necessary to investigate in detail the discrete-time game (see their Lemma 3).
However, the role of the discretization goes well beyond selecting the ``right'' boundary condition; see Sadzik and Stacchetti (2015).  }
Moreover, we allow for non-Markovian strategies.

The expected discounted payoff increment from using the safe arm for the length of time $\Delta$ is
$\int_0^\Delta r\, e^{-r\,t}\, s \, dt = (1-\delta) s$
with $\delta = e^{-r\,\Delta}$.
Conditional on $\theta$, the expected discounted payoff increment from using the risky arm is
$\int_0^\Delta r\, e^{-r\,t}\, m_\theta \, dt = (1-\delta) m_\theta$.
Given the probability $p$ assigned to $\theta=1$, the expected discounted payoff increment from the risky arm conditional on all available information is
$(1-\delta) m(p)$.

A history of length $t=\Delta,2\Delta,\ldots$ is a sequence
$$
h_t = \left(
\big(k_{n,0},\widetilde Y^n_{[0,\Delta)}\big)_{n=1}^N,
\big(k_{n,\Delta},\widetilde Y^n_{[\Delta,2\Delta)}\big)_{n=1}^N,
\ldots,
\big(k_{n,t-\Delta},\widetilde Y^n_{[t-\Delta,t)}\big)_{n=1}^N
\right),
$$
where $k_{n,\ell \Delta}=1$ if player $n$ uses the risky arm on the time interval $[\ell \Delta, (\ell+1)\Delta)$;
$k_{n,\ell\Delta}=0$ if player $n$ uses the safe arm on this interval;
$\widetilde Y^n_{[\ell \Delta, (\ell+1)\Delta)}$ is the observed sample path $Y^n_{[\ell \Delta, (\ell+1)\Delta)}$ on the interval $[\ell \Delta, (\ell+1)\Delta)$ of the payoff process associated with player $n$'s risky arm if $k_{n,\ell \Delta}=1$;
and $\widetilde Y^n_{[\ell \Delta, (\ell+1)\Delta)}$ equals the empty set if $k_{n,\ell \Delta}=0$.
We write $H_t$ for the set of all histories of length $t$, set $H_0=\{\emptyset\}$, and let $H = \bigcup_{t=0,\Delta,2\Delta,\ldots}^\infty H_t$.

In addition, we assume that players have access to a public randomization device in every period, namely, a draw from the uniform distribution on $[0,1]$, which is assumed to be independent of $\theta$ and across periods. Following standard practice, we omit its realizations from the description of histories.

A behavioral strategy $\sigma_n$ for player $n$ is a sequence $(\sigma_{n,t})_{t=0,\Delta,2\Delta,\ldots}$, where $\sigma_{n,t}$ is a measurable map from $H_t$ to the set of probability distributions on $\{0,1\}$;
a pure strategy takes values in the set of degenerate distributions only.

Along with the prior probability $p$ assigned to $\theta = 1$, each profile of strategies induces a distribution over $H$.
Given his opponents' strategies $\sigma_{-n}$, player $n$ seeks to maximize
$$
(1-\delta)\, \Exp^{\ \sigma_{-n},\sigma_n}\left[\sum_{\ell=0}^\infty \delta^\ell \left\{ \astrut{2.75}
[1-k_{n,\ell\Delta}] s
+ k_{n,\ell\Delta} m_\theta
\right\}\right].
$$
By the law of iterated expectations, this equals
$$
(1-\delta)\, \Exp^{\ \sigma_{-n},\sigma_n}\left[\sum_{\ell=0}^\infty \delta^\ell \left\{ \astrut{2.75}
[1-k_{n,\ell\Delta}] s
+ k_{n,\ell\Delta} m(p_{\ell\Delta})
\right\}\right].
$$

Nash equilibrium, PBE and MPE, with actions after history $h_t$ depending only on the associated posterior belief $p_t$, are defined in the usual way.
Imposing the standard ``no signaling what you don't know'' refinement, beliefs are pinned down after all histories, on and off path.\footnote{
While we could equivalently define this Bayesian game as a stochastic game with the common posterior belief as a state variable, no characterization or folk theorem applies to our setup, as the Markov chain (over consecutive states) does not satisfy the sufficient ergodicity assumptions;
see Dutta (1995) and H\"{o}rner, Sugaya, Takahashi and Vieille (2011).
}

An SSE is a PBE in which all players use the same strategy: $\sigma_n(h_t)=\sigma_{n'}(h_t)$ for all $n,n'$ and $h_t \in H$.
This implies symmetry of behavior after \emph{any} history, not just on the equilibrium path of play. By definition, any symmetric MPE is an SSE, and any SSE is a PBE.

\section{Main Results}
\label{sec:result}

Fix $\Delta > 0$.
For $p \in [0,1]$, let $\WsupPBE(p)$ and $\WinfPBE(p)$ denote the supremum and infimum, respectively, of the set of average payoffs (per player) over all PBE, given prior belief $p$.
Let $\WsupSSE(p)$ and $\WinfSSE(p)$ be the corresponding supremum and infimum over all SSE. If such equilibria exist,
\begin{equation}\label{eq:payoffs}
\WsupPBE(p) \ge \WsupSSE(p) \ge \WinfSSE(p) \ge \WinfPBE(p).
\end{equation}
Given that we assume a public randomization device, these upper and lower bounds define the corresponding equilibrium average payoff sets.

As any player can choose to ignore the information contained in the
other players' experimentation results, the value function
$W_1^\Delta$ of a single agent experimenting in isolation constitutes
a lower bound on a player's payoff in any PBE.
Lemma \ref{lem:convergence-single-agent} establishes that this lower bound converges to $V_1^*$ as $\Delta \rightarrow 0$. Hence, we obtain a lower bound to the limits of all terms in \eqref{eq:payoffs}, namely $\liminf_{\Delta\rightarrow 0}\WinfPBE \geq V_1^*$.

An upper bound is also easily found. As any discrete-time strategy profile is feasible for the continuous-time planner from the previous section,
it holds that $\WsupPBE \leq V_N^*$.

The main theorem provides an exact characterization of the limits of
all four functions. It requires introducing a new family of
payoffs. Namely, we define the players' common payoff in continuous time when they all use the
risky arm if, and only if, the belief exceeds a given threshold $\phat$. This
function admits a closed form that generalizes the first-best payoff $V_N^*$ (cf.\ Section \ref{sec:continuous-time}).
It is given by
\[
\VNphat(p)=m(p)+\frac{c(\hat{p})}{u(\hat{p};\mu_N)}\ u(p;\mu_N)
\]
for $p > \hat{p}$, and by $\VNphat(p)=s$ otherwise.
For $\phat=p_N^*$, $\VNphat$ coincides with the cooperative value function $V_N^*$.
For $\phat>p_N^*$, it satisfies $\VNphat < V_N^*$ on $(p_N^*,1)$, is continuous, strictly increasing and strictly convex on $[\hat{p},1]$, and continuously differentiable except for a convex kink at $\phat$.

\begin{thm}\label{thm}
{\rm (i)} There exists $\phat \in [p_N^*,p_1^*]$ such that
$$\lim_{\Delta \rightarrow 0} \WsupPBE  = \lim_{\Delta \rightarrow 0} \WsupSSE = \VNphat,$$
and
$$\lim_{\Delta \rightarrow 0} \WinfPBE = \lim_{\Delta \rightarrow 0} \WinfSSE = V_1^*,$$
uniformly on $[0,1]$.

{\rm (ii)} If $\rho > 0$, then $\phat = p_N^*$ (and hence $\VNphat= V_N^*$).

{\rm (iii)} If $\rho = 0$, then $\phat$ is the unique belief in $[p_N^*,p_1^*]$ satisfying
\begin{equation} \label{eq:phat}
N \lambda(\phat) \left[ \VNphat(j(\phat)) - s\right] -  (N-1) \lambda(\phat) \left[V_1^*(j(\phat)) - s \right]=rc(\phat);
\end{equation}
moreover,
$\hat{p} = p_N^*$ if, and only if, $j(p_N^*) \leq p_1^*$,
and
$\hat{p} = p_1^*$ if, and only if, $\lambda_0 = 0$.
\end{thm}

This result is proved in Section \ref{sec:construction}, where we construct SSEs that get arbitrarily close to the highest and lowest possible average PBE payoffs for sufficiently short discretization intervals.
Given the fundamental difference between learning from a Brownian component ($\rho > 0$) and learning from jumps only ($\rho = 0)$, we treat these scenarios separately: the former case is covered by Propositions \ref{prop:SSE-Brownian}--\ref{prop:limit-Brownian} in Section \ref{sec:construction-Brownian}, the latter by Propositions \ref{prop:thresh}--\ref{prop:limit-Poisson} in Section \ref{sec:construction-Poisson}.
Pure Poisson learning needs two more intermediate results than the case with a Brownian component: one to identify the highest possible average PBE payoff in the frequent-action limit (Proposition \ref{prop:thresh}), another (Proposition \ref{prop:SSE-exponential}) to cover the case of fully conclusive Poisson news ($\rho = 0$ and $\lambda_0 = 0$), which requires a different approach to equilibrium construction than Brownian and inconclusive Poisson learning, respectively.

When $\rho > 0$ or $\lambda_0 > 0$, in fact, we can construct SSEs of the discrete game via two-state automata with a ``normal'' and a ``punishment'' state.
In the normal state, players are supposed to use the risky arm at all beliefs above some threshold $\pr$;
in the punishment state, the players are again supposed to use a cutoff strategy, but with a higher threshold $\pp$.
The idea here is that the normal state has all players experiment over as large a range of beliefs as possible,
whereas the punishment state has all players refrain from experimentation---and thus from the production of valuable information---at all beliefs except the most optimistic ones.
A unilateral deviation in the normal state triggers a transition to the punishment state; otherwise, the normal state persists.
The punishment state persists after a unilateral deviation there; otherwise, a public randomization device determines whether play reverts to the normal state.
When $\rho = 0$ and $\lambda_0 = 0$, by contrast, our proof relies on the existence of two symmetric mixed-strategy equilibria of the discrete game for beliefs close to the single-agent cutoff.
Choosing continuation play as a function of history in the appropriate way, we can then construct SSEs with suitable properties at higher beliefs.

Turning to part (i) of the theorem, the fact that the best SSE payoff and the best average PBE payoff coincide---and equal the payoff of a cutoff strategy---in the frequent-action limit is plausible (though not obvious) because efficiency in continuous time requires symmetric play of a cutoff strategy; cf.\ Section \ref{sec:continuous-time}.
As to the worst payoffs, the requisite punishments can also be implemented in a strongly symmetric fashion.
At a belief below the threshold $\pp$ in the punishment state of the above automaton, for instance, either everybody playing safe forever is already an equilibrium of the game, or a unilateral deviation to the risky arm provides a higher payoff.
In the latter case, the promise to revert to joint risky play at a later time serves to compensate the players for the flow payoff deficit that playing safe causes in the meantime.

Part (ii) of the theorem states that efficiency can be achieved in the frequent-action limit whenever the Brownian component of risky payoffs is informative about the true state ($\rho > 0$).
The reason is that the resulting diffusion component in the stochastic process of posterior beliefs is the dominant force in belief updating for small discretization steps $\Delta$.
To gain some intuition, consider the above two-state automaton with belief thresholds $\pr \in (p_N^*,p_1^*)$ and $\pp \in (p_1^*,1)$ in the normal and punishment state, respectively, and think of $\pp$ as being very close to 1, so that punishment essentially means autarky.
At a belief $p$ to the immediate right of $\pr$, a player contemplating a deviation from the risky to the safe arm in the normal state then faces the following trade-off.
On the one hand, the deviation saves the player the opportunity cost of experimentation, $(1-\delta) c(\pi)$, which is $O(\Delta)$ as $\Delta$ vanishes.
On the other hand, the deviation changes the lottery over continuation values to the player's disadvantage.
In fact, use of the safe arm triggers a transition to the punishment state, with an expected continuation value of at most the safe payoff level $s$ plus a term that is linear in $\Delta$.
This is because the probability that the opponents' experiments lift the posterior belief close to $p_1^*$ (the only scenario in which all experimentation does not stop for good) within the length of time $\Delta$ is $O(\Delta)$.\footnote{
In the absence of any lump-sum payoff, this is a consequence of Chebysheff's inequality; the probability of a lump-sum payoff within $\Delta$ units of time, moreover, is itself of order $\Delta$ for small $\Delta$.}
Staying with the risky arm, by contrast, would mean that continuation values above $s$ are always within immediate reach---no matter how small $\Delta$ becomes and even if one takes $p$ all the way down to $\pr$---implying an expected continuation value of at least $s$ plus a term that is $O(\Delta^\gamma)$ with $\gamma < 1$.\footnote{
The proof of Proposition \ref{prop:SSE-Brownian} shows that one can take $\gamma = \divn{3}{4}$.}
Roughly speaking, this is because the payoff function $V_{N,\pr}$ (to which continuation payoffs in the normal state converge as $\Delta$ vanishes) has a convex kink at $\pr$ and, owing to the diffusion part of the belief dynamics, the probability of reaching its upward-sloping part stays bounded away from 0 as $\Delta$ vanishes.
For sufficiently small $\Delta$, therefore, the loss in expected continuation value after a deviation from risky to safe play outweighs the saved opportunity cost, and the deviation is unprofitable.
As this argument works as long as $V_{N,\pr}$ has a convex kink, one can take $\pr$ arbitrarily close to $p_N^*$.
It thus follows that $\phat$, the infimum of possible thresholds $\pr$, equals $p_N^*$.\footnote{
This argument is reminiscent of the intuitive explanation for smooth pasting in stopping problems for diffusion processes as given in Dixit and Pindyck (1994), for example.}

We can therefore reinterpret Figure 1 as depicting the best and worst SSE and average PBE payoffs in the frequent-action limit for the given parameter values, with the payoffs of the symmetric MPE of the continuous-time game sandwiched in between them.

Part (iii) of the theorem characterizes $\phat$ when all updating is driven by the Poisson component of risky payoffs ($\rho = 0$).
The fundamental difference to Brownian learning (and the reason that asymptotic efficiency may be out of reach) is that the cost of a deviation is of the \emph{same} order in $\Delta$ as its benefit.
To see this, consider a two-state automaton with thresholds $\pr$ and $\pp$ as above.
In the normal state, the players' temptation to deviate to the safe arm is strongest when the belief is so close to $\pr$ that the lack of good news over a period of $\Delta$ makes the belief drop below $\pr$, and thus into a region where safe prevails in either state---whether a single player has deviated or not.
Absent a success in the current round, therefore, deviations cannot be punished in the future.
The cost of deviating thus arises only if good news arrives.
Starting out from $\pr$, this is expected to happen with probability $N \lambda(\pr) \Delta + o(\Delta)$ if no player deviates; a deviation reduces this probability to $(N-1) \lambda(\pr) \Delta + o(\Delta)$.
Without a deviation, moreover, a player's continuation payoff then amounts at most to the cooperative payoff---evaluated at the posterior belief after the news event---given no use of the risky arm below $\pr$; the resulting payoff improvement relative to the case that no news arrives converges to $\VNpr(j(\pr)) - s$ as $\Delta$ vanishes.
In the event of a deviation, the continuation payoff is at least the single-player payoff, and the corresponding payoff increment converges to $V_1^*(j(\pr)) - s$.
The cost of the deviation in terms of expected continuation value foregone is at most
$
\left\{ N \lambda(\pr) [\VNpr(j(\pr)) - s] - (N-1) \lambda(\pr) [V_1^*(j(\pr)) - s] \right\} \Delta + o(\Delta),
$
therefore.
A necessary condition for equilibrium is that this again exceed the saved opportunity cost of playing risky, $(1-\delta) c(\pi) = r c(\pr) \Delta + o(\Delta)$.
At the infimum of possible thresholds $\pr$, the leading (that is, first-order) terms in the cost and benefit of a deviation are just equalized, hence the equation \eref{eq:phat} for $\phat$.

Asymptotic efficiency means that $\phat$ coincides with the efficient continuous-time cutoff $p_N^*$,
which equates the \emph{social} benefit of an experiment with its opportunity cost.
When $\rho = 0$, this benefit is $N \lambda(p_N^*) [V_N^*(j(p_N^*)) - s]$ as the experiment contributes to the arrival of news at rate $\lambda(p_N^*)$, with all $N$ players then reaping the gain $V_N^*(j(p_N^*)) - s$;
this is also the first term on the left-hand side of \eref{eq:phat} when $\phat = p_N^*$ and hence $\VNphat = V_N^*$.
The opportunity cost is $r c(p_N^*)$, the term on the right-hand side of \eref{eq:phat}.
So, $p_N^*$ solves \eref{eq:phat} if, and only if, the second term on the left-hand side of \eref{eq:phat} vanishes at $\phat = p_N^*$.
As $V_1^* = s$ on $[0,p_1^*]$, this is tantamount to the inequality $j(p_N^*) \le p_1^*$.
Intuitively, this condition means that a deviation from efficient play can be punished by a complete stop to all experimentation---even after good news.
A player's incentives are then perfectly aligned with the social planner's: the individual action effectively dictates the collective action choice.

When $\rho = 0$ and $\lambda_0 = 0$, finally, the arrival of good news freezes the belief at 1, and the resulting cooperative and single-player payoffs both equal $m_1 = \lambda_1 h$.
In this case, \eref{eq:phat} reduces to $\lambda(\phat) [m_1 - s] = r c(\phat)$, which equates the benefit of an experiment to a \emph{single} agent with its opportunity cost.
Hence, the solution is $\phat = p_1^*$.
The intuition is straightforward: when a player's continuation payoffs coincide with those of a single agent whether he deviates or not, it is impossible to sustain experimentation below the single-agent cutoff.

Figure 2 shows the cooperative value function $V_N^*$,
the supremum $\VNphat$ and infimum $V_1^*$ of the limit SSE and average PBE payoffs,
and the payoff function $\tilde{V}_N$ of the symmetric continuous-time MPE for a parameter configuration that implies $p_N^* < \phat < p_1^*$.
For $\rho = 0$ and $(r,s,h,\lambda_1,\lambda_0,N)=(1,1,1.5,1,0.2,5)$, we indeed have $(p_N^*,\phat,\ptilde_N,p_1^*,p_N^\dagger)\simeq (0.270,0.399,0.450,0.455,0.571)$.

\begin{figure}[h]
\centering
\begin{picture}(175.00,100.00)(0,0)
\put(10,00){\scalebox{1.45}{\includegraphics{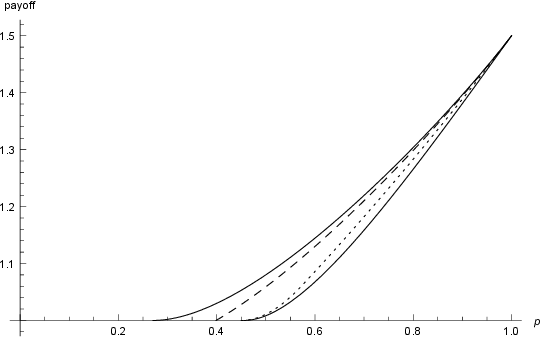}}}
\end{picture}
\begin{quote}
\caption{Payoffs $V_N^*$ (upper solid curve), $\VNphat$ (dashed), $\tilde{V}_N$ (dotted) and $V^*_1$ (lower solid curve) for $\rho = 0$ and $(r,s,h,\lambda_1,\lambda_0,N)=(1,1,1.5,1,0.2,5)$.}
\end{quote}
\end{figure}

The figure suggests that, even when the first-best is not achievable in the frequent-action limit, the best SSE performs strictly better than the symmetric continuous-time MPE, maintaining experimentation---at maximal intensity---on a larger set of beliefs.
The following result confirms this ordering of belief thresholds.
It is formulated for pure Poisson learning with inconclusive news since $\phat = \ptilde_N = p_1^*$ when $\rho = 0$ and $\lambda_0 = 0$.

\begin{prop}\label{prop:comparison_sym_MPE}
Let $\rho = 0$ and $\lambda_0>0$. Then $\phat < \ptilde_N$.
\end{prop}

Figure 2 further shows that relative to the symmetric MPE of the continuous-time game, the best SSE internalizes a much larger part of the informational externality that the players exert on each other.
If we take the payoff difference $V_N^*(p)-V_1^*(p)$ as a measure of the size of that externality,
we can interpret the ratios
$[\VNphat(p)-V_1^*(p)]/[V_N^*(p)-V_1^*(p)]$
and
$[\tilde{V}(p)-V_1^*(p)]/[V_N^*(p)-V_1^*(p)]$
as the fraction of the externality that is internalized by the best SSE and the symmetric MPE, respectively.\footnote{
We set both ratios to 100\% whenever the denominator vanishes, that is, whenever it is efficient for all players to play safe.}
At $\ptilde_N$, for example, the latter ratio is 0\% whereas the former is 52.5\% for the parameters underlying Figure 2---
in a scenario with $\phat = p_N^*$, it would even be 100\%.

It was said in Section \ref{sec:continuous-time} that the efficient cutoff $p_N^*$ decreases with the number of players.
For $\rho = 0 $ and $\lambda_0=0$, the threshold $\phat$ is independent of $N$; for inconclusive news, however, it behaves like $p_N^*$.

\begin{prop}\label{prop:comp-statics-N}
For $\rho = 0$ and $\lambda_0>0$, $\hat{p}$ is decreasing in $N$.
\end{prop}

The last result of this section characterizes the area (in the $(\lambda_1, \lambda_0)$-plane) where asymptotic efficiency obtains under pure Poisson learning.

\begin{prop} \label{prop:efficiency-region-Poisson}
Let $\rho = 0$.
Then, $j(p_N^*) > p_1^*$ whenever $\lambda_0 \leq \lambda_1/N$.
On any ray in $\mathbbm{R}_+^2$ emanating from the origin $(0,0)$ with a slope strictly between $1/N$ and 1, there is a unique critical point $(\lambda_1^*,\lambda_0^*)$ at which $j(p_N^*) = p_1^*$;
moreover, $j(p_N^*) > p_1^*$ at all points of the ray that are closer to the origin than $(\lambda_1^*,\lambda_0^*)$,
and $j(p_N^*) < p_1^*$ at all points that are farther from the origin than $(\lambda_1^*,\lambda_0^*)$.
These critical points form a continuous curve that is bounded away from the origin and asymptotes to the ray of slope $1/N$.
The curve shifts downward as $r$ falls or $N$ rises.
\end{prop}

This result is illustrated in Figure 3.

\setlength {\unitlength} {1mm}
\begin{figure}[t]
\begin{picture}(175.00,100.00)(0,0)
\put(30,00){\scalebox{.5}{\includegraphics{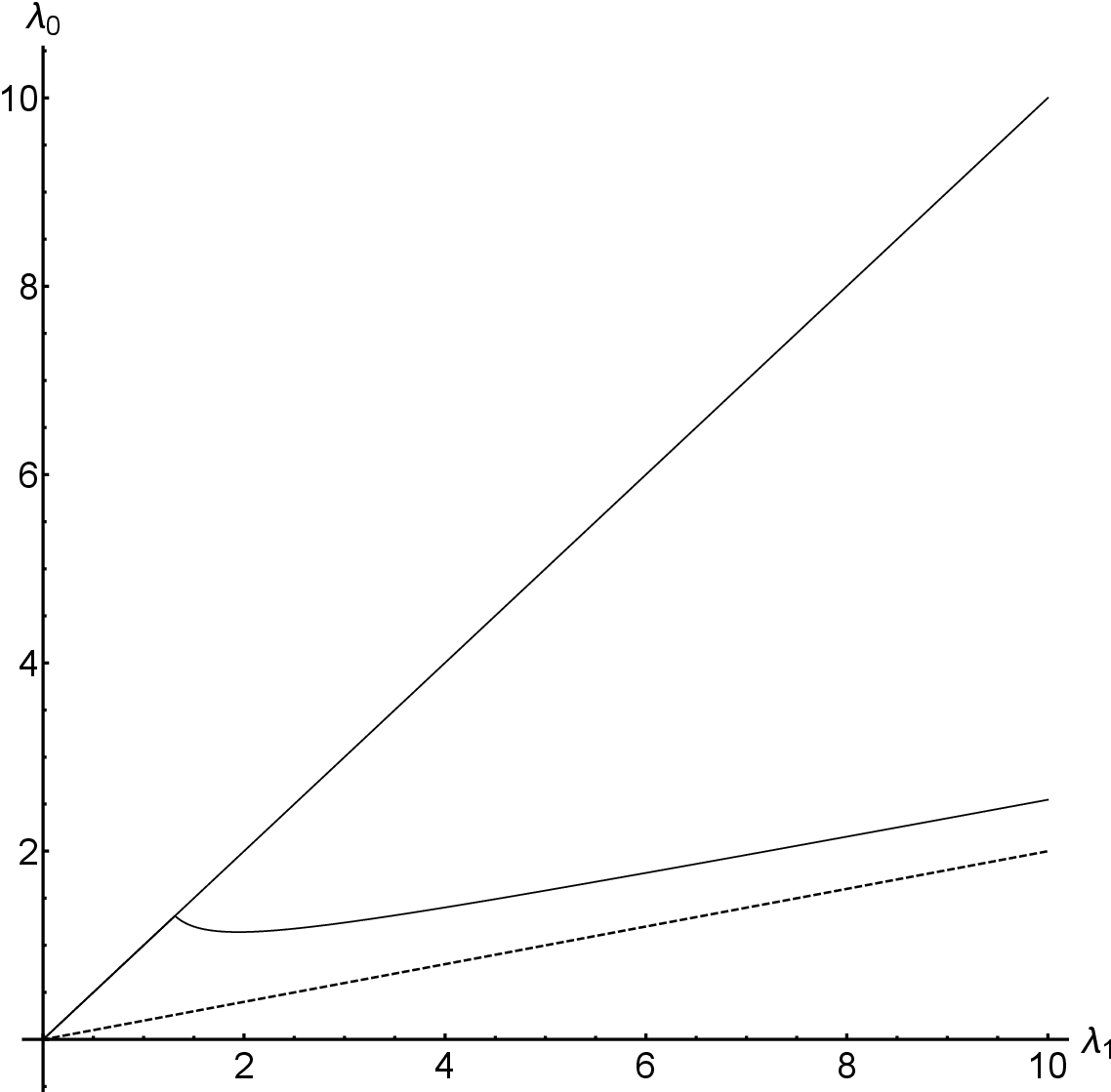}}}
\put(100,40){\makebox(0,0)[cc]{$j(p_N^*) < p_1^*$}}
\put(100,10){\makebox(0,0)[cc]{$j(p_N^*) > p_1^*$}}
\end{picture}
\begin{quote}
\caption{
Asymptotic efficiency is achieved for parameter combinations $(\lambda_1,\lambda_0)$ between the diagonal and the curve but not below the curve. The dashed line is the ray of slope $1/N$.
Parameter values: $r=1$, $N=5$.
}
\end{quote}
\end{figure}

As is intuitive, having more players, or more patience, increases the scope for the first-best.
When $r \to 0$, the curve in Figure 3 converges to the ray of slope $1/N$; given Poisson rates $(\lambda_1,\lambda_0)$, therefore, asymptotic efficiency can always be achieved with a sufficiently large number of players.
When $r \to \infty$, by contrast, the curve in Figure 3 shifts ever higher: as myopic players do not react to future rewards and punishments,
it is no surprise that asymptotic efficiency cannot be attained then.

\section{Construction of Equilibria}
\label{sec:construction}

We first consider the case of a diffusion component (Section \ref{sec:construction-Brownian}) and then turn to the case of pure jump processes (Section \ref{sec:construction-Poisson}).

We need the following notation.
Let $F_K^\Delta(\cdot\vert p)$ denote the cumulative distribution function of the posterior belief $p_\Delta$ when $p_0 = p$ and $K$ players use the risky arm on the time interval $[0,\Delta)$.
For any measurable function $w$ on $[0,1]$ and  $p \in [0,1]$, we write
$$
\E_K w(p) = \int_0^1 w(p') \, F_K^\Delta(dp'\vert p),
$$
whenever this integral exists.
Thus, $\E_K w(p)$ is the expectation of $w(p_\Delta)$ given the prior $p$ and $K$ experimenting players.

\subsection{Learning with a Brownian Component ($\rho > 0$)
\label{sec:construction-Brownian}}

For a sufficiently small $\Delta > 0$, we specify an SSE that can be summarized by two functions,
$\kr$ 
and
$\kp$, 
which do not depend on $\Delta$.
The equilibrium strategy is characterized by a two-state automaton.
In the ``good'' state, play proceeds according to $\kr$, and the equilibrium payoff satisfies
\begin{equation} \label{eq:SSE-good-state}
\wr(p) = (1-\delta)[(1-\kr(p))s + \kr(p) m(p)]+ \delta \E_{N\kr(p)} \wr(p),
\end{equation}
while in the ``bad'' state, play proceeds according to $\kp$, and the payoff satisfies
\begin{equation} \label{eq:SSE-bad-state}
\wp(p) = \max_{k \in \{0,1\}} \left\{\astrut{2.5}
         (1-\delta)[(1-k)s + k m(p)] + \delta \E_{(N-1)\kp(p)+k} \wp(p)
         \right\}.
\end{equation}
That is, $\wp$ is the value from the best response to all other players following $\kp$.

A unilateral deviation from $\kr$ in the good state is punished by a transition to the bad state in the following period; otherwise, we remain in the good state.
If there is a unilateral deviation from $\kp$ in the bad state, we remain in the bad state.
Otherwise, a draw of the public randomization device determines whether the state next period is good or bad; this probability is chosen such that the expected payoff is indeed given by $\wp$ (see below).

With continuation payoffs given by $\wr$ and $\wp$, the common action $\kappa \in \{0,1\}$
is incentive compatible at a belief $p$ if, and only if,
\begin{eqnarray} \label{eq:SSE-incentive-constraint}
\lefteqn{(1-\delta)[(1-\kappa)s + \kappa m(p)] + \delta \E_{N\kappa} \wr(p)} \\
& \geq & \astrut{3} (1-\delta)[\kappa s + (1-\kappa) m(p)] + \delta \E_{(N-1)\kappa + 1 - \kappa} \wp(p). \nonumber
\end{eqnarray}
Therefore, the functions $\kr$ and $\kp$ define an SSE if, and only if, \eref{eq:SSE-incentive-constraint} holds for $\kappa = \kr(p)$ and $\kappa = \kp(p)$ at all $p$.

The probability $\eta^\Delta(p)$ of a transition from the bad to the good state in the absence of a unilateral deviation from $\kp(p)$ is pinned down by the requirement that
\begin{eqnarray} \label{eq:SSE-randomization}
\wp(p) & = & (1-\delta)[(1-\kp(p))s + \kp(p) m(p)] \\
       &   & \astrut{3.5} \mbox{} + \delta \left\{ \astrut{2.5}
               \eta^\Delta(p)\, \E_{N\kp(p)} \wr(p)
               + [1-\eta^\Delta(p)]\, \E_{N\kp(p)} \wp(p)
             \right\}. \nonumber
\end{eqnarray}
If $k = \kp(p)$ is optimal in \eref{eq:SSE-bad-state}, we simply set $\eta^\Delta(p) = 0$.
Otherwise, \eref{eq:SSE-bad-state} and \eref{eq:SSE-incentive-constraint} imply
$$
\ \delta \E_{N\kr(p)} \wr(p)
\geq \wp(p) - (1-\delta)[(1-\kp(p))s + \kp(p) m(p)]
> \delta \E_{N\kp(p)} \wp(p),
$$
so \eref{eq:SSE-randomization} holds with
$$\eta^\Delta(p)
= \frac{\wp(p) - (1-\delta)[(1-\kp(p))s + \kp(p) m(p)] - \delta \E_{N\kp(p)} \wp(p)}
{\delta \E_{N\kp(p)} \wr(p) - \delta \E_{N\kp(p)} \wp(p)}
\in (0,1].$$

It remains to specify $\kr$ and $\kp$.
Let
$$
p^m = \frac{s-m_0}{m_1 - m_0}\,.
$$
As $m(p^m) = s$, this is the belief at which a myopic agent is indifferent between the two arms.
It is straightforward to verify that $p_1^* < p^m$.
Fixing $\pr \in (p_N^*,p_1^*)$ and $\pp \in (p^m,1)$, we let
$\kr(p)=\mathbbm{1}_{p > \pr}$
and $\kp(p)=\mathbbm{1}_{p > \pp}$.\footnote{
$\mathbbm{1}_A$ denotes the indicator function of the event $A$.}
Note that punishment and reward strategies coincide outside of $(\pr,\pp)$.
Also note that $\pp>p^m$, \textit{i.e.}, in the punishment state,
less experimentation is enforced than would be myopically optimal.
In fact, for Proposition \ref{prop:limit-Brownian}, we are letting $\pp\rightarrow 1$.
This way, we can, in a strongly symmetric way, exact the harshest conceivable punishment, which, for any given player, entails a vanishing information spill-over from the other players.
Indeed, this construction, in the limit, pushes players' payoffs down to their single-agent value $V_1^*$.
As players cannot be prevented from best-responding to the vanishing information spill-overs, it is not possible to push their continuation values below $V_1^*$.

\begin{prop} \label{prop:SSE-Brownian}
For $\rho > 0$, there are beliefs $p^\flat \in (p_N^*, p_1^*)$ and $p^\sharp \in (p^m,1)$ such that for all $\pr \in (p_N^*,p^\flat)$ and $\pp \in (p^\sharp,1)$,
there exists $\bar{\Delta} > 0$ such that for all $\Delta \in (0,\bar{\Delta})$, the two-state automaton with functions $\kr$ and $\kp$ defines an SSE of the experimentation game with period length $\Delta$.
\end{prop}

The proof consists of verifying that, for a sufficiently small $\Delta$, the actions $\kr(p)$ and $\kp(p)$ satisfy the incentive-compatibility constraint \eref{eq:SSE-incentive-constraint} at all $p$.
First, we find $\varepsilon>0$ small enough that $\wp=s$ in a neighborhood of $\pr+\varepsilon$.
The payoff functions $\wr$ and $\wp$ resulting from the two-state automaton are
then bounded away from one another on $[\pr+\varepsilon,\pp]$ for small $\Delta$.
In this range, therefore, the difference in expected continuation values across states does not vanish as $\Delta$ tends to 0, whereas the difference in current expected payoffs across actions is of order $\Delta$, rendering deviations unattractive for small enough $\Delta$.
On $(\pp,1]$ and $[0,\pr]$, $\kr$ and  $\kp$ both prescribe the myopically optimal action.
Given that continuation payoffs are weakly higher in the good state, it is easy to show that there are no incentives to deviate on these intervals.
For beliefs in $(\pr,\pr+\varepsilon)$, $\kp$ again prescribes the myopically optimal action.
The proof of incentive compatibility of $\kr$ on this interval crucially relies on the fact that, for small $\Delta$, $\wr$ is bounded below by $\VNulp$\hspace{0.05em}, which has a convex kink at $\pr$.
This, together with the fact that, conditional on no lump sum arriving, the log-likelihood ratio of posterior beliefs is Gaussian, allows us to demonstrate the existence of some constant $C_1 > 0$ such that,
for $\Delta$ small enough,
$
\E_N \wr(p) \geq s + C_1 \Delta^{\frac{3}{4}}
$
to the immediate right of $\pr$, whereas
$\E_{N-1}\wp(p) \leq s + C_0 \Delta$
with some constant $C_0>0$.
For small $\Delta$, therefore, the linearly vanishing current-payoff advantage of the safe over the risky arm is dominated by the incentives provided through continuation payoffs.

The next result is the last remaining step in the proof of Theorem \ref{thm} for the case $\rho>0$; it
essentially follows from letting $\pr\rightarrow p_N^*$ and $\pp\rightarrow 1$ in Proposition \ref{prop:SSE-Brownian}.

\begin{prop} \label{prop:limit-Brownian} For $\rho > 0$,
$\lim_{\Delta \rightarrow 0} \WsupSSE  = V_N^*$
and
$\lim_{\Delta \rightarrow 0} \WinfSSE = V_1^* $, uniformly on $[0,1]$.
\end{prop}

\subsection{Pure Poisson Learning ($\rho = 0$)
\label{sec:construction-Poisson}}

Let $\rho = 0$, and take $\phat$ as in
part (iii)
of Theorem \ref{thm}.

\begin{prop} \label{prop:thresh}
Let $\rho = 0$.
For any $\varepsilon > 0$, there is a $\Delta_\varepsilon > 0$ such that for all $\Delta \in (0,\Delta_\varepsilon)$, the set of beliefs at which experimentation can be sustained in a PBE of the discrete game with period length $\Delta$ is contained in the interval $(\hat{p}-\varepsilon,1]$.
In particular,
$\limsup_{\Delta \rightarrow 0} \WsupPBE(p) \leq \VNphat(p)$.
\end{prop}

For a heuristic explanation of the logic behind this result, consider a sequence of pure-strategy PBEs for vanishing $\Delta$ such that the infimum of the set of beliefs at which at least one player experiments converges to some limit $\ptilde$.
Selecting a subsequence of $\Delta$s and relabeling players, if necessary, we can assume without loss of generality that players $1,\ldots,L$ play risky immediately to the right of $\ptilde$, while players $L+1,\ldots,N$ play safe.
In the limit, players' individual continuation payoffs are bounded below by the single-agent value function $V_1^*$ and cannot sum to more than $N\VNplow$, so the sum of the continuation payoffs of players $1, \ldots, L$ is bounded above by $N\VNplow-(N-L)V_1^*$.
Averaging these players' incentive-compatibility constraints thus yields
$$
L\lambda(\ptilde) \left[ \frac{N \VNplow(j(\ptilde))-(N-L)V_1^*(j(\ptilde))}{L}- s \right] - rc(\ptilde)
\geq (L-1)\lambda(\ptilde) \left[ V_1^*(j(\ptilde))- s \right].
$$
Simplifying the left-hand side, adding $(N-L)\lambda(\ptilde) \left[ V_1^*(j(\ptilde))- s \right]$ to both sides and re-arranging, we obtain
$$
N\lambda(\ptilde) \left[ \VNplow(j(\ptilde))- s \right] - rc(\ptilde)
\geq (N-1)\lambda(\ptilde) \left[ V_1^*(j(\ptilde))- s \right],
$$
which in turn implies $\ptilde \geq \phat$, as we show in Lemma \ref{lem:thresh} in the appendix.
The proof of Proposition \ref{prop:thresh} makes this heuristic argument rigorous and extends it to mixed equilibria.

For non-revealing jumps ($\lambda_0 > 0$), the construction of SSEs that achieve the above bounds in the limit relies on the same two-state automaton as in Proposition \ref{prop:SSE-Brownian}, the only difference being that the threshold $\pr$ is now restricted to exceed $\phat$.

\begin{prop} \label{prop:SSE-Poisson}
Let $\rho = 0$ and $\lambda_0 >0$.
There are beliefs $p^\flat \in (\phat, p_1^*)$ and $p^\sharp \in (p^m,1)$ such that for all $\pr \in (\phat,p^\flat)$ and $\pp \in (p^\sharp,1)$,
there exists $\bar{\Delta} > 0$ such that for all $\Delta \in (0,\bar{\Delta})$, the two-state automaton with functions $\kr$ and $\kp$ defines an SSE of the experimentation game with period length $\Delta$.
\end{prop}

The strategy for the proof of this proposition is the same as that of Proposition \ref{prop:SSE-Brownian}, except for the belief region to the immediate right of $\pr$,
where incentives are now provided through terms of first order in $\Delta$,
akin to those in equation \eref{eq:phat}.

In the case $\lambda_0>0$, we are able to provide incentives in the
potentially last round of experimentation by threatening punishment
conditional on there being a success (that is, a successful
experiment). This option is no longer available in the case of
$\lambda_0=0$. Indeed, any success now takes us to a posterior of one, so that everyone plays risky forever after.
This means that, irrespective of whether a success occurs in that round, continuation strategies are independent of past behavior, conditional on the players' belief.
This raises the possibility of unravelling.
If incentives just above the candidate threshold at which players give up on the risky arm cannot be provided, can this threshold be lower than in the MPE?

To settle whether unravelling occurs requires us to study the discrete game in considerable detail.
We start by noting that for $\lambda_0 = 0$, we can strengthen Proposition \ref{prop:thresh} as follows:
there is no PBE with any experimentation at beliefs below the discrete-time single-agent cutoff $p_1^\Delta = \inf\{p\!: W_1^\Delta(p) > s\}$; see Heidhues et al. (2015).\footnote{
In particular, this excludes the possibility that the asymmetric MPE of \KRC\ with an infinite number of switches between the two arms below $p_1^*$ can be approximated in the discrete game.}
The highest average payoff that can be hoped for, then, involves all players experimenting above $p_1^\Delta$.

Unlike in the case of $\lambda_0>0$ (see Proposition \ref{prop:SSE-Poisson}), an explicit description of a two-state automaton implementing SSEs whose payoffs converge to the obvious upper and lower bounds appears elusive.
This is partly because equilibrium strategies are, as it turns out, necessarily mixed for beliefs that are arbitrarily close to (but above) $p_1^\Delta$.
The proof of the next proposition establishes that the length of the interval of beliefs for which this is the case vanishes as $\Delta \rightarrow 0$.
In particular, for higher beliefs (except for beliefs arbitrarily close to 1, when playing risky is strictly dominant), both pure actions can be enforced in some equilibrium.

\begin{prop}\label{prop:SSE-exponential}
Let $\rho = 0$ and $\lambda_0 = 0$.
For any beliefs $\pr$ and $\pp$ such that $p_1^* < \pr < p^m < \pp < 1$,
there exists a $\bar{\Delta} > 0$ such that for all $\Delta \in (0,\bar{\Delta})$, there exists
\begin{itemize}
\vspace{-2ex}
\item[-] an SSE in which, starting from a prior above $\pr$, all players
use the risky arm
on the path of play as long as the belief remains above $\pr$ and use
the safe arm for beliefs below $p_1^*$; and
\vspace{-1ex}
\item[-] an SSE in which, given a prior between $\pr$ and $\pp$, the players' payoff is no larger than their best-reply payoff against opponents who use the risky arm
if, and only if, the belief lies in $[p_1^*,\pr] \cup [\pp,1]$.
\end{itemize}
\end{prop}

While this is somewhat weaker than Proposition \ref{prop:SSE-Poisson}, its implications for limit payoffs as $\Delta \rightarrow 0$ are the same.
Intuitively, given that the interval $[p_1^*,\pr]$ can be chosen arbitrarily small (actually, of the order $\Delta$, as the proof establishes), its impact on equilibrium payoffs starting from priors above $\pr$ is of order $\Delta$.
This suggests that for the equilibria whose existence is stated in Proposition \ref{prop:SSE-exponential}, the payoff converges to the payoff from all players experimenting above $p_1^*$ and to the best-reply payoff against none of the opponents experimenting.
Indeed, we have the following result, covering both inconclusive and conclusive jumps.

\begin{prop} \label{prop:limit-Poisson} For $\rho = 0$,
$\lim_{\Delta \rightarrow 0} \WsupSSE  = \VNphat$
and
$\lim_{\Delta \rightarrow 0} \WinfSSE = V_1^* $, uniformly on $[0,1]$.
\end{prop}


\section{Functional Equations for SSE Payoffs}
\label{sec:functional-equations}

While it is possible to derive explicit solutions to the equilibrium payoff sets of interest, at least asymptotically, note that, already in the discrete game, a characterization in terms of optimality equations can be obtained, which defines the correspondence of SSE payoffs. As discussed in the introduction, these generalize the familiar equation characterizing the value function of the symmetric MPE. Instead of a single (HJB) equation, the characterization of SSE payoffs involves two coupled functional equations, whose solution delivers the highest and lowest equilibrium payoff. Proposition \ref{prop:chara-discrete} states this in the discrete game, while Proposition \ref{prop:chara-continuous} gives the continuous-time limit. As these propositions do not   heavily rely on the specific structure of our game, we believe that they might be useful for analyzing SSE payoffs for more general processes or other stochastic games.

Fix $\Delta > 0$.
For $p \in [0,1]$, let $\Wsup(p)$ and $\Winf(p)$ denote the supremum
and infimum, respectively, of the set of payoffs over
\emph{pure-strategy} SSEs, given prior belief $p$.\footnote{For the existence of various types of equilibria in discrete-time stochastic games, see Mertens, Sorin and Zamir (2015), Chapter 7.}
If such an equilibrium exists, these extrema are achieved, and $\Wsup(p) \ge \Winf(p)$.
For $\rho > 0$ or $\lambda_0 > 0$, we have shown in Sections \ref{sec:construction-Brownian}--\ref{sec:construction-Poisson} that in the limit as $\Delta \rightarrow 0$, the best and worst average payoffs (per player) over all PBEs are achieved by SSE in pure strategies.
The following result characterizes $\Wsup$ and $\Winf$ via a pair of coupled functional equations.

\begin{prop}\label{prop:chara-discrete}
Suppose that the discrete game with time increment $\Delta > 0$ admits a pure-strategy SSE for any prior belief.
Then, the pair of functions $(\wbar,\wlow) = (\Wsup, \Winf)$ solves the functional equations
\begin{eqnarray}
\wbar(p) & = & \max_{\kappa\in\Kp} \left\{ \astrut{2.5}
(1-\delta)[(1-\kappa)s + \kappa m(p)] + \delta \E_{N\kappa}\wbar(p)
\right\}, \label{eq:wbar} \\
\wlow(p) & = & \min_{\kappa\in\Kp} \max_{k\in\{0,1\}} \left\{ \astrut{2.5}
(1-\delta)[(1-k)s + k m(p)] + \delta \E_{(N-1)\kappa+k} \wlow(p) \right\}, \label{eq:wlow}
\end{eqnarray}
where $\Kp \subseteq \{0, 1\}$ denotes the set of all $\kappa$ such that
\begin{eqnarray}
\lefteqn{(1-\delta)[(1-\kappa)s + \kappa m(p)] + \delta \E_{N\kappa}\wbar(p)} \label{eq:kappa} \\
& \geq & \astrut{3} \max_{k\in\{0,1\}} \left\{ \astrut{2.5}
(1-\delta)[(1-k)s + k m(p)] + \delta \E_{(N-1)\kappa+k} \wlow(p) \right\}. \nonumber
\end{eqnarray}
Moreover, $\Winf \leq \wlow \leq \wbar \leq \Wsup$ for any solution $(\wbar,\wlow)$ of \eref{eq:wbar}--\eref{eq:kappa}.
\end{prop}

This result relies on arguments that are familiar from Cronshaw and Luenberger (1994).
We briefly sketch them here.

The above equations can be understood as follows. The ideal condition for a given (symmetric) action profile to be incentive compatible is that if each player conforms to it, the continuation payoff is the highest possible, while a deviation triggers the lowest possible continuation payoff. These actions are precisely the elements of $\Kp$, as defined by equation \eref{eq:kappa}. Given this set of actions, equation \eqref{eq:wlow} provides the recursion that characterizes the constrained minmax payoff under the assumption that if a player were to deviate to his myopic best reply to the constrained minmax action profile, the punishment would be restarted next period, resulting in a minimum continuation payoff. Similarly, equation \eqref{eq:wbar} yields the highest payoff under this constraint, but here, playing the best action (within the set) is on the equilibrium path.

Note that in \textit{any} SSE, given $p$, the action $\kappa(p)$ must be an element of $\mathcal{K}(p;\Wsup,\Winf)$. This is because the left-hand side of \eqref{eq:kappa} with $\wbar=\Wsup$ is an upper bound on the continuation payoff if no player deviates, and the right-hand side with $\wlow=\Winf$ a lower bound on the continuation payoff after a unilateral deviation. Consider the equilibrium that achieves $\Wsup$. Then,
\[
\Wsup(p) \le  \max_{\kappa \in \mathcal{K}(p;\Wsup,\Winf)} \left\{ \astrut{2.5}
(1-\delta)[(1-\kappa)s + \kappa m(p)] + \delta \E_{N\kappa}\Wsup(p)
\right\},
\]
as the action played must be in $\mathcal{K}(p;\Wsup,\Winf)$, and the continuation payoff is at most given by $\Wsup$. Similarly, $\Winf$ must satisfy \eref{eq:wlow} with ``$\ge$'' instead of ``$=$.'' Suppose now that the ``$\le$'' were strict. Then, we can define a strategy profile given prior $p$ that (i) in period 0, plays the maximizer of the right-hand side, and (ii) from $t=\Delta$ onward, abides by the continuation strategy achieving $\Wsup(p_\Delta)$. Because the initial action is in $\mathcal{K}(p;\Wsup,\Winf)$, this constitutes an equilibrium, and it achieves a payoff strictly larger than $\Wsup(p)$, a contradiction. Hence, \eref{eq:wbar} must hold with equality for $\Wsup$. The same reasoning applies to $\Winf$ and \eref{eq:wlow}.

Fix a pair $(\wbar,\wlow)$ that satisfies
\eref{eq:wbar}--\eref{eq:kappa}. Note that this implies $\wlow \le
\wbar$. Given such a pair and any prior $p$, we specify two SSEs whose
payoffs are $\wbar$ and $\wlow$, respectively. It then follows that
$\Winf \leq \wlow \leq \wbar \leq \Wsup$. Let $\kr$ and $\kp$ denote a
selection of the maximum and minimum of
\eref{eq:wbar}--\eref{eq:wlow}. The equilibrium strategies are
described by a two-state automaton, whose states are referred to as
``good'' or ``bad.'' The difference between the two equilibria lies in
the initial state: $\wbar$ is achieved when the initial state is good,
$\wlow$ is achieved when it is bad.
In the good state, play proceeds according to $\kr$; in the bad state,
it proceeds according to $\kp$. Transitions are exactly as in the equilibria described in Sections \ref{sec:construction-Brownian}--\ref{sec:construction-Poisson}.
This structure precludes profitable one-shot deviations in either state, so that the automaton describes equilibrium strategies, and the desired payoffs are obtained.

Figure 4 presents the result of a numerical computation of $\Wsup$ and $\Winf$ based on Proposition \ref{prop:chara-discrete}.\footnote{
We thank Kai Echelmeyer and Martin Rumpf from the Institute for Numerical Simulation at the University of Bonn for the implementation of the underlying algorithm.
Starting from the pair of functions $(\wbar^0,\wlow^0)=(V_N^*, V_1^*)$, it computes $(\wbar^{k+1},\wlow^{k+1})$ by evaluating the right-hand sides of \eref{eq:wbar}--\eref{eq:wlow} at $(\wbar^k,\wlow^k)$.
Because of the incentive-compatibility constraint \eref{eq:SSE-incentive-constraint}, the corresponding value-iteration operator does not appear to be a contraction mapping.
While we do not have a convergence proof for this algorithm, it converged reliably for sufficiently small $\Delta$.
}
In between the thresholds $\pr^\Delta$ and $\pp^\Delta$, both risky and safe play can be sustained in an SSE; the former is chosen in the best SSE, the latter in the worst.
Only risky play can be sustained above $\pp^\Delta$, and only safe play below $\pr^\Delta$.
These changes in the set of enforceable actions manifest themselves in jump discontinuities of $\Wsup$ and $\Winf$ at $\pr^\Delta$ and $\pp^\Delta$, respectively.
Note also that $\Winf$ dips below $V_1^*$ to the immediate right of $p_1^*$.
The worst punishment can thus be harsher for positive $\Delta$ than in the frequent-action limit, and convergence of the corresponding payoff function is non-monotonic.
While the example shown is one of pure Brownian learning, these patterns also emerge when conclusive lump-sums are added to the payoff process.

\begin{figure}[h]
\centering
\begin{picture}(175.00,110.00)(0,0)
\put(10,-45){\scalebox{1.45}{\includegraphics{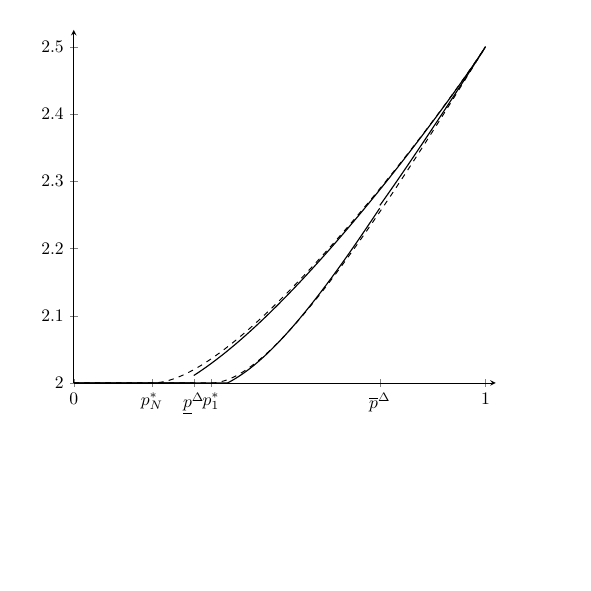}}}
\put(28,97){\makebox(0,0)[cc]{payoff}}
\put(135,6){\makebox(0,0)[cc]{$p$}}
\end{picture}
\begin{quote}
\caption{Payoffs $V_N^*$ (upper dashed curve), $\Wsup$ (upper solid curve), $\Winf$ (lower solid curve) and $V^*_1$ (lower dashed curve) for $\Delta = 0.1$ and $(r,s,\sigma,\alpha_1,\alpha_0,h,\lambda_1,\lambda_0,N)=(1,2,1,2.5,1.5,0,0,0,5)$.}
\end{quote}
\end{figure}

As $\Delta$ tends to 0, equations \eref{eq:wbar}--\eref{eq:wlow} transform into differential-difference equations involving terms that are familiar from the continuous-time analysis in Section \ref{sec:continuous-time}.
A formal Taylor approximation shows that for any $\kappa \in \{0,1\}$, $K \in \{0,1,\ldots,N\}$ and a sufficiently regular function $w$ on the unit interval,
\begin{eqnarray*}
\lefteqn{(1-\delta)[(1-\kappa)s + \kappa m(p)] + \delta \E_K w(p)} \\
& = & \astrut{3} w(p) + r \left\{ \astrut{2.5}
(1-\kappa)s + \kappa m(p) + K\, b(p,w) - w(p) \right\} \Delta
+ o(\Delta).
\end{eqnarray*}
Applying this approximation to \eref{eq:wbar}--\eref{eq:wlow}, cancelling the terms of order 0 in $\Delta$, dividing through by $\Delta$, letting $\Delta \rightarrow 0$ and recalling the notation
$c(p) = s - m(p)$
for the opportunity cost of playing risky, we obtain the coupled differential-difference equations that appear in the following result.

\begin{prop}\label{prop:chara-continuous}
Let $\rho > 0$ or $\lambda_0 > 0$.
As $\Delta \rightarrow 0$, the pair of functions $(\Wsup,\Winf)$ converges uniformly (in $p$) to a pair of functions $(\wbar,\wlow)$ solving
\begin{eqnarray}
\wbar(p) & = & s + \max_{\kappa \in \overline{\K}(p)} \kappa \left[N b(p,\wbar) - c(p) \right], \label{eq:wbar-ct-b} \\
\wlow(p) & = & s + \min_{\kappa \in \overline{\K}(p)} (N-1) \kappa\, b(p,\wlow) + \max_{k\in\{0,1\}} k \left[ b(p,\wlow) - c(p) \right] \label{eq:wlow-ct-b},
\end{eqnarray}
where
\begin{equation} \label{eq:K-ct}
\overline{\K}(p) = \left\{
\begin{array}{lll}
\{0\} & \mbox{for} & p \leq \phat, \\
\{0,1\} & \mbox{for} & \phat < p < 1, \\
\{1\} & \mbox{for} & p = 1,
\end{array} \right.
\end{equation}
and
$\phat$ is as in parts (ii) and (iii) of Theorem \ref{thm}.
\end{prop}

This result is an immediate consequence of the previous results.
It follows from Sections \ref{sec:construction-Brownian}--\ref{sec:construction-Poisson} that,
except when $\rho = \lambda_0 = 0$, there exist pure-strategy SSEs and
the pair $(\Wsup,\Winf)$ converges uniformly to $(\VNphat,V_1^*)$.
It is straightforward to verify that $(\wbar,\wlow)=(\VNphat,V_1^*)$ solves \eref{eq:wbar-ct-b}--\eref{eq:K-ct}.
First, as $V_N^*$ satisfies\footnote{
This equation follows from the HJB equation in Section \ref{sec:continuous-time}: because the maximand is linear in $K$, the continuous-time planner finds it optimal to set $K=0$ or $K=N$ at any given belief.}
$$
V_N^*(p) = s + \max_{\kappa \in \{0,1\}} \kappa \left[N b(p,V_N^*) - c(p) \right],
$$
with $N b(p,V_N^*) - c(p) > 0$ to the right of $p_N^*$, \eref{eq:wbar-ct-b} is trivially solved by $V_N^*$ whenever $\phat = p_N^*$.
Second, for $\phat > p_N^*$, the function $\VNphat$
satisfies
$$
\VNphat(p) = s + \mathbbm{1}_{p > \phat} \left[Nb(p,\VNphat) - c(p) \right],
$$
with $N b(p;\VNphat) - c(p) > 0$ on $(\phat, 1)$.
This implies that $\VNphat$ solves \eref{eq:wbar-ct-b} when $\phat > p_N^*$.
Third, $V_1^*$ always solves \eref{eq:wlow-ct-b}.
In fact, as $b(p;V_1^*) \geq 0$ everywhere, we have $\min_{\kappa \in \{0,1\}} (N-1) \kappa\, b(p,V_1^*) = 0$, and \eref{eq:wlow-ct-b} with this minimum set to zero is just the HJB equation for $V_1^*$.

Note that the continuous-time functional equations \eref{eq:wbar-ct-b}--\eref{eq:wlow-ct-b} would be equally easy to solve for any \emph{arbitrary} $\phat$ in \eref{eq:K-ct}.
However, only the solution with $\phat$ as in Theorem \ref{thm} captures the asymptotics of our discretization of the experimentation game.

\section{Concluding Comments}
\label{sec:conclu}
We have shown that the inefficiencies arising in strategic bandit problems are driven by the solution concept, MPE.
Inefficiencies entirely disappear when news has a Brownian component or good news events are not too informative. The best PBE can be achieved with an SSE, specifying a simple rule of conduct (unlike in an MPE), namely on-path play of the cutoff type.

Of course, we do not expect the finding that SSE and PBE payoffs coincide to generalize to all symmetric stochastic games.
For instance, SSE can be restrictive when actions are imperfectly monitored, as shown by Fudenberg, Levine and Takahashi (2007). Nonetheless, SSE is a class of equilibria that both allows for ``stick-and-carrot'' incentives, as in standard discrete-time repeated (or stochastic) games, but is also amenable to continuous-time optimal control techniques, as illustrated by Proposition \ref{prop:chara-continuous} (for a given transversality condition that must be derived from independent considerations, such as a discretized version of the game).

The information/payoff processes that we consider are a subset of those in Cohen and Solan (2013).
There, the size of a lump-sum payoff is allowed to contain information about the state of world, so that the arrival of a discrete payoff increment makes the belief jump to a posterior that depends on the size of the increment.
As lump sums of any size are assumed to arrive more frequently in state $\theta = 1$, however, they are always good news.
For processes with an informative Brownian component, our proof that risky play is incentive compatible immediately to the right of the threshold $p_N^*$ only exploits the properties of the posterior belief process \emph{conditional on no lump sum arriving}.
As these properties are the same whether lump sums are informative or not, asymptotic efficiency in the presence of a Brownian component should obtain more generally---and even when lump sums of certain sizes are bad news (meaning that they are less frequent in state $\theta = 1$).
When learning is driven by lump-sum payoffs only, inspection of equation \eref{eq:phat} suggests that efficiency requires that a lump sum \emph{of any size} arriving at the initial belief $p_N^*$ lead to a posterior belief no higher than $p_1^*$.
This is a constraint on the maximal amount of good news that a lump sum can carry; lump-sum sizes that carry bad news should again be innocuous here.

The ``breakdowns'' variant of pure Poisson learning in Keller and Rady (2015) is one of cost minimization.
In our payoff-maximization setting, this corresponds to letting both the safe flow payoff and the average size of lump-sum payoffs be negative with $\lambda_1 h < s < \lambda_0 h \leq 0$.
Now, $\theta = 1$ is the \emph{bad} state of the world, and the efficient and single-player solution cutoffs in continuous time satisfy $p_N^* > p_1^*$, with the stopping region lying to the \emph{right} of the cutoff in either case.
The associated value functions $V_1^*$ and $V_N^*$ solve the same HJB equations as in Section \ref{sec:continuous-time}; except for the case $\lambda_0 = 0$, they are not available in closed form, however.
Starting from $p_N^*$, the belief now always remains in the single-agent stopping region for small $\Delta$: either there is a breakdown and the belief jumps up to $j(p_N^*) > p_N^* > p_1^*$, or there is no breakdown and the belief slides down to somewhere close to $p_N^*$, and hence still above $p_1^*$.
This means that the harshest possible punishment, consisting of all other players playing safe forever, can be meted out to any potential deviator, whether there is a breakdown or not.
Thus, we conjecture that asymptotic efficiency also obtains in this framework.

The intricacies of our proofs derive to some extent from specific features of the benchmark models in the literature on strategic bandit problems: these are games of informational externalities only, and safe play halts learning.
In many applications, payoff externalities are present as well, as might be background learning. Extraneous instruments that players might have at their disposal in order to provide incentives will certainly facilitate cooperation, and thus facilitate efficiency. Similarly, exogenous learning, such as background learning, presumably helps with efficiency, as a deviating player can never be too sure “what the future holds”.
Nonetheless, such changes to the model would blur our point that informational externalities by themselves suffice for first-best to be achievable.

While the environment (perfect monitoring and lack of commitment) called for a discretization of the game, and a meticulous analysis of the convergence of payoffs and strategies as the mesh vanishes, one might wonder whether there is no shortcut to perform the analysis in continuous time directly.
For the case of pure Poisson learning, an attempt to describe continuous-time strategies which achieve the extremal payoffs can be found in H\"{o}rner et al. (2014, Appendix A).
It involves an independent ``Poisson clock'' that determines the random times at which play reverts to the normal state of the automaton.
However, we do not believe that there is an easy way to determine optimality of such a strategy profile, as the key boundary condition that determines how much experimentation takes place (independent of whether efficiency obtains) seems difficult to identify in continuous time; and indeed, as discussed in footnote \ref{fn:discretization}, we cannot rule out that other discretizations might yield other boundary conditions.

Finally, our study is limited to solving for the best---and worst---average equilibrium payoff across players. As we have stressed, the symmetry in the strategy profiles that achieve these payoffs is a result, not an assumption. Nonetheless, it would be interesting to characterize the entire equilibrium payoff set, especially in view of a potential generalization to asymmetric games. For instance, what is the equilibrium that maximizes player 1's payoff, say? A careful analysis of this problem would take us too far astray, but we note that the findings of this paper provide a foundation for it, as we identify the minmax value of the game, and equilibrium strategies that support it. Since the game has observable actions, this is the punishment that should be used to support any Pareto efficient equilibrium, leaving us with the task of identifying the equilibrium paths of such equilibria.

\subsubsection*{Data Availability Statement (DAS)}

The code underlying Figure 4 is available on Zenodo at https://dx.doi.org/10.5281/zenodo.5516947.

\AppendixIn                         
\setcounter{equation}{0}

\section{Auxiliary Results} \label{app:auxiliary}

\subsection{Evolution of Beliefs} \label{app:beliefs}

For the description of the evolution of beliefs, it is convenient to work with the log odds ratio
$$
\ell_t = \ln \frac {p_t}{1-p_t}\,.
$$
Suppose that starting from $\ell_0=\ell$, the players use the fixed action profile $(k_1,\ldots,k_N) \in \{0,1\}^N$.
By Peskir and Shiryayev (2006, pp.\ 287--289 and 334--338), the log odds ratio at time $t>0$ is then
$$
\ell_t = \ell + \sum_{\{n: k_n=1\}} \left\{ \frac{\alpha_1 - \alpha_0}{\sigma^2} \left( X^n_t - \alpha_0 t - h N^n_t \right) - \left[ \frac{(\alpha_1 - \alpha_0)^2}{2 \sigma^2} + \lambda_1 - \lambda_0 \right] t +  \ln \frac{\lambda_1}{\lambda_0} \, N^n_t \right\},
$$
where $X^n$ and $N^n$ are the payoff and Poisson processes, respectively, associated with player $n$'s risky arm.
The terms involving $\alpha_1, \alpha_0$ and $\sigma$ capture learning from the continuous component, $X^n_t - h N^n_t$, of the payoff process, with higher realizations making the players more optimistic.
The terms involving $\lambda_1$ and $\lambda_0$ capture learning from lump-sum payoffs, with the players becoming more pessimistic on average as long as no lump-sum arrives,
and each arrival increasing the log odds ratio by the fixed increment $\ln (\lambda_1/\lambda_0)$.\footnote{
Here, $\lambda_1/\lambda_0$ is understood to be 1 when $\lambda_1 = \lambda_0 = 0$.
When $\lambda_1 > \lambda_0 = 0$, we have $\ell_t = \infty$ and $p_t = 1$ from the arrival time of the first lump-sum on.}

Under the probability measure $\mathbb{P}_\theta$ associated with state $\theta \in \{0,1\}$,
$X^n_t - \alpha_0 t - h N^n_t$ is Gaussian with mean $(\alpha_\theta-\alpha_0) t$ and variance $\sigma^2 t$,
so that
$\sum_{\{n: k_n=1\}} (\alpha_1 - \alpha_0) \sigma^{-2} \left( X^n_t - \alpha_0 t - h N^n_t \right)$
is Gaussian with mean $K (\alpha_1 - \alpha_0) (\alpha_\theta-\alpha_0) \sigma^{-2}  t$ and variance $K \rho t$,
where
$K = \sum_{n=1}^{N} k_n$ and $\rho = (\alpha_1-\alpha_0)^2 \sigma^{-2}$.
Conditional on the event that $\sum_{\{n: k_n=1\}} N^n_t = J$, therefore, $\ell_t$ is normally distributed with mean
$\ell - K \left(\lambda_1 - \lambda_0 - \frac{\rho}{2}\right) t + J \ln (\lambda_1/\lambda_0)$
and variance $K \rho t$ under $\mathbb{P}_1$,
and normally distributed with mean
$\ell - K \left(\lambda_1 - \lambda_0 + \frac{\rho}{2}\right) t + J \ln (\lambda_1/\lambda_0)$
and variance $K \rho t$ under $\mathbb{P}_0$.
Finally, the probability under measure $\mathbb{P}_\theta$ that $\sum_{\{n: k_n=1\}} N^n_t = J$ equals
$\frac{(K \lambda_\theta t)^J}{J!} e^{-K \lambda_\theta t}$
by the sum property of the Poisson distribution.

Taken together, these facts make it possible to explicitly compute the distribution of
$$
p_t = \frac{e^{\ell_t}}{1+e^{\ell_t}}
$$
under the players' measure $\mathbb{P}_p = p \mathbb{P}_1 +(1-p) \mathbb{P}_0$.
As this explicit representation is not needed in what follows, we omit it here.

Instead, we turn to the characterization of infinitesimal changes of $p_t$, once more assuming a fixed action profile with $K$ players using the risky arm.
Arguing as in Cohen and Solan (2013, Section 3.3), one shows that, with respect to the players' information filtration, the process of posterior beliefs is a Markov process whose infinitesimal generator ${\cal L}^K$ acts as follows on real-valued functions $v$ of class $C^2$ on the open unit interval:
$$
{\cal L}^K v(p)
= K \left\{ \astrut{3}
\frac{\rho}{2} p^2(1-p)^2 v''(p)
- (\lambda_1-\lambda_0) p(1-p) v'(p)
+ \lambda(p) \left[ v(j(p)) - v(p) \right]
\right\}.
$$
In particular, instantaneous changes in beliefs exhibit linearity in $K$ in the sense that ${\cal L}^K = K {\cal L}^1$.

By the very nature of Bayesian updating, finally, the process of posterior beliefs is a martingale with respect to the players' information filtration.

\subsection{Payoff Functions}\label{app:payoffs}

Our first auxiliary result concerns the function $u(\cdot; \mu_N)$ defined in Section \ref{sec:continuous-time}.

\begin{lem} \label{lem:expectation-u}
$\delta \E_K u(\cdot;\mu_N)(p) = \delta^{1-\frac{K}{N}} u(p;\mu_N)$
for all $\Delta > 0$, $K \in \{1,\ldots,N\}$ and $p \in (0,1]$.
\end{lem}
\proof{
We simplify notation by writing $u$ for $u(\cdot; \mu_N)$.
Consider the process $(p_t)$ of posterior beliefs in continuous time when $p_0 = p > 0$ and $K$ players use the risky arm.
By Dynkin's formula,
\begin{eqnarray*}
\mathbb{E} \left[e^{-rK\Delta/N} u(p_\Delta)\right]
& = & u(p) + \mathbb{E} \left[ \int_0^\Delta e^{-rKt/N} \left\{ {\cal L}^K u(p_t) - \frac{rK}{N} u(p_t) \right\} dt \right] \\
& = & u(p) + K\, \mathbb{E} \left[ \int_0^\Delta e^{-rKt/N} \left\{ {\cal L}^1 u(p_t) - \frac{r}{N} u(p_t) \right\} dt \right] \\
& = & u(p),
\end{eqnarray*}
where the last equality follows from the fact that ${\cal L}^1 u = r u/N$ on $(0,1]$.\footnote{
To verify this identity, note that
$$
u'(p) = -\frac{\mu_N + p}{p (1-p)} u(p), \quad u''(p) = \frac{\mu_N (\mu_N+1)}{p^2 (1-p)^2} u(p), \quad u(j(p)) = \frac{\lambda_0}{\lambda(p)} \left( \frac{\lambda_0}{\lambda_1} \right)^{\mu_N} u(p),
$$
and use the equation defining $\mu_N$.}
Thus, $\delta^{K/N} \E_K u(p) = u(p)$.
}

We further note that $\E_K m(p) = m(p)$ for all $K$ by the martingale property of beliefs and the linearity of $m$ in $p$.

These properties are used repeatedly in what follows.
Their first application is in the proof of uniform convergence of the discrete-time single-agent value function to its continuous-time counterpart.

Let $(\mathcal{W},\|\cdot\|)$ be the Banach space of bounded real-valued functions on $[0,1]$ equipped with the supremum norm.
Given $\Delta > 0$, and any $w \in \mathcal{W}$, define a function $\Tone w \in \mathcal{W}$ by
$$
\Tone w(p)
= \max\left\{\astrut{2.5}
(1-\delta) m(p) + \delta \E_1 w(p),\
(1-\delta) s + \delta w(p)
\right\}.
$$
The operator $\Tone$ satisfies Blackwell's sufficient conditions for being a contraction mapping with modulus $\delta$ on $(\mathcal{W},\|\cdot\|)$: monotonicity ($v \leq w$ implies $\Tone v \leq \Tone w$) and discounting ($\Tone(w + c) = \Tone w + \delta c$ for any real number $c$).
By the contraction mapping theorem, $\Tone$ has a unique fixed point in $\mathcal{W}$; this is the value function $W_1^\Delta$ of an agent experimenting in isolation.

The corresponding continuous-time value function is $V_1^*$ as introduced in Section \ref{sec:continuous-time}.
As any discrete-time strategy is feasible in continuous time, we trivially have $W_1^\Delta \leq V_1^*$.

\begin{lem} \label{lem:convergence-single-agent}
$W_1^\Delta \to V_1^*$ uniformly as $\Delta \rightarrow 0$.
\end{lem}
\proof{
A lower bound for $W_1^\Delta$ is given by the payoff function $W_*^\Delta$ of a single agent who uses the cutoff $p_1^*$ in discrete time; this function is the unique fixed point in $\mathcal{W}$ of the contraction mapping $\Tstar$ defined by
\[\Tstar w(p)
= \left\{ \begin{array}{lll}
(1-\delta) m(p) + \delta \E_1 w(p)
& \mbox{if} & p > p_1^*, \\
(1-\delta) s + \delta w(p)
& \mbox{if} & p \leq p_1^*.
\end{array}\right.\]
Next, choose $\breve{p} < p_1^*$, and define $p^\natural=\frac{\breve{p}+p_1^*}{2}$ and the function $v=m+Cu(\cdot;\mu_1)+\mathbbm{1}_{[0,p^\natural]}(s-m-Cu(\cdot;\mu_1))$, where the constant $C$ is chosen so that $s=m(\breve{p})+Cu(\breve{p};\mu_1)$.

Fix $\varepsilon > 0$.
As $v$ converges uniformly to $V_1^*$ as $\breve p \to p_1^*$, we can choose $\breve p$ such that $v \geq V_1^* - \varepsilon$.
It suffices now to show that there is a $\bar \Delta > 0$ such that $\Tstar v \geq v$ for $\Delta < \bar \Delta$.
In fact, the monotonicity of $\Tstar$ then implies $W_*^\Delta \geq v$ and hence
$V_1^* - \varepsilon \leq v \leq W_*^\Delta \leq W_1^\Delta \leq V_1^*$
for all $\Delta < \bar \Delta$.

For $p \leq p_1^*$, we have $\Tstar v(p) = (1-\delta) s + \delta v (p) \geq v(p)$ for all $\Delta$, because $v\leq s$ in this range.
For $p>p_1^*$,
\begin{eqnarray*}
\Tstar v(p)
& = & (1-\delta) m(p)+\delta \mathcal{E}_1^\Delta v(p)\\
& = & (1-\delta) m(p)+ \delta \E_1\left[ m + Cu + \mathbbm{1}_{[0,p^\natural]}(s-m-Cu) \right](p) \\
& = & v(p) + \delta \E_1\left[  \mathbbm{1}_{[0,p^\natural]}(s-m-Cu) \right](p),
 \end{eqnarray*}
where the last equation uses that $\E_1 m(p) = m(p)$ and $\delta\E_1 u(p)=u(p)$.
In particular, $\Tstar v(1) = v(1)$.

The function $s-m-Cu$ is negative on the interval $(0,\breve p)$ and positive on $(\breve p,p^\sharp)$, for some $p^\sharp>p_1^*$.
The expectation of $s - m(p_\Delta)- C u(p_\Delta)$ conditional on $p_0 = p$ and $p_\Delta \leq p^\natural$ is continuous in
$(p,\Delta) \in [p_1^*,1) \times (0,\infty)$
and converges to $s - m(p^\natural)- C u(p^\natural) > 0$ as $p \rightarrow 1$ or $\Delta \rightarrow 0$ because the conditional distribution of $p_\Delta$ becomes a Dirac measure at $p^\natural$ in either limit.
This implies existence of $\bar \Delta > 0$ such that this conditional expectation is positive for all $(p,\Delta) \in [p_1^*,1) \times (0,\bar \Delta)$.
For these $(p,\Delta)$, we thus have
$$
\E_1\left[ \mathbbm{1}_{[0,p^\natural]}(s-m-Cu) \right](p)
\geq \E_1\left[  \mathbbm{1}_{[p^\flat,p^\natural]}(s-m-Cu) \right](p)
\geq 0,
$$
where $p^\flat=\frac{\check p+p^\natural}{2}$.
As a consequence, $\Tstar v \ge v$ for all $(p,\Delta) \in (p_1^*,1) \times (0,\bar \Delta)$.
}

Next, we turn to the payoff function associated with the good state of the automaton defined in Section \ref{sec:construction}.
By the same arguments as invoked immediately before Lemma \ref{lem:convergence-single-agent}, $\wr$ is the unique fixed point in $\mathcal{W}$ of the operator $\Tr$ defined by
\[\Tr w(p)
= \left\{ \begin{array}{lll}
(1-\delta) m(p) + \delta \E_N w(p)
& \mbox{if} & p > \pr, \\
(1-\delta) s + \delta w(p)
& \mbox{if} & p \leq \pr.
\end{array}\right.\]

\begin{lem} \label{lem:lower-bound-reward}
Let $\pr > p_N^*$. Then $\wr \geq \VNpr$ for $\Delta$ sufficiently small.
\end{lem}
\proof{
Because of the monotonicity of the operator $\Tr$, it suffices to show that $\Tr\VNpr \geq \VNpr$ for sufficiently small $\Delta$.
Recall that for $p > \pr$, $\VNpr(p) = m(p) + C u(p;\mu_N)$ where the constant $C > 0$ is chosen to ensure continuity at $\pr$.

For $p \leq \pr$, we use exactly the same argument as in the penultimate paragraph of the proof of Lemma \ref{lem:convergence-single-agent};
for $p > \pr$, the argument is the same as in the last paragraph of that proof.
}

The next two results concern the payoff function associated with the bad state of the automaton defined in Section \ref{sec:construction}.
Fix a cutoff $\pp \in (p^m,1)$ and let $K(p)=N-1$ when $p>\pp$, and $K(p)=0$ otherwise.
Given $\Delta > 0$, and any bounded function $w$ on $[0,1]$, define a bounded function $\Tp w$ by
$$
\Tp w(p)
= \max\left\{\astrut{2.5}
(1-\delta) m(p) + \delta \E_{K(p)+1}w(p),\
(1-\delta) s + \delta \E_{K(p)}w(p)
\right\}.
$$
The operator $\Tp$ again satisfies Blackwell's sufficient conditions for being a contraction mapping with modulus $\delta$ on $\mathcal{W}$.
Its unique fixed point in this space is the payoff function $\wp$ (introduced in Section \ref{sec:construction}) from playing a best response against $N-1$ opponents who all play risky when $p > \pp$, and safe otherwise.

\begin{lem} \label{lem:upper-bound-punishment}
Let $\pr\in(p_N^*,p_1^*)$.
Then there exists $p^\diamond\in[p^m,1)$ such that for all $\pp\in(p^\diamond,1)$,
the inequality $\wp\leq V_{N,(\pr+p_1^*)/2}$ holds for $\Delta$ sufficiently small.
\end{lem}
\proof{
Let $\tilde p = (\pr+p_1^*)/2$.
For $p > \tilde p$, we have $V_{N,\tilde p}(p) = m(p) + C u(p;\mu_N)$ where the constant $C > 0$ is chosen to ensure continuity at $\tilde p$.
To simplify notation, we write $\tilde v$ instead of $V_{N,\tilde p}$ and $u$ instead of $u(\cdot;\mu_N)$.

For $x > 0$, we define
\[p^*_x=\frac{\mu_x(s-m_0)}{(\mu_x+1)(m_1-s)+\mu_x(s-m_0)}\,,\]
where $\mu_x$ is the unique positive root of
\[f(\mu;x)=
\frac{\rho}{2}\mu(\mu+1)+(\lambda_1-\lambda_0)\mu+\lambda_0\left(\frac{\lambda_0}{\lambda_1}\right)^\mu-\lambda_0-\frac{r}{x}\,;\]
existence and uniqueness of this root follow from continuity and monotonicity of $f(\cdot;x)$ together with the fact that $f(0;x) < 0$ while $f(\mu;x) \to \infty$ as $\mu \to \infty$.\footnote{\emph{Cf.}\ Lemma 6 in Cohen \& Solan (2013).}
This extends our previous definitions of $\mu_N$ and $p_N^*$ to non-integer numbers.
It is immediate to verify now that $\frac{d \mu_x}{dx}<0$ and hence $\frac{d p^*_x}{d x}<0$.
Thus, there exists $\breve{x}\in (1,N)$ such that $p^*_{\breve{x}}\in (\tilde p, p_1^*)$.

Having chosen such an $\breve{x}$, we fix a belief $\breve{p}\in(\tilde p, p^*_{\breve{x}})$ and, on the open unit interval, consider the function $\breve{v}$ that solves
\[{\cal L}^1 v -\frac{r}{\breve{x}}(v-m)=0\]
subject to the conditions $\breve{v}(\breve{p})=s$ and $\breve{v}'(\breve{p})=0$.
This function has the form
\[\breve{v}(p)=m(p)+\breve{u}(p),\]
with
\[\breve{u}(p)=A(1-p)\left(\frac{1-p}{p}\right)^{\breve{\mu}}+Bp\left(\frac{p}{1-p}\right)^{\hat{\mu}}=A u(p;\breve{\mu})+Bu(1-p;\hat{\mu}).\]
Here, $\breve{\mu} = \mu_{\breve{x}}$ and $\hat{\mu}$ is the unique positive root of
\[g(\mu;x)=
\frac{\rho}{2}\mu(\mu+1)-(\lambda_1-\lambda_0)\mu+\lambda_1\left(\frac{\lambda_1}{\lambda_0}\right)^\mu-\lambda_1-\frac{r}{x}\,;\]
existence and uniqueness of this root follow along the same lines as above.

The constants of integration $A$ and $B$ are pinned down by the conditions $\breve{v}(\breve{p})=s$ and $\breve{v}'(\breve{p})=0$.
One calculates that $B>0$ if, and only if, $\breve{p}<p^*_{\breve{x}}$, which holds by construction,
and that $A>0$ if, and only if,
\[\breve{p}<\frac{(1+\hat{\mu})(s-m_0)}{\hat{\mu}(m_1-s)+(1+\hat{\mu})(s-m_0)}\,.\]
The right-hand side of this inequality is decreasing in $\hat{\mu}$ and tends to $p^m$ as $\hat{\mu}\rightarrow\infty$. Therefore, we can conclude that the inequality holds, and $A>0$ as well.
Moreover, $A+B>0$ implies that $\breve{v}$ is strictly increasing and strictly convex on $(\breve{p},1)$; as $B > 0$, finally, $\breve{v}(p) \to \infty$ for $p \to 1$.

So there exists a belief $p^\natural\in(\breve{p},1)$ such that $\breve{v}(p^\natural)=\tilde v(p^\natural)$ and $\breve{v}>\tilde v$ on $(p^\natural,1)$.
We  now show that $\breve{v}<\tilde v$ in $(\breve{p},p^\natural)$.
Indeed, if this is not the case, then $\breve{v}-\tilde v$ assumes a non-negative local maximum at some $p^\sharp \in (\breve{p},p^\natural)$.
This implies:

\noindent (i) $\breve{v}(p^\sharp)\geq\tilde v(p^\sharp)$, \emph{i.e.},
\begin{equation}\label{eq:C8_0}
A u(p^\sharp;\breve{\mu})+B u(1-p^\sharp;\hat{\mu})\geq C u(p^\sharp;\mu_N);
\end{equation}
(ii) $\breve{v}'(p^\sharp)=\tilde v'(p^\sharp)$, \emph{i.e.},
\begin{equation}\label{eq:C8_1}
-(\breve{\mu}+p^\sharp)A u(p^\sharp;\breve{\mu})+(\hat{\mu}+1-p^\sharp)B u(1-p^\sharp;\hat{\mu})=-(\mu_N+p^\sharp)C u(p^\sharp;\mu_N);
\end{equation}
and (iii) $\breve{v}''(p^\sharp)\leq\tilde v''(p^\sharp)$, \emph{i.e.},
\begin{equation}\label{eq:C8_2}
\breve{\mu}(\breve{\mu}+1)A u(p^\sharp;\breve{\mu})+\hat{\mu}(1+\hat{\mu})B u(1-p^\sharp;\hat{\mu})\leq \mu_N(\mu_N+1)C u(p^\sharp;\mu_N).
\end{equation}
Solving for $B u(1-p^\sharp;\hat{\mu})$ in \eref{eq:C8_1} and inserting the result into \eref{eq:C8_0} and \eref{eq:C8_2}, we obtain, respectively,
\[\frac{C u(p^\sharp;\mu_N)}{A u(p^\sharp;\breve{\mu})}\leq \frac{\breve{\mu}+\hat{\mu}+1}{\mu_N+\hat{\mu}+1},\]
and
\[\frac{C u(p^\sharp;\mu_N)}{A u(p^\sharp;\breve{\mu})}\geq\frac{\breve{\mu}(\breve{\mu}+1)(\hat{\mu}+1-p^\sharp)+\hat{\mu}(\hat{\mu}+1)(\breve{\mu}+p^\sharp)} {\mu_N(\mu_N+1)(\hat{\mu}+1-p^\sharp)+\hat{\mu}(\hat{\mu}+1)(\mu_N+p^\sharp)}\,.\]
This implies that
\[\frac{\breve{\mu}+\hat{\mu}+1}{\mu_N+\hat{\mu}+1}\geq \frac{\breve{\mu}(\breve{\mu}+1)(\hat{\mu}+1-p^\sharp)+\hat{\mu}(\hat{\mu}+1)(\breve{\mu}+p^\sharp)} {\mu_N(\mu_N+1)(\hat{\mu}+1-p^\sharp)+\hat{\mu}(\hat{\mu}+1)(\mu_N+p^\sharp)}\,,\]
which one shows to be equivalent to $\breve{\mu} \leq \mu_N$.
But $\breve{x}<N$ and $\frac{d\mu_x}{dx}<0$ imply $\breve{\mu} > \mu_N$.
This is the desired contradiction.

Now let $p^\diamond=\max\{p^m,p^\natural\}$, fix $\pp\in(p^\diamond,1)$ and define
\[v(p)=\left\{\begin{array}{lll}
\tilde v(p) & \mbox{if} & p>p^\natural,\\
\breve{v}(p) & \mbox{if} & \breve{p}\leq p\leq p^\natural,\\
s &\mbox{if} & p<\breve{p}.\end{array}\right.\]
By construction, $s\leq v\leq \min\{\tilde v, \breve v\}$.
This immediately implies that $(1-\delta)s+\delta v\leq v$.
We now show that $\underline{T}^\Delta v\leq v$, and hence $\wp\leq v$, for $\Delta$ sufficiently small.

First, let $p \in (\pp,1]$.
We have
\begin{eqnarray*}
(1-\delta)m(p)+\delta\E_N v(p)
& \leq & (1-\delta)m(p)+\delta\E_N\left[ m + Cu + \mathbbm{1}_{(0,\breve{p})}(s-m-Cu)\right](p) \\
& = & m(p) + Cu(p)+\delta\E_N\left[\mathbbm{1}_{(0,\breve{p})}(s-m-Cu)\right](p) \\
& \leq & m(p)+Cu(p) \\
& = & v(p),
\end{eqnarray*}
for $\Delta$ small enough that $\E_N\left[\mathbbm{1}_{(0,\breve{p})}(s-m-Cu)\right](\pp) \leq 0$; that this inequality holds for small $\Delta$ follows from the fact that $s<m+Cu$ on $(\tilde p,\breve{p})$.
By the same token,
\begin{eqnarray*}
(1-\delta)s+\delta\E_{N-1} v(p)
& \leq & (1-\delta)s+\delta\E_{N-1} (m+Cu)(p)+\delta\E_{N-1}\left[\mathbbm{1}_{(0,\breve{p})}(s-m-Cu)\right](p)\\
& = &(1-\delta)s+\delta m(p)+\delta^{\frac{1}{N}}Cu(p)+\delta\E_{N-1}\left[\mathbbm{1}_{(0,\breve{p})}(s-m-Cu)\right](p)\\
&\leq & m(p)+Cu(p)\\
&=& v(p),
\end{eqnarray*}
for $\Delta$ small enough that $\E_{N-1}\left[\mathbbm{1}_{(0,\breve{p})}(s-m-Cu)\right](\pp) \leq 0$, as $Cu(p)>0$ and $s<m(p)$ for $p>p^m$.

Second, let $p\in(p^\natural,\pp]$. Now, we have
\begin{eqnarray*}
(1-\delta)m(p)+\delta\E_1 v(p)
& \leq & m(p)+\delta^{1-\frac{1}{N}} C u(p) + \delta\E_1\left[\mathbbm{1}_{(0,\breve{p})}(s-m-Cu)\right](p)\\
& \leq & m(p) + Cu(p) \\
& = & v(p),
\end{eqnarray*}
for $\Delta$ small enough that $\E_1\left[\mathbbm{1}_{(0,\breve{p})}(s-m-Cu)\right](p^\natural) \leq 0$.

Third, let $p\in[\breve{p},p^\natural]$. In this case,
\begin{eqnarray*}
(1-\delta)m(p)+\delta\E_1v(p)
& \leq & (1-\delta)m(p)+\delta\E_1\breve{v}(p)\\
& = & m(p)+\delta\E_1\breve{u}(p)\\
& = & m(p)+\breve{u}(p) + \mathbb{E}\left[\left. \int_0^\Delta e^{-rt}\left\{ \astrut{2} {\cal L}^1\breve{u}(p_t) - r \breve{u}(p_t) \right\} dt\, \right| p_0=p \right] \\
& \leq & m(p)+\breve{u}(p) + \mathbb{E}\left[\left. \int_0^\Delta e^{-rt}\left\{ \astrut{2} {\cal L}^1\breve{u}(p_t) - \frac{r}{\breve{x}} \breve{u}(p_t) \right\} dt\, \right| p_0=p \right] \\
& = & m(p)+\breve{u}(p) \\
& = & v(p),
\end{eqnarray*}
where the second equality follows from Dynkin's formula, the second inequality holds because $\breve{u}(p_t)>0$ and $\breve{x}>1$, and the third equality is a consequence of the identity
${\cal L}^1\breve{u} - r\breve{u}/\breve{x} = 0$.

Finally, let $p\in [0,\breve{p})$. By monotonicity of $m$ and $v$ (and the previous step), we see that $(1-\delta)m(p)+\delta\E_1v(p)\leq(1-\delta)m(\breve{p})+\delta\E_1v(\breve{p})\leq v(\breve{p})=s=v(p)$.
}

\begin{lem}\label{lem:flat-part}
There exist $\check{p} \in (p^m,1)$ and $p^\ddagger \in (p_N^*,p_1^*)$ such that $\wp(p) = s$ for all $\pp \in (\check{p},1)$, $p \leq p^\ddagger$ and $\Delta > 0$.
For any $\varepsilon > 0$, moreover, there exists $\check p_\varepsilon \in (\check p,1)$ such that $\wp \leq V_1^* + \varepsilon$ for all $\Delta > 0$.
\end{lem}
\proof{
Consider any $\pp \in (p^m,1)$ and an initial belief $p < \pp$.
We obtain an upper bound on $\wp(p)$ by considering a modified problem in which (i) the player can choose a best response in continuous time and (ii) the game is stopped with continuation payoff $m_1$ as soon as the belief $\pp$ is reached.
This problem can be solved in the standard way, yielding an optimal cutoff $p^\ddagger$.
By construction, $\wp = s$ on $[0,p^\ddagger]$.
As we take $\pp$ close to 1, $p^\ddagger$ approaches $p_1^*$ from the left and thus gets to lie strictly in between $p_N^*$ and $p_1^*$.
This proves the first statement.

The second follows from the fact that the value function of the modified problem converges uniformly to $V_1^*$ as $\pp \to 1$.
}

In the case of pure Poisson learning ($\rho = 0$), we  need a sharper characterization of the payoff function $\wp$ as $\Delta$ becomes small.
To this end, we define $\VIpp$ as the continuous-time counterpart to $\wp$.
The methods employed in \KR\ can be used to establish that $\VIpp$ has the following properties for $\rho = 0$.
First, there is a cutoff $p^\dagger < p^m$ such that $\VIpp=s$ on $[0,p^\dagger]$, and $\VIpp>s$ everywhere else.
Second, $\VIpp$ is continuously differentiable everywhere except at $\pp$.
Third, $\VIpp$ solves the Bellman equation
$$
v(p) = \max\left\{\astrut{2.5}
m(p) + [K(p)+1] b(p,v),\
s + K(p) b(p,v)\right\},
$$
where
$$
b(p,v)
= \frac{\lambda(p)}{r} \, \left[ v(j(p)) - v(p) \right]
- \frac{\lambda_1-\lambda_0}{r} \, p(1-p) \, v'(p),
$$
and
$v'(p)$ is taken to mean the left-hand derivative of $v$.
Fourth,
$b(p,\VIpp) \geq 0$ for all $p$.
Fifth, because of smooth pasting at $p^\dagger$, the term $m(p) + b(p,\VIpp) - s$ is continuous in $p$ except at $\pp$;
it has a single zero at $p^\dagger$, being positive to the right of it and negative to the left.
Finally, we note that $\VIpp = V_1^*$ and $p^\dagger = p_1^*$ for $\pp = 1$.

\begin{lem} \label{lem:convergence-punish-pbar}
Let $\rho = 0$.
Then $\VIpp \rightarrow V_1^*$ uniformly as $\pp \rightarrow 1$.
The convergence is monotone in the sense that $\pp' > \pp$ implies
$V_{1,\pp'} < \VIpp$ on $\{p\!: s < \VIpp(p) < m_1\}$.
\end{lem}

As the closed-form solutions for the functions in question make it straightforward to establish this result, we omit the proof.

A key ingredient in the analysis of the pure Poisson case is uniform convergence of $\wp$ to $\VIpp$ as $\Delta \to 0$, which we establish by means of the following result.\footnote{
To the best of our knowledge, the earliest appearance of this result in the economics literature is in Biais et al.\ (2007).
A related approach is taken in Sadzik and Stacchetti (2015).}

\begin{lem} \label{lem:convergence-fixed-point}
Let $\{T^\Delta\}_{\Delta > 0}$ be a family of contraction mappings on the Banach space $(\mathcal{W}; \|\cdot\|)$ with moduli $\{\beta^\Delta\}_{\Delta > 0}$ and associated fixed points $\{w^\Delta\}_{\Delta > 0}$.
Suppose that there is a constant $\nu > 0$ such that $1-\beta^\Delta = \nu \Delta + o(\Delta)$ as $\Delta \rightarrow 0$.
Then, a sufficient condition for $w^\Delta$ to converge in $(\mathcal{W}; \|\cdot\|)$ to the limit $v$ as $\Delta \rightarrow 0$ is that
$\|T^\Delta v - v\| = o(\Delta)$.
\end{lem}
\proof{
As
$$
\|w^\Delta - v\|
= \| T^\Delta w^\Delta - v \|
\leq \| T^\Delta w^\Delta - T^\Delta v \| + \| T^\Delta v - v \|
\leq \beta^\Delta \|w^\Delta - v\| + \|T^\Delta v - v\|,
$$
the stated conditions on $\beta^\Delta$ and $\|T^\Delta v - v\|$ imply
$$
\|w^\Delta - v\|
\leq \frac{\|T^\Delta v - v\|}{1-\beta^\Delta}
= \frac{\Delta f(\Delta)}{\nu \Delta + \Delta g(\Delta)}
= \frac{f(\Delta)}{\nu + g(\Delta)},
$$
with $\lim_{\Delta\rightarrow0} f(\Delta) = \lim_{\Delta\rightarrow0} g(\Delta) = 0$.
}

In our application of this lemma, $\cal W$ is again the Banach space of bounded real-valued functions on the unit interval, equipped with the supremum norm.
The operator in question is $\Tp$ as defined above; the corresponding moduli are $\beta^\Delta = \delta = e^{-r\Delta}$, so that $\nu = r$.

\begin{lem} \label{lem:convergence-punish-Delta}
Let $\rho = 0$.
Then $\wp \rightarrow \VIpp$ uniformly as $\Delta \rightarrow 0$.
\end{lem}
\proof{
To simplify notation, we write $v$ instead of $\VIpp$.
For $K \in \{0,1,\ldots,N\}$, a straightforward Taylor expansion of $\E_K v$ with respect to $\Delta$ yields
\begin{equation} \label{eq:Taylor}
\lim_{\Delta \rightarrow 0} \frac{1}{\Delta} \left\| \delta\, \E_K v - v - r [K b(\cdot,v) - v] \Delta \right\| = 0.
\end{equation}

For $p > \pp$, we have $K(p)=N-1$, and \eref{eq:Taylor} implies
\begin{eqnarray*}
(1-\delta) m(p) + \delta \E_Nv(p) & = & v(p) + r \left[m(p) + N b(p,v) - v(p)\right] \Delta +  o(\Delta), \\
(1-\delta) s            + \delta \E_{N-1}v(p)   & = & v(p) + r \left[s + (N-1) b(p,v) - v(p)\right] \Delta +  o(\Delta).
\end{eqnarray*}
As $m(p) > s$ on $[\pp,1]$ and $b(p,v) \geq 0$, there exists $\xi > 0$ such that
$$m(p) + N b(p,v) - \left[s + (N-1) b(p,v)\right] > \xi,$$
on $(\pp,1]$.
Thus,
$\Tp v(p) = (1-\delta) m(p) + \delta \E_N v(p)$ for $\Delta$ sufficiently small,
and the fact that
$v(p) = m(p) + N b(p,v)$ now implies
$\Tp v(p) = v(p) + o(\Delta)$ on $(\pp,1]$.

On $[0,\pp]$, we have $K(p)=0$, and \eref{eq:Taylor} implies
\begin{eqnarray}
\label{eq:taylor1}
\left\|(1-\delta)m+\delta\E_1 v-v-r[m+b(\cdot,v)-v)\Delta\right\| & = & \Delta \psi_R(\Delta), \\
\label{eq:taylor2}
\left\|(1-\delta)s+\delta\E_0 v-v-r[s-v]\Delta\right\| & = & \Delta \psi_S(\Delta),
\end{eqnarray}
for some functions $\psi_R,\psi_S\!:(0,\infty)\to[0,\infty)$ that satisfy $\psi_R(\Delta)\rightarrow 0$ and $\psi_S(\Delta)\rightarrow 0$ as $\Delta\rightarrow 0$.

First, let $p\in (p^\dagger,\pp]$.
We note that $\Tp v(p)\geq (1-\delta)m(p)+\delta\E_1 v(p)\geq v(p)-\Delta \psi_R(\Delta)$ in this range,
where the first inequality follows from the definition of $\Tp$, and the second inequality is implied by \eref{eq:taylor1} and $v(p) = m(p) + b(p,v) $ for $p\in (p^\dagger,\pp]$.
If the maximum in the definition of $\Tp v(p)$ is achieved by the risky action, the first in the previous chain of inequalities holds as an equality, and \eref{eq:taylor1} immediately implies that $\Tp v(p) = v(p) + o(\Delta)$.
If the maximum in the definition of $\Tp v(p)$ is achieved by the safe action, however, we have $\Tp v(p)=(1-\delta)s+\delta\E_0 v(p)\leq v(p)+r[s-v(p)]\Delta+\Delta \psi_S(\Delta)\leq v(p)+\Delta\psi_S(\Delta)$,
where the second inequality follows from $v>s$ on $(p^\dagger,\pp]$.
Thus $v(p)-\Delta\psi_R(\Delta)\leq \Tp v(p)\leq v(p)+\Delta \psi_S(\Delta)$, and we can conclude that $\Tp v(p) = v(p) + o(\Delta)$ in this case as well.

Now, let $p \leq p^\dagger$.
We note that $\Tp v(p)\geq (1-\delta)s+\delta\E_0 v(p)\geq v(p)-\Delta \psi_S(\Delta)$ in this range,
where the first inequality follows from the definition of $\Tp$, and the second inequality is implied by \eref{eq:taylor2} and $v(p) = s $ for $p\leq p^\dagger$.
If the maximum in the definition of $\Tp v(p)$ is achieved by the safe action, the first in the previous chain of inequalities holds as an equality, and \eref{eq:taylor2} immediately implies that $\Tp v(p) = v(p) + o(\Delta)$.
If the maximum in the definition of $\Tp v(p)$ is achieved by the risky action, however, we have $\Tp v(p)=(1-\delta)m(p)+\delta\E_1 v(p)\leq v(p)+r[m(p)+b(p,v)-v(p)]\Delta+\Delta \psi_R(\Delta)\leq v(p)+\Delta\psi_R(\Delta)$,
where the second inequality follows from $v=s\geq m(p)+b(p,v)$ on $[0,p^\dagger]$.
Thus $v(p)-\Delta\psi_S(\Delta)\leq \Tp v(p)\leq v(p)+\Delta \psi_R(\Delta)$, and we can again conclude that $\Tp v(p) = v(p) + o(\Delta)$ in this case as well.
}

Our last two auxiliary results pertain to the case of pure Poisson learning.

\begin{lem}\label{lem:thresh}
Let $\rho = 0$.
There is a belief $\phat \in [p_N^*,p_1^*]$ such that
$$
\lambda(\underline{p}) \left[ N \VNulp(j(\underline{p})) - (N-1) V_1^*(j(\underline{p})) - s \right] - rc(\underline{p})
$$
is negative if $0 < \underline{p} < \phat$, zero if $\underline{p} = \phat$, and positive if $\phat < \underline{p} <1$.
Moreover, $\hat{p} = p_N^*$ if, and only if, $j(p^*_N)\leq p_1^*$,
and $\hat{p} = p_1^*$ if, and only if, $\lambda_0 = 0$.
\end{lem}
\proof{
We start by noting that given the functions $V_1^*$ and $V_N^*$, the cutoffs $p_1^*$ and $p_N^*$ are uniquely determined by
\begin{equation} \label{eq:p_1^*}
\lambda(p_1^*)[V_1^*(j(p_1^*))-s] = rc(p_1^*),
\end{equation}
and
\begin{equation} \label{eq:p_N^*}
\lambda(p_N^*)[NV_N^*(j(p_N^*))-Ns] = rc(p_N^*),
\end{equation}
respectively.

Consider the differentiable function $f$ on $(0,1)$ given by
\[
f(\underline{p})=\lambda(\underline{p})[N\VNulp(j(\underline{p}))-(N-1)V_1^*(j(\underline{p}))-s]-rc(\underline{p}).
\]

For $\lambda_0 = 0$, we have $j(\underline{p})=1$ and $\VNulp(j(\underline{p}))=V_1^*(j(\underline{p}))=m_1$ for all $p$, so
$f(\underline{p})=\lambda(\underline{p})[V_1^*(j(\underline{p}))-s] - rc(\underline{p})$,
which is zero at $\underline{p}=p_1^*$ by \eref{eq:p_1^*}, positive for $\underline{p}>p_1^*$, and negative for $\underline{p}<p_1^*$.

Assume $\lambda_0 > 0$.
For $0 < \underline{p} < p \leq 1$, we have $\VNulp(p)=m(p)+c(\underline{p}) u(p;\mu_N)/u(\underline{p};\mu_N)$.
Moreover, we have $V_1^*(p)=s$ when $p \leq p_1^*$, and $V_1^*(p)=m(p)+Cu(p;\mu_1)$ with a constant $C>0$ otherwise.
Using the fact that
$$u(j(p);\mu)=\frac{\lambda_0}{\lambda(p)}\left(\frac{\lambda_0}{\lambda_1}\right)^\mu u(p;\mu),$$
we see that the term $\lambda(\underline{p})N\VNulp(j(\underline{p}))$ is actually linear in $\underline{p}$.
When $j(\underline{p}) \leq p_1^*$, the term $-\lambda(\underline{p})(N-1)V_1^*(j(\underline{p}))$ is also linear in $\underline{p}$;
when $j(\underline{p}) > p_1^*$, the nonlinear part of this term simplifies to
$-(N-1) C \lambda_0^{\mu_1+1} u(\underline{p};\mu_1)/\lambda_1^{\mu_1}$.
This shows that $f$ is concave, and strictly concave on the interval of all $\underline{p}$ for which $j(\underline{p}) > p_1^*$.
As $\lim_{\underline{p} \rightarrow 1} f(\underline{p}) > 0$, this in turn implies that $f$ has at most one root in the open unit interval; if so, $f$ assumes negative values to the left of the root, and positive values to the right.

As $V_{N,p_1^*}(j(p_1^*))>V_1^*(j(p_1^*))$, moreover, we have
$
f(p_1^*) > \lambda(p_1^*)[V_1^*(j(p_1^*))-s]-rc(p_1^*) = 0
$
by \eref{eq:p_1^*}.
Any root of $f$ must thus lie in $[0,p_1^*)$.
If $j(p_N^*) \leq p_1^*$, then $V_1^*(j(p_N^*)) = s$ and
$
f(p_N^*)=\lambda(p_N^*)[NV_N^*(j(p_N^*))-Ns]-rc(p_N^*) = 0
$
by \eref{eq:p_N^*}.
If $j(p_N^*) > p_1^*$, then $V_1^*(j(p_N^*)) > s$ and $f(p_N^*) < 0$, so $f$ has a root in $(p_N^*,p_1^*)$.
}

The following result is used in the proof of Proposition \ref{prop:efficiency-region-Poisson}.

\begin{lem} \label{lem:comparison-mu1-muN}
 Let $\rho = 0$.
Then $\mu_1 (\mu_1+1) > N \mu_N (\mu_N+1)$.
\end{lem}
\proof{
We change variables to $\beta =\lambda_0/\lambda_1$ and $x=r/\lambda_1$, so that $\mu_N$ and $\mu_1$ are implicitly defined as the positive solutions of the equations
\begin{eqnarray*}
\frac{x}{N} + \beta - (1-\beta) \mu_N & = & \beta^{\mu_N+1}, \\
x + \beta - (1-\beta) \mu_1 & = & \beta^{\mu_1+1}.
\end{eqnarray*}
Fixing $\beta \in [0,1)$ and considering $\mu_N$ and $\mu_1$ as functions of $x \in (0,\infty)$, we obtain
\begin{eqnarray*}
\mu_N' & = & \frac{N^{-1}}{1-\beta+\beta^{\mu_N+1}\ln\beta} \ \ = \ \ \frac{N^{-1}}{1-\beta+\left[\divn{x}{N} + \beta - (1-\beta) \mu_N\right] \ln\beta} \,, \\
\mu_1' & = & \frac{1}{1-\beta+\beta^{\mu_1+1}\ln\beta} \ \ = \ \ \frac{1}{1-\beta+\left[x + \beta - (1-\beta) \mu_1\right] \ln\beta} \,.
\end{eqnarray*}
(All denominators are positive because $1-\beta+\beta^{\mu+1}\ln\beta \geq 1-\beta+\beta \ln\beta > 0$ for all $\mu \geq 0$.)

Let $d = \mu_1 (\mu_1+1) - N \mu_N (\mu_N+1)$.
As $\lim_{x \rightarrow 0} \mu_N = \lim_{x \rightarrow 0} \mu_1 = 0$, we see that $\lim_{x \rightarrow 0} d = 0$ as well.
It is thus enough to show that $d' > 0$ at any $x > 0$.
This is the case if, and only if, $(2\mu_1+1)\mu_1' > N (2\mu_N+1) \mu_N'$, that is,
$$
(2\mu_1+1) \left\{1-\beta+\left[\divn{x}{N} + \beta - (1-\beta) \mu_N\right]\ln\beta\right\} > (2\mu_N+1) \left\{1-\beta+\left[x + \beta - (1-\beta) \mu_1\right]\ln\beta\right\}.
$$
This inequality reduces to
$$
(\mu_1-\mu_N) \left\{ 2(1-\beta) + \left[\divn{2x}{N} + 1 + \beta\right]\ln\beta\right\} > (2\mu_N+1) \left[x-\divn{x}{N}\right] \ln\beta.
$$
It is straightforward to show that $\mu_1 > \mu_N + \frac{1}{1-\beta}\left[x-\divn{x}{N}\right]$.
So $d' > 0$ if
$$
2(1-\beta) + \left[\divn{2x}{N} + 1 + \beta\right]\ln\beta > (2\mu_N+1) (1-\beta) \ln\beta,
$$
which simplifies to $1-\beta+\left[\divn{x}{N} + \beta - (1-\beta) \mu_N\right] \ln\beta > 0$ -- an inequality that we have already established.
}

\section{Proofs} \label{app:proofs}

\subsection{Main Results (Theorem 1 and Propositions \ref{prop:comparison_sym_MPE}--\ref{prop:efficiency-region-Poisson})}

\proofof{Theorem}{thm}{
For $\rho > 0$, this result is an immediate consequence of inequalities \eref{eq:payoffs},
the fact that $\liminf_{\Delta \rightarrow 0} \WinfPBE \geq V_1^*$ and $\WsupPBE \leq V_N^*$,
and Proposition \ref{prop:limit-Brownian}.
For $\rho = 0$, the result follows from inequalities \eref{eq:payoffs},
the fact $\liminf_{\Delta \rightarrow 0} \WinfPBE \geq V_1^*$,
and Propositions \ref{prop:thresh} and \ref{prop:limit-Poisson}.
}

\proofof{Proposition}{prop:comparison_sym_MPE}{
Arguing as in \KR, one establishes that in the unique symmetric MPE of the continuous-time game, all experimentation stops at the belief $\ptilde_N$ implicitly defined by
$rc(\ptilde_N) =  \lambda(\ptilde_N)[\tilde{V}_N(j(\ptilde_N))-s]$,
where $\tilde{V}_N$ is the players' common equilibrium payoff function.
The equilibrium construction along the lines of \KR\ further implies that $V_{N,\ptilde_N}(j(\ptilde_N)) > \tilde{V}_N(j(\ptilde_N)) > V_1^*(j(\ptilde_N))$,
so that
$NV_{N,\ptilde_N}(j(\ptilde_N))-(N-1)V_1^*(j(\ptilde_N)) > \tilde{V}_N(j(\ptilde_N))$,
and hence $\hat{p} < \ptilde_N$ by Lemma \ref{lem:thresh}.
}

\proofof{Proposition}{prop:comp-statics-N}{
We only need to consider the case that $\hat{p}>p_N^*$.

Recall the defining equation for $\hat{p}$ from Lemma \ref{lem:thresh},
\[
 \lambda(\hat{p} )NV_{N,\hat{p}}(j(\hat{p} ))-\lambda (\hat{p} )s-rc(\hat{p} )=(N-1) \lambda (\hat{p} )V_1^*(j(\hat{p} )).
 \]
We make use of the closed-form expression for $V_{N,\hat{p}}$ to rewrite its left-hand side as
\[
N\lambda (\hat{p})\lambda(j(\hat{p}))h+Nc(\hat{p})[\lambda_0-\mu_N(\lambda_1-\lambda_0)]-\lambda(\hat{p})s.
\]
Similarly, by noting that $\hat{p}>p_N^*$ implies $j(\hat{p})> j(p_N^*)>p_1^*$, we can make use of the closed-form expression for $V_1^*$ to rewrite the right-hand side as
\[
(N-1)\lambda (\hat{p})\lambda(j(\hat{p}))h+(N-1) c(p_1^*)\frac{u(\hat{p};\mu_1)}{u(p_1^*;\mu_1)}[r+\lambda_0-\mu_1(\lambda_1-\lambda_0)].
\]
Combining, we have
\[
\frac{\lambda (\hat{p})\lambda(j(\hat{p}))h+Nc(\hat{p})[\lambda_0-\mu_N(\lambda_1-\lambda_0)]-\lambda(\hat{p})s}{(N-1) [r+\lambda_0-\mu_1(\lambda_1-\lambda_0)] c(p_1^*)}=\frac{u(\hat{p};\mu_1)}{u(p_1^*;\mu_1)}.
\]

It is convenient to change variables to
\[
\beta=\frac{\lambda_0}{\lambda_1} \quad \mbox{and} \quad y=\frac{\lambda_1}{\lambda_0} \, \frac{m_1-s}{s-m_0} \, \frac{\hat{p}}{1-\hat{p}}\,.
\]
The implicit definitions of $\mu_1$ and $\mu_N$ imply
\[
N=\frac{\beta^{1+\mu_1}-\beta+\mu_1(1-\beta)}{\beta^{1+\mu_N}-\beta+\mu_N(1-\beta)}\,,
\]
allowing us to rewrite the defining equation for $\hat{p}$ as the equation $F(y,\mu_N) = 0$ with
\begin{eqnarray*}
F(y,\mu)
& = & 1 - y + [\beta(1+\mu)y-\mu]\,\frac{1-\beta}{\beta}\,\frac{\beta^{1+\mu_1}-\beta+\mu_1(1-\beta)}{(\mu_1-\mu)(1-\beta)+\beta^{1+\mu_1}-\beta^{1+\mu}} \\
&   & \mbox{} - \frac{\mu_1^{\mu_1}}{(1+\mu_1)^{1+\mu_1}} \, y^{-\mu_1}.
\end{eqnarray*}
As $y$ is a strictly increasing function of $\hat{p}$, we know from Lemma \ref{lem:thresh} that $F(\cdot,\mu_N)$ admits a unique root, and that it is strictly increasing in a neighborhood of this root.

A straightforward computation shows that
\[
\frac{\partial F(y,\mu_N)}{\partial \mu}
= \frac{1-\beta}{\beta}\,\frac{\beta^{1+\mu_1}-\beta+\mu_1(1-\beta)}{((\mu_1-\mu_N)(1-\beta)+\beta^{1+\mu_1}-\beta^{1+\mu_N})^2}\ \zeta(y,\mu_N),
\]
with
\[
\zeta(y,\mu) = \beta(1-\beta)(1+\mu_1)y - (1-\beta)\mu_1 + (1-\beta y)(\beta^{1+\mu}-\beta^{1+\mu_1}) + \beta^{1+\mu}\,(\beta(1+\mu)y-\mu)\ln \beta.
\]
As $p_N^* < \hat{p} < p_1^*$, we have
\[
\frac{\mu_N}{1+\mu_N} < \beta y < \frac{\mu_1}{1+\mu_1}\,,
\]
which implies
\[
\zeta(y,\mu_1) = (\beta(1+\mu_1)y-\mu_1)\,(1-\beta+\beta^{1+\mu_1}\ln \beta) < 0,
\]
and
\[
\frac{\partial \zeta(y,\mu)}{\partial \mu} = \beta^{1+\mu} [\beta(1+\mu)y-\mu] (\ln \beta)^2 > 0,
\]
for all $\mu \in [\mu_N,\mu_1]$.
This establishes $\zeta(y,\mu_N) < 0$.

By the implicit function theorem, therefore, $y$ is increasing in $\mu_N$.
Recalling from Section \ref{sec:continuous-time} that $\mu_N$ is decreasing in $N$, we have shown that $y$ (and hence $\hat{p}$) are decreasing in $N$.
}

\proofof{Proposition}{prop:efficiency-region-Poisson}{
There is nothing to show for $\lambda_0 = 0$.
Using the same change of variables as in the proof of Lemma \ref{lem:comparison-mu1-muN}, we fix $\beta \in (0,1)$, therefore, and define
$$
q = \beta \cdot \frac{1+\mu_N^{-1}}{1+\mu_1^{-1}}\,,
$$
so that $j(p_N^*) \leq p_1^*$ if, and only if, $q \geq 1$.
As $\lim_{x \rightarrow \infty} \mu_N = \lim_{x \rightarrow \infty} \mu_1 = \infty$,
we have $\lim_{x \rightarrow \infty} q = \beta < 1$.
As $\lim_{x \rightarrow 0} \mu_N = \lim_{x \rightarrow 0} \mu_1 = 0$, moreover,
$$\lim_{x \rightarrow 0} q
= \beta \lim_{x \rightarrow 0} \frac{\mu_1}{\mu_N}
= \beta \lim_{x \rightarrow 0} \frac{\mu_1'}{\mu_N'}
= \beta N$$
by l'H\^{o}pital's rule.
Finally, $q'$ is easily seen to have the same sign as
$$
- \mu_1 (\mu_1+1) (1-\beta+\beta^{\mu_1+1}\ln\beta) + N \mu_N (\mu_N+1) (1-\beta+\beta^{\mu_N+1}\ln\beta).
$$
As $\beta^{\mu_1+1}\ln\beta > \beta^{\mu_N+1}\ln\beta$, Lemma \ref{lem:comparison-mu1-muN} implies that $q$ decreases strictly in $x$.
This in turn implies that $q < 1$ at all $x \in (0,\infty)$ when $\beta N \leq 1$, which proves the first part of the proposition.
Otherwise, there exists a unique $x^* \in (0,\infty)$ at which $q = 1$.
The second part of the proposition thus holds with $(\lambda_1^*,\lambda_0^*) = (r/x^*,\beta r/x^*)$.

It is straightforward to see that $x$ varies continuously with $\beta$ and that $\lim_{\beta \rightarrow 1/N} x^* = 0$.
So it remains to show that $x^*$ remains bounded as $\beta \rightarrow 1$.
Rewriting the defining equation for $x^*$ as
$$
1+\frac{1}{(1-\beta)\mu_1(x^*(\beta),\beta)}=\frac{1}{(1-\beta)\mu_N(x^*(\beta),\beta)}\,,
$$
we see that $(1-\beta)\mu_N(x^*(\beta),\beta)$ must stay bounded as $\beta \rightarrow 1$.
By the defining equation for $\mu_N$, $x^*(\beta)$ must then also stay bounded.
}

\subsection{Learning with a Brownian Component (Propositions \ref{prop:SSE-Brownian}--\ref{prop:limit-Brownian})}

The proof of Proposition \ref{prop:SSE-Brownian} rests on a sequence of lemmas that prove incentive compatibility of the proposed strategies on various subintervals of $[0,1]$.
When no assumption on the signal-to-noise ratio $\rho$ is stated, the respective result holds irrespectively of whether $\rho > 0 $ or $\rho = 0$.

In view of Lemmas
\ref{lem:upper-bound-punishment} and \ref{lem:flat-part},
we take $\pr$ and $\pp$ such that
\begin{equation} \label{eq:thresholds-Brownian}
p_N^* < \pr < p^\ddagger < p_1^* < p^m < \max\{p^\diamond,\check{p}\} < \pp < 1.
\end{equation}

The first two lemmas deal with the safe action ($\kappa = 0$) on the interval $[0,\pp]$.

\begin{lem}\label{lem:k=0;0-pddagger}
For all $p \leq p^\ddagger$,
$$
(1-\delta) s + \delta \wr(p) \geq (1-\delta) m(p) + \delta \E_1\wp(p).
$$
\end{lem}
\proof{
As $\wr(p) \geq s = \wp(p)$ for $p \leq p^\ddagger$,
we have
$(1-\delta) s + \delta \wr(p) \geq s$
whereas $s \geq (1-\delta) m(p) + \delta \E_1\wp(p)$ by the functional equation for $\wp$.
}

\begin{lem}\label{lem:k=0;pddagger-pp}
There exists $\Delta_{(p^\ddagger,\pp]} > 0$ such that
$$
(1-\delta) s + \delta \wr(p) \geq (1-\delta) m(p) + \delta \E_1\wp(p),
$$
for all $p \in (p^\ddagger,\pp]$ and $\Delta < \Delta_{(p^\ddagger,\pp]}$.
\end{lem}
\proof{
By Lemmas \ref{lem:lower-bound-reward} and \ref{lem:upper-bound-punishment}, there exist $\nu > 0$ and $\Delta_0 > 0$ such that
$\wr(p) - \wp(p) \geq \nu$ for all $p \in [p^\ddagger, \pp]$ and $\Delta < \Delta_0$.
Further, there is a $\Delta_1 \in (0,\Delta_0]$ such that
$|\E_1 \wp(p) - \wp(p)| \leq \frac{\nu}{2}$ for all $p \in [p^\ddagger, \pp]$ and $\Delta < \Delta_1$.
For these $p$ and $\Delta$, we thus have
$$
(1-\delta) s + \delta \wr(p) - \left[ (1-\delta) m(p) + \delta \E_1\wp(p) \right]
\geq (1-\delta) [s-m(p)] + \delta \, \frac{\nu}{2}.
$$
Finally, there is a $\Delta_{(p^\ddagger,\pp]} \in (0,\Delta_1]$ such that the right-hand side of this inequality is positive for all $p \in (p^\ddagger,\pp]$ and $\Delta < \bar{\Delta}$.
}

We establish incentive compatibility of the risky action ($\kappa = 1$) to the immediate right of $\pr$ by means of the following result.

\begin{lem}\label{lem:Gaussian-variable}
Let $X$ be a Gaussian random variable with mean $m$ and variance $V$.
\begin{enumerate}
\item For all $\eta > 0$,
\[
\mathbb{P}\!\left[X - m > \eta\right] < \frac{V}{\eta^2}\,.
\]
\item There exists $\overline{V} \in (0,1)$ such that for all $V < \overline{V}$,
\[
\mathbb{P}\!\left[ V^{\frac{3}{4}} \le X - m \le V^{\frac{1}{4}}\right] \ge \frac{1}{2}-V^{\frac{1}{4}}\,.
\]
\end{enumerate}
\end{lem}
\proof{
The first statement is a trivial consequence of Chebysheff's inequality. The proof of the second relies on the following inequality (13.48) of Johnson et al.\ (1994) for the standard normal cumulative distribution function:
\[
\frac{1}{2}\left[1+(1-e^{-x^2/2})^{\frac{1}{2}}\right] \le \Phi(x) \le \frac{1}{2}\left[1+(1-e^{-x^2})^{\frac{1}{2}}\right].
\]
Letting $\Phi^{V}$ denote the cdf of the Gaussian distribution with variance $V$ (and mean 0),
and using the above upper and lower bounds, we have
\begin{equation*}
\frac{\frac{1}{2}+\Phi^{V}(V^{\frac{3}{4}})-\Phi^{V}(V^{\frac{1}{4}})}{ \sqrt[4]{V }}
\le \frac{1-\sqrt{1-e^{-\frac{1}{2 \sqrt{V}}}}+\sqrt{1-e^{-\sqrt{V }}}}{2 \sqrt[4]{V }}\,.
\end{equation*}
Writing $x=\sqrt{V}$ and using the fact that $1 - \sqrt{1-y} \le \sqrt{y}$ for $0 \leq y \le 1$, moreover, we have
\[
\frac{1-\sqrt{1-e^{-\frac{1}{2 x}}}+\sqrt{1-e^{-x}}}{2 \sqrt{x}}\le \frac{1}{2}\sqrt{\frac{e^{-\frac{1}{2x}}}{x}}+\frac{1}{2} \sqrt{\frac{1-e^{-x}}{x}}\rightarrow \frac{1}{2},
\]
as $x\rightarrow 0$.
Thus,
\begin{equation*}
\frac{\frac{1}{2}+\Phi^{V}(V^{\frac{3}{4}})-\Phi^{V}(V^{\frac{1}{4}})}{ \sqrt[4]{V }}\le 1,
\end{equation*}
for sufficiently small $V$, which is the second statement of the lemma.
}
We  apply this lemma to the log odds ratio
$\ell$ associated with the current belief $p$.
For later use, we note that
$dp/d\ell = p\,(1-p)$.

\begin{lem}\label{lem:k=1;pr-pr+epsilon-Brownian}
Let $\rho > 0$. There exist $\varepsilon \in (0,p^\ddagger-\pr)$ and $\Delta_{(\pr,\pr+\varepsilon]} > 0$ such that
$$
(1-\delta) m(p) + \delta \E_{N}\wr(p) \geq (1-\delta) s + \delta \E_{N-1}\wp(p),
$$
for all $p \in (\pr,\pr+\varepsilon]$ and $\Delta < \Delta_{(\pr,\pr+\varepsilon]}$.
\end{lem}
\proof{
Consider a belief $p_0 = p$ and the corresponding log odds ratio $\ell$.
Let $K$ players use the risky arm on the time interval $[0,\Delta)$ and consider the resulting belief $p^{(K)}_\Delta$ and the associated log odds ratio $\ell^{(K)}_\Delta$.

Let $\mathbb{P}_\theta$ denote the probability measure associated with state $\theta \in \{0,1\}$.
Expected continuation payoffs are computed by means of the measure $\mathbb{P}_p = p \mathbb{P}_1 +(1-p) \mathbb{P}_0$.

Let $J_0^\Delta$ denote the event that no lump-sum arrives by time $\Delta$.
The probability of $J_0^\Delta$ under the measure $\mathbb{P}_\theta$ is $e^{-\lambda_\theta \Delta}$.
Note that
$$
e^{-\lambda_\theta \Delta} \mathbb{P}_\theta[A \mid J_0^\Delta]
\le \mathbb{P}_\theta[A]
\le e^{-\lambda_\theta \Delta} \mathbb{P}_\theta[A \mid J_0^\Delta] + 1 - e^{-\lambda_\theta \Delta},
$$
for any event $A$.

As we have seen in Appendix \ref{app:beliefs}, conditional on $J_0^\Delta$, the random variable $\ell^{(K)}_\Delta$ is normally distributed with mean
$\ell - K \left(\lambda_1 - \lambda_0 - \frac{\rho}{2}\right) \Delta$ and variance $K \rho \Delta$ under $\mathbb{P}_1$,
and normally distributed with mean
$\ell - K \left(\lambda_1 - \lambda_0 + \frac{\rho}{2}\right) \Delta$ and variance $K \rho \Delta$ under $\mathbb{P}_0$.

Now choose $\varepsilon > 0$ such that $\pr + \varepsilon < p^\ddagger$.
Write $\underline{\ell}$, $\ell_\varepsilon$, $\ell^\ddagger$ and $\bar{\ell}$ for the log odds ratios associated with $\pr$, $\pr + \varepsilon$, $p^\ddagger$ and $\pp$, respectively.
Choose $\Delta_0 > 0$ such that
$$
\nu_0 = \min_{(\Delta,\ell)\in[0,\Delta_0]\times[\underline{\ell},\ell_\varepsilon]} \left[\ell^\ddagger - \ell + (N-1) \left(\lambda_1 - \lambda_0 - \frac{\rho}{2}\right) \Delta\right]^2 > 0.
$$
For all $p\in(\pr,\pr+\varepsilon]$ and $\Delta \in (0,\Delta_0)$, the first part of Lemma \ref{lem:Gaussian-variable} now implies
\begin{eqnarray*}
\mathbb{P}_p\!\left[ p^{(N-1)}_\Delta > p^\ddagger \right]
& = & \mathbb{P}_p\!\left[ \ell^{(N-1)}_\Delta > \ell^\ddagger \right] \\
& \le & p \left\{ e^{-\lambda_1 \Delta} \mathbb{P}_1\!\left[ \left. \ell^{(N-1)}_\Delta > \ell^\ddagger \right| J_0^\Delta \right] + 1 - e^{-\lambda_1 \Delta} \right\} \\
&     & \mbox{} + (1-p) \left\{ e^{-\lambda_0 \Delta} \mathbb{P}_0\!\left[ \left. \ell^{(N-1)}_\Delta > \ell^\ddagger \right| J_0^\Delta \right] + 1 - e^{-\lambda_0 \Delta} \right\} \\
& \le & p \left\{ \frac{e^{-\lambda_1 \Delta} (N-1) \rho \Delta}{\nu_0} + 1 - e^{-\lambda_1 \Delta} \right\} \\
&     & \mbox{} + (1-p) \left\{ \frac{e^{-\lambda_0 \Delta} (N-1) \rho \Delta}{\nu_0} + 1 - e^{-\lambda_0 \Delta} \right\} \\
& \le & \frac{(N-1) \rho \Delta}{\nu_0} + 1 - e^{-\lambda_1 \Delta} \\
& \le & \left\{ \frac{(N-1) \rho}{\nu_0} + \lambda_1 \right\} \Delta.
\end{eqnarray*}
As $\wp\leq s+ (m_1-s) \mathbbm{1}_{(p^\ddagger,1]}$, moreover,
$$
\E_{N-1}\wp(p) \leq s + (m_1-s) \, \mathbb{P}_p\!\!\left[ p^{(N-1)}_\Delta > p^\ddagger\right].
$$
So there exists $C_0 > 0$ such that
$\E_{N-1}\wp(p) \leq s + C_0 \Delta$
for all $p\in(\pr,\pr+\varepsilon]$ and $\Delta \in (0,\Delta_0)$.

Next, define
$\nu_1 = \min_{\pr \leq p \leq \pp} p \, (1-p)$
and note that for $\pr \leq p \leq \pp$ (and thus for $\underline{\ell} \leq \ell \leq \bar{\ell}$),
$$
\VNpr(p) \geq s + \max\left\{0, \VNpr'(\pr+) (p - \pr)\right\} \geq s + \max\left\{0, \VNpr'(\pr+) \nu_1 (\ell - \underline{\ell})\right\}.
$$
By the second part of Lemma \ref{lem:Gaussian-variable}, there exists $\Delta_1 > 0$ such that $N \rho \Delta_1 < 1$ and
$$
\mathbb{P}_1\!\left[ \left. (N \rho \Delta)^{\frac{3}{4}} \le \ell^{(N)}_\Delta - \ell + N \left(\lambda_1 - \lambda_0 - \frac{\rho}{2}\right) \Delta \le (N \rho \Delta)^{\frac{1}{4}} \right| J_0^\Delta \right]
\ge \frac{1}{2} - (N \rho \Delta)^{\frac{1}{4}},
$$
for arbitrary $\ell$ and all $\Delta \in (0,\Delta_1)$.
In particular,
\begin{eqnarray*}
\lefteqn{\mathbb{P}_p\!\left[ (N \rho \Delta)^{\frac{3}{4}} \le \ell^{(N)}_\Delta - \ell + N \left(\lambda_1 - \lambda_0 - \frac{\rho}{2}\right) \Delta \le (N \rho \Delta)^{\frac{1}{4}} \right]} \\
& \ge & p \mathbb{P}_1\!\left[ (N \rho \Delta)^{\frac{3}{4}} \le \ell^{(N)}_\Delta - \ell + N \left(\lambda_1 - \lambda_0 - \frac{\rho}{2}\right) \Delta \le (N \rho \Delta)^{\frac{1}{4}} \right] \\
& \ge & p e^{-\lambda_1 \Delta} \mathbb{P}_1\!\left[ \left. (N \rho \Delta)^{\frac{3}{4}} \le \ell^{(N)}_\Delta - \ell + N \left(\lambda_1 - \lambda_0 - \frac{\rho}{2}\right) \Delta \le (N \rho \Delta)^{\frac{1}{4}} \right| J_0^\Delta \right] \\
& \ge & p e^{-\lambda_1 \Delta} \left(\frac{1}{2} - (N \rho \Delta)^{\frac{1}{4}}\right),
\end{eqnarray*}
for these $\Delta$.
Taking $\Delta_1$ smaller if necessary, we can also ensure that
$$
\underline{\ell}
< \ell - N \left(\lambda_1 - \lambda_0 - \frac{\rho}{2}\right) \Delta + (N \rho \Delta)^{\frac{3}{4}}
< \ell - N \left(\lambda_1 - \lambda_0 - \frac{\rho}{2}\right) \Delta + (N \rho \Delta)^{\frac{1}{4}}
< \bar{\ell},
$$
for all $\ell \in (\underline{\ell},\ell_\varepsilon]$ and all $\Delta \in (0,\Delta_1)$.

By Lemma \ref{lem:lower-bound-reward}, there exists $\Delta_2 \in (0,\Delta_1)$ such that $\wr \geq \VNpr$ for $\Delta \in (0,\Delta_2)$.
For such $\Delta$ and $p \in (\pr, \pr+\varepsilon]$, we now have
\begin{eqnarray*}
\E_N \wr(p)
\!\! & \geq & \!\! s + p e^{-\lambda_1 \Delta} \left(\frac{1}{2} - (N \rho \Delta)^{\frac{1}{4}}\right) \VNpr'(\pr+) \, \nu_1 \left[ \ell -  N \left(\lambda_1 - \lambda_0 - \frac{\rho}{2}\right) \Delta  + (N \rho \Delta)^{\frac{3}{4}} - \underline{\ell} \right] \\
& \geq & \!\! s + \pr (1 - \lambda_1 \Delta) \left(\frac{1}{2} - (N \rho \Delta)^{\frac{1}{4}}\right) \VNpr'(\pr+) \, \nu_1 \left[ - N \left(\lambda_1 - \lambda_0 - \frac{\rho}{2}\right) \Delta  + (N \rho \Delta)^{\frac{3}{4}} \right].
\end{eqnarray*}
This implies the existence of $\Delta_3 \in (0,\Delta_2)$ and $C_1 > 0$ such that
$$
\E_N \wr(p) \geq s + C_1 \Delta^{\frac{3}{4}},
$$
for all $p\in(\pr,\pr+\varepsilon]$ and $\Delta \in (0,\Delta_3)$.

For $p\in(\pr,\pr+\varepsilon]$ and $\Delta \in (0,\min\{\Delta_0,\Delta_3\})$, finally,
\begin{eqnarray*}
\lefteqn{(1-\delta) m(p) + \delta \E_{N}\wr(p) - \left[(1-\delta) s + \delta \E_{N-1}\wp(p)\right]}\\
& \ge & (1-\delta) [m(\pr)-s] + \delta \left\{ C_1 \Delta^{\frac{3}{4}} - C_0 \Delta \right\} \\
& = & C_1 \Delta^{\frac{3}{4}} - \left\{r [s-m(\pr)] + C_0 \right\} \Delta + o(\Delta).
\end{eqnarray*}
As the term in $\Delta^{\frac{3}{4}}$ dominates as $\Delta$ becomes small, there exists
$\Delta_{(\pr,\pr+\varepsilon]} \in (0,\min\{\Delta_0,\Delta_3\})$ such that this expression is positive for all $p \in (\pr,\pr+\varepsilon]$ and $\Delta < \Delta_{(\pr,\pr+\varepsilon]}$.
}

\begin{lem}\label{lem:k=1;pr+epsilon-pp}
For all $\varepsilon \in (0,p^\ddagger-\pr)$, there exists $\Delta_{(\pr+\varepsilon,\pp]} > 0$ such that
$$
(1-\delta) m(p) + \delta \E_N\wr(p) \geq (1-\delta) s + \delta \E_{N-1}\wp(p),
$$
for all $p \in (\pr+\varepsilon,\pp]$ and $\Delta < \Delta_{(\pr+\varepsilon,\pp]}$.
\end{lem}
\proof{
First, by Lemma \ref{lem:lower-bound-reward}, there exists $\Delta_0 > 0$ such that $\wr \geq \VNpr$ on the unit interval.
Second, by Lemma \ref{lem:upper-bound-punishment}, there exist $\nu > 0$, $\eta > 0$ and $\Delta_1 \in (0, \Delta_0)$ such that
$\VNpr(p) - \wp(p) \geq \nu$ for all $p \in [\pr+\frac{\varepsilon}{2}, \pp+\eta]$ and $\Delta < \Delta_1$.
For these $p$ and $\Delta$, and by convexity of $\VNpr$, we then have
\begin{eqnarray*}
\E_N \wr(p) - \E_{N-1} \wp(p)
& \geq & \E_N \VNpr(p) - \E_{N-1} \wp(p) \\
& \geq & \E_{N-1} \VNpr(p) - \E_{N-1} \wp(p) \\
& \geq & \chi^\Delta(p) \nu + [1 - \chi^\Delta(p)] (s - m_1),
\end{eqnarray*}
where $\chi^\Delta(p)$ denotes the probability that the belief $p_{t+\Delta}$ lies in $[\pr+\frac{\varepsilon}{2}, \pp+\eta]$ given that $p_t = p$ and $N-1$ players use the risky arm for a length of time $\Delta$.
Next, there exists $\Delta_2 \in (0, \Delta_1)$ such that
$$
\chi^\Delta(p) \geq \frac{\frac{\nu}{2}+m_1-s}{\nu+m_1-s},
$$
for all $p \in (\pr+\varepsilon,\pp]$ and $\Delta < \Delta_2$.
For these $p$ and $\Delta$, we thus have
$$
(1-\delta) m(p) + \delta \E_N\wr(p) - \left[ (1-\delta) s + \delta \E_{N-1}\wp(p) \right]
\geq (1-\delta) [m(p)-s] + \delta \, \frac{\nu}{2}.
$$
Finally, there is a $\Delta_{(\pr+\varepsilon,\pp]} \in (0, \Delta_2)$ such that the right-hand side of this inequality is positive for all $p \in (\pr+\varepsilon,\pp]$ and $\Delta < \Delta_{(\pr+\varepsilon,\pp]}$.}

\begin{lem}\label{lem:k=1;pp-1}
There exists $\Delta_{(\pp,1]} > 0$ such that
$$
(1-\delta) m(p) + \delta \E_N\wr(p) \geq (1-\delta) s + \delta \E_{N-1}\wp(p),
$$
for all $p > \bar{p}$ and $\Delta < \Delta_{(\pp,1]}$.
\end{lem}
\proof{
By Lemmas \ref{lem:lower-bound-reward} and \ref{lem:upper-bound-punishment}, there exists $\Delta_{(\pp,1]} > 0$ such that $\wr \geq \wp$ for all $\Delta < \Delta_{(\pp,1]}$.
For such $\Delta$ and all $p > \bar{p}$, we thus have
$$
(1-\delta) m(p) + \delta \E_N\wr(p) = \wr(p) \geq \wp(p)\geq (1-\delta) s + \delta \E_{N-1}\wp(p),
$$
with the last inequality following from the functional equation for $\wp$.
}

\proofof{Proposition}{prop:SSE-Brownian}{
Given $\pr$ and $\pp$ as in \eref{eq:thresholds-Brownian}, choose $\varepsilon > 0$ and $\Delta_{(\pr,\pr+\varepsilon]}$ as in Lemma \ref{lem:k=1;pr-pr+epsilon-Brownian},
and
$\Delta_{(p^\ddagger,\pp]}$,
$\Delta_{(\pr+\varepsilon,\pp]}$
and
$\Delta_{(\pp,1]}$
as in Lemmas
\ref{lem:k=0;pddagger-pp},
\ref{lem:k=1;pr+epsilon-pp}
and
\ref{lem:k=1;pp-1}.
The two-state automaton is an SSE for all
$$\Delta < \min\left\{
\Delta_{(p^\ddagger,\pp]},
\Delta_{(\pr,\pr+\varepsilon]},
\Delta_{(\pr+\varepsilon,\pp]},
\Delta_{(\pp,1]}
\right\}.
$$
So the statement of the proposition holds with $p^\flat=p^\ddagger$ and $p^\sharp=\max\{\check{p},p^\diamond\}$.
}

\proofof{Proposition}{prop:limit-Brownian}{
Let $\varepsilon > 0$ be given.
First, the explicit representation for $V_{N,\underline{p}}$ in Section \ref{sec:result} and Lemma \ref{lem:flat-part} allow us to choose $\pr \in (p_N^*,p^\flat)$ and $\pp \in (p^\sharp,1)$ such that
$\VNpr > V_N^* - \varepsilon$
and
$\wp < V_1^* + \varepsilon$ for all $\Delta > 0$.
Second,
Lemmas \ref{lem:convergence-single-agent} and \ref{lem:lower-bound-reward}
and Proposition \ref{prop:SSE-Brownian}
imply the existence of a $\Delta^\dagger > 0$ such that for all $\Delta \in (0,\Delta^\dagger)$:
$W_1^\Delta > V_1^* - \varepsilon$,
$\wr \geq \VNpr$,
and $\wr$ and $\wp$ are SSE payoff functions of the game with period length $\Delta$.
Third, $\WsupPBE \leq V_N^*$ for all $\Delta > 0$ because any discrete-time strategy profile is feasible for a planner who maximizes the players' average payoff in continuous time.

For $\Delta \in (0,\Delta^\dagger)$, we thus have
$$
V_N^* - \varepsilon < \VNpr \leq \wr \le \WsupSSE \leq \WsupPBE \leq V_N^*,
$$
and
$$
V_1^* - \varepsilon < W_1^\Delta \leq \WinfPBE \leq \WinfSSE \leq \wp < V_1^* + \varepsilon,
$$
so that
$\|\WsupPBE-V_N^*\|$,
$\|\WsupSSE-V_N^*\|$,
$\|\WinfPBE-V_1^*\|$
and
$\|\WinfSSE-V_1^*\|$
are all smaller than $\varepsilon$, which was to be shown.
}

\subsection{Pure Poisson Learning (Propositions \ref{prop:thresh}--\ref{prop:limit-Poisson})}

\proofof{Proposition}{prop:thresh}{
For any given $\Delta > 0$, let
$\plow^\Delta$ be the infimum of the set of beliefs at which there is some PBE that gives a payoff $w_n(p) > s$ to at least one player.
Let $\plow = \liminf_{\Delta \rightarrow 0} \plow^\Delta$.

For any fixed $\varepsilon > 0$ and $\Delta > 0$, consider the problem of maximizing the players' average payoff subject to no use of the risky arm at beliefs $p \leq \plow-\varepsilon$.
Denote the corresponding value function by $\Wt$.
By the definition of $\plow$, there exists a $\Dt >0$ such that for $\Delta \in (0,\Dt)$, the function $\Wt$ provides an upper bound on the players' average payoff in any PBE, and so $\WsupPBE \leq \Wt$.
The value function of the continuous-time version of this maximization problem is $\VNpeps$ with $p_\varepsilon = \max\{\plow - \varepsilon, p_N^*\}$.
As the discrete-time solution is also feasible in continuous time, we have $\Wt \leq \VNpeps$, and hence $\WsupPBE \leq \VNpeps$ for $\Delta < \Dt$.

Consider a sequence of such $\Delta$'s converging to 0 such that the corresponding beliefs $\plow^\Delta$ converge to $\plow$.
For each $\Delta$ in this sequence, select a belief $p^\Delta > \plow^\Delta$ with the following two properties: (i) starting from $p^\Delta$, a single failed experiment takes us below $\plow^\Delta$; (ii) given the initial belief $p^\Delta$, there exists a PBE for reaction lag $\Delta$ in which at least one player plays risky with positive probability in the first round.
Select such an equilibrium for each $\Delta$ in the sequence and let $L^\Delta$ be the number of players in this equilibrium who, at the initial belief $p^\Delta$, play risky with positive probability.
Let $L$ be an accumulation point of the sequence of $L^\Delta$'s.
After selecting a subsequence of $\Delta$'s, we can assume without loss of generality that player $n=1,\ldots,L$ plays risky with probability
$\pi_n^\Delta > 0$ at $p^\Delta$, while player $n=L+1,\ldots,N$ plays safe; we can further assume that $(\pi_n^\Delta)_{n=1}^L$ converges to a limit $(\pi_n)_{n=1}^L$ in $[0,1]^L$.

For player $n = 1,\ldots,L$ to play optimally at $p^\Delta$, it must be the case that
\begin{eqnarray*}
\lefteqn{
(1-\delta)\left[\pi_n^\Delta\lambda(p^\Delta)h+(1-\pi_n^\Delta)s\right]
+\delta \left\{
  \Pr^\Delta(\emptyset)w_{n,\emptyset}^\Delta
  + \sum_{K=1}^L\sum_{\vert I\vert =K}\Pr^\Delta(I)\sum_{J=0}^\infty \Lambda_{J,K}^\Delta(p^\Delta) w_{n,I,J}^\Delta
\right\}
} \\
& \geq &
(1-\delta)s
+\delta \left\{
  \Pr_{-n}^\Delta(\emptyset)w_{n,\emptyset}^\Delta
  + \sum_{K=1}^{L-1}\sum_{\vert I\vert =K, \, n\not\in I}\Pr_{-n}^\Delta(I)\sum_{J=0}^\infty \Lambda_{J,K}^\Delta(p^\Delta) w_{n,I,J}^\Delta
\right\}, \hspace{5em}
\end{eqnarray*}
where we write $\Pr^\Delta(I)$ for the probability that the set of players experimenting is $I \subseteq \{1,\ldots,L\}$,
$\Pr_{-n}^\Delta(I)$ for the probability that among the $L-1$ players in $\{1,\cdots,L\}\setminus\{n\}$ the set of players experimenting is $I$,
and $w_{n,I,J}^\Delta$ for the conditional expectation of player $n$'s continuation payoff given that exactly the players in $I$ were experimenting and had $J$ successes ($w_{n,\emptyset}^\Delta$ is player $n$'s continuation payoff if no one was experimenting).
As $\Pr^\Delta(\emptyset) = (1-\pi_n^\Delta) \Pr_{-n}^\Delta(\emptyset) \leq \Pr_{-n}^\Delta(\emptyset)$, the inequality continues to hold when we replace $w_{n,\emptyset}^\Delta$ by its lower bound $s$.
After subtracting $(1-\delta)s$ from both sides, we then have
\begin{eqnarray*}
\lefteqn{
(1-\delta)\pi_n^\Delta \left[\lambda(p^\Delta)h-s\right]
+\delta \left\{
  \Pr^\Delta(\emptyset) s
  + \sum_{K=1}^L\sum_{\vert I\vert =K}\Pr^\Delta(I)\sum_{J=0}^\infty \Lambda_{J,K}^\Delta(p^\Delta) w_{n,I,J}^\Delta
\right\}
}\\
& \geq &
\delta \left\{
  \Pr_{-n}^\Delta(\emptyset) s
  + \sum_{K=1}^{L-1}\sum_{\vert I\vert =K, \, n\not\in I}\Pr_{-n}^\Delta(I)\sum_{J=0}^\infty \Lambda_{J,K}^\Delta(p^\Delta) w_{n,I,J}^\Delta
\right\}. \hspace{5em}
\end{eqnarray*}

Summing up these inequalities over $n=1,\ldots,L$ and writing $\bar{\pi}^\Delta=\frac{1}{L}\sum_{n=1}^L \pi_n^\Delta$ yields
\begin{eqnarray*}
\lefteqn{
(1-\delta)L\bar{\pi}^\Delta \left[\lambda(p^\Delta)h-s\right]
+\delta \left\{
  \Pr^\Delta(\emptyset) L s
  + \sum_{K=1}^L\sum_{\vert I\vert =K}\Pr^\Delta(I)\sum_{J=0}^\infty \Lambda_{J,K}^\Delta(p^\Delta) \sum_{n=1}^L w_{n,I,J}^\Delta
\right\}
}\\
& \geq &
\delta \left\{
  \sum_{n=1}^L \Pr_{-n}^\Delta(\emptyset) s
  + \sum_{n=1}^L \sum_{K=1}^{L-1} \sum_{\vert I\vert =K, \, n\not\in I}\Pr_{-n}^\Delta(I)\sum_{J=0}^\infty \Lambda_{J,K}^\Delta(p^\Delta) w_{n,I,J}^\Delta
\right\}. \hspace{5em}
\end{eqnarray*}
By construction, $w_{n,I,0}^\Delta = s$ whenever $I \neq \emptyset$.
For $\vert I \vert = K > 0$ and $J > 0$, moreover, we have $w_{n,I,J}^\Delta \geq W_1^\Delta(B_{J,K}^\Delta(p^\Delta))$ for \emph{all} players $n=1,\ldots,N$, and hence
\begin{eqnarray*}
\sum_{n=1}^L w_{n,I,J}^\Delta
& \leq & N \WsupPBE(B_{J,K}^\Delta(p^\Delta)) - (N-L) W_1^\Delta(B_{J,K}^\Delta(p^\Delta)) \\
& \leq & N \VNpeps(B_{J,K}^\Delta(p^\Delta)) - (N-L) W_1^\Delta(B_{J,K}^\Delta(p^\Delta)).
\end{eqnarray*}
So, for the preceding inequality to hold, it is necessary that
\begin{eqnarray*}
\lefteqn{
(1-\delta)L\bar{\pi}^\Delta \left[\lambda(p^\Delta)h-s\right]
+\delta \left\{ \astrut{4.5}
  \Pr^\Delta(\emptyset) L s + \sum_{K=1}^L \sum_{\vert I\vert =K} \Pr^\Delta(I) \Lambda_{0,K}^\Delta(p^\Delta) L s
  \right.}\\
& & \left.
  + \sum_{K=1}^L \sum_{\vert I\vert =K} \Pr^\Delta(I) \sum_{J=1}^\infty \Lambda_{J,K}^\Delta(p^\Delta) \left[ N \VNpeps(B_{J,K}^\Delta(p^\Delta)) - (N-L) W_1^\Delta(B_{J,K}^\Delta(p^\Delta)) \right]
\right\} \\
& \geq & \astrut{6}
\delta \left\{ \astrut{4.5}
  \sum_{n=1}^L \Pr_{-n}^\Delta(\emptyset) s
  + \sum_{n=1}^L \sum_{K=1}^{L-1} \sum_{\vert I\vert =K, \, n\not\in I} \Pr_{-n}^\Delta(I) \Lambda_{0,K}^\Delta(p^\Delta) s
  \right. \\
& & \left. \hspace{1.75em}
  + \sum_{n=1}^L \sum_{K=1}^{L-1} \sum_{\vert I\vert =K, \, n\not\in I} \Pr_{-n}^\Delta(I) \sum_{J=1}^\infty \Lambda_{J,K}^\Delta(p^\Delta) W_1^\Delta(B_{J,K}^\Delta(p^\Delta))
\right\}.
\end{eqnarray*}

As
$$
\Pr^\Delta(\emptyset) + \sum_{K=1}^L \sum_{\vert I\vert =K} \Pr^\Delta(I) = 1
\quad \mbox{and} \quad
\sum_{K=1}^L \sum_{\vert I\vert =K} \Pr^\Delta(I) K = L \bar{\pi}^\Delta,
$$
we have the first-order expansions
\begin{eqnarray*}
\lefteqn{
\Pr^\Delta(\emptyset) + \sum_{K=1}^L \sum_{\vert I\vert =K} \Pr^\Delta(I) \Lambda_{0,K}^\Delta(p^\Delta)
} \\
& = & \Pr^\Delta(\emptyset) + \sum_{K=1}^L \sum_{\vert I\vert =K} \Pr^\Delta(I) \left( 1 - K \lambda(p^\Delta) \Delta \right) + o(\Delta) \\
& = & \astrut{3} 1 - L \bar{\pi}^\Delta \lambda(p^\Delta) \Delta + o(\Delta),
\end{eqnarray*}
and
$$
\sum_{K=1}^L \sum_{\vert I\vert =K} \Pr^\Delta(I) \Lambda_{1,K}^\Delta(p^\Delta)
= \sum_{K=1}^L \sum_{\vert I\vert =K} \Pr^\Delta(I) K \lambda(p^\Delta) \Delta + o(\Delta)
= L \bar{\pi}^\Delta \lambda(p^\Delta) \Delta + o(\Delta),
$$
so, by uniform convergence $W_1^\Delta \to V_1^*$ (Lemma \ref{lem:convergence-single-agent}), the left-hand side of the last inequality expands as
$$
Ls + L \left\{ \astrut{3} r \bar{\pi} \left[\lambda(\plow)h-s\right] - r s
+ \bar{\pi} \lambda(\plow) \left[N \VNpeps(j(\plow)) - (N\!-\!L) V_1^*(j(\plow)) - L s\right] \right\} \Delta
+ o(\Delta),
$$
with $\bar{\pi} = \lim_{\Delta \rightarrow 0} \bar{\pi}^\Delta$.
In the same way, the identities
$$\Pr_{-n}^\Delta(\emptyset) + \sum_{K=1}^{L-1} \sum_{\vert I\vert =K, \, n\not\in I} \Pr_{-n}^\Delta(I) = 1
\quad \mbox{and} \quad
\sum_{K=1}^{L-1} \sum_{\vert I\vert =K, \, n\not\in I} \Pr_{-n}^\Delta(I) K = L \bar{\pi}^\Delta - \pi_n^\Delta
$$
imply
$$
\sum_{n=1}^L \Pr_{-n}^\Delta(\emptyset) + \sum_{n=1}^L \sum_{K=1}^{L-1} \sum_{\vert I\vert =K, \, n\not\in I} \Pr_{-n}^\Delta(I) \Lambda_{0,K}^\Delta(p^\Delta)
= \astrut{3} L - L (L-1) \bar{\pi}^\Delta \lambda(p^\Delta) \Delta + o(\Delta)
$$
and
$$
\sum_{n=1}^L \sum_{K=1}^{L-1} \sum_{\vert I\vert =K, \, n\not\in I} \Pr_{-n}^\Delta(I) \Lambda_{1,K}^\Delta(p^\Delta)
= L (L-1) \bar{\pi}^\Delta \lambda(p^\Delta) \Delta + o(\Delta),
$$
and so the right-hand side of the inequality expands as
$$
Ls + L \left\{ \astrut{2.5} - r s + (L-1) \bar{\pi} \lambda(\plow) \left[ V_1^*(j(\plow)) - s \right]  \right\} \Delta
+ o(\Delta).
$$
Comparing terms of order $\Delta$, dividing by $L$ and letting $\varepsilon \rightarrow 0$, we obtain
$$
\bar{\pi} \left\{ \astrut{2.5} \lambda(\plow) \left[ \astrut{2} N \VNpb(j(\plow)) - (N\!-\!1) V_1^*(j(\plow)) - s \right] - r c(\plow) \right\} \geq 0.
$$
By Lemma \ref{lem:thresh}, this means $\plow \geq \hat{p}$ whenever $\bar{\pi} > 0$.

For the case that $\bar{\pi}=0$, we write the optimality condition for player $n \in \{1,\ldots, L\}$ as
\begin{eqnarray*}
\lefteqn{
(1-\delta) \lambda(p^\Delta)h
+\delta \left\{\sum_{K=0}^{L-1}\sum_{\vert I\vert =K,\, n\not\in I}\Pr_{-n}^\Delta(I)\sum_{J=0}^\infty \Lambda_{J,K+1}^\Delta(p^\Delta) w_{n,I\dot{\cup}\{n\},J}^\Delta\right\}
} \\
& \geq &
(1-\delta)s
+\delta \left\{
  \Pr_{-n}^\Delta(\emptyset)w_{n,\emptyset}^\Delta
  + \sum_{K=1}^{L-1}\sum_{\vert I\vert =K, \, n\not\in I}\Pr_{-n}^\Delta(I)\sum_{J=0}^\infty \Lambda_{J,K}^\Delta(p^\Delta) w_{n,I,J}^\Delta
\right\}. \hspace{5em}
\end{eqnarray*}
As above, $w_{n,\emptyset}^\Delta \geq s$, and $w_{n,I,0}^\Delta = s$ whenever $I \neq \emptyset$.
For $\vert I \vert = K > 0$ and $J > 0$, moreover, we have
$w_{n,I,J}^\Delta \geq W_1^\Delta(B_{J,K}^\Delta(p^\Delta))$,
$w_{n,I\dot{\cup}\{n\},J}^\Delta \geq W_1^\Delta(B_{J,K+1}^\Delta(p^\Delta))$ and
$w_{n,I\dot{\cup}\{n\},J}^\Delta \leq N \VNpeps(B_{J,K+1}^\Delta(p^\Delta)) - (N-1) W_1^\Delta(B_{J,K+1}^\Delta(p^\Delta))$.
So, for the optimality condition to hold, it is necessary that
\begin{eqnarray*}
\lefteqn{
(1-\delta) \lambda(p^\Delta)h
+\delta \left\{ \astrut{4.5}
  \sum_{K=0}^{L-1} \sum_{\vert I\vert =K,\, n\not\in I}\Pr_{-n}^\Delta(I) \Lambda_{0,K+1}^\Delta(p^\Delta) s
  \right.}\\
\lefteqn{
\left.
   + \sum_{K=0}^{L-1} \sum_{\vert I\vert =K,\, n\not\in I}\Pr_{-n}^\Delta(I) \sum_{J=1}^\infty \Lambda_{J,K+1}^\Delta(p^\Delta) \left[ N \VNpeps(B_{J,K+1}^\Delta(p^\Delta)) - (N\!-\!1) W_1^\Delta(B_{J,K+1}^\Delta(p^\Delta)) \right]
\right\}
} \\
& \geq & \astrut{6}
(1-\delta)s
+\delta \left\{ \astrut{4.5}
  \Pr_{-n}^\Delta(\emptyset) s
  + \sum_{K=1}^{L-1} \sum_{\vert I\vert =K, \, n\not\in I} \Pr_{-n}^\Delta(I) \Lambda_{0,K}^\Delta(p^\Delta) s
  \right. \\
& & \left. \hspace{6em}
  + \sum_{K=1}^{L-1} \sum_{\vert I\vert =K, \, n\not\in I} \Pr_{-n}^\Delta(I)
    \sum_{J=1}^\infty \Lambda_{J,K}^\Delta(p^\Delta) W_1^\Delta(B_{J,K}^\Delta(p^\Delta))
\right\}. \hspace{5em}
\end{eqnarray*}
Now,
$$
\sum_{K=1}^{L-1} \sum_{\vert I\vert =K, \, n\not\in I} \Pr_{-n}^\Delta(I) K = L \bar{\pi}^\Delta - \pi_n^\Delta \rightarrow 0
$$
as $\Delta$ vanishes.
Therefore, the left-hand side of the above inequality expands as
$$
s + \left\{ \astrut{3} r \left[\lambda(\plow)h-s\right]
+ \lambda(\plow) \left[N \VNpeps(j(\plow)) - (N\!-\!1) V_1^*(j(\plow)) - s\right] \right\} \Delta
+ o(\Delta),
$$
and the right-hand side as $s + o(\Delta)$.
Comparing terms of order $\Delta$, letting $\varepsilon \rightarrow 0$ and using Lemma \ref{lem:thresh} once more, we again obtain $\plow \geq \hat{p}$.

The statement about the range of experimentation now follows immediately from the fact that
for $\Delta < \Dt$, we have $\WsupPBE \leq \VNpeps$,
and hence $\WsupPBE = \VNpeps = s$ on $[0,\plow - \varepsilon] \supseteq [0,\phat - \varepsilon]$.

The statement about the supremum of equilibrium payoffs follows from
the inequality $\WsupPBE \leq \VNpeps$ for $\Delta < \Dt$,
convergence $\VNpeps \rightarrow V_{N,\plow}$ as $\varepsilon \rightarrow 0$,
and the inequality $V_{N,\plow} \leq \VNphat$.
}

We now turn to the proof of Proposition \ref{prop:SSE-Poisson}.
The only difference to the case with a Brownian component is the proof of incentive compatibility to the immediate right of $\pr$.

In view of Lemmas
\ref{lem:thresh}, \ref{lem:upper-bound-punishment} and \ref{lem:flat-part},
we consider $\pr$ and $\pp$ such that
\begin{equation} \label{eq:thresholds-Poisson}
\phat < \pr < p^\ddagger < p_1^* < p^m < \max\{p^\diamond,\check{p}\} < \pp < 1.
\end{equation}

\begin{lem}\label{lem:k=1;pr-pr+epsilon-Poisson}
Let $\rho = 0$ and $\lambda_0 > 0$.
There exists $p^\sharp \in (\max\{p^\diamond,\check{p}\},1)$ such that for all $\pp \in (p^\sharp,1)$, there exist $\varepsilon \in (0,p^\ddagger-\pr)$ and $\Delta_{(\pr,\pr+\varepsilon]} > 0$ such that
$$
(1-\delta) m(p) + \delta \E_{N}\wr(p) \geq (1-\delta) s + \delta \E_{N-1}\wp(p),
$$
for all $p \in (\pr,\pr+\varepsilon]$ and $\Delta < \Delta_{(\pr,\pr+\varepsilon]}$.
\end{lem}
\proof{
By Lemma \ref{lem:lower-bound-reward}, there exists $\Delta_0 > 0$ such that $\wr \geq \VNpr$ for $\Delta \in (0,\Delta_0)$.

By Lemma \ref{lem:thresh},
\[
\lambda(p)[NV_{N,p}(j(p))-(N-1)V_1^*(j(p))-s]-rc(p) > 0
\]
on $[\pr,1]$.
As $V_{N,p}(j(p)) \leq \VNpr(j(p))$ for $p \geq \pr$, this implies
\[
\lambda(p)[N\VNpr(j(p))-(N-1)V_1^*(j(p))-s]-rc(p) > 0
\]
on $[\pr,1]$.
By Lemma \ref{lem:convergence-punish-pbar}, there exists a belief $p^\sharp > \max\{p^\diamond,\check{p}\}$ such that for all $\pp > p^\sharp$,
$$
\lambda(p)[N\VNpr(j(p))-(N-1)\VIpp(j(p))-s]-rc(p) > 0
$$
on $[\pr,1]$.
Fix a $\pp \in (p^\sharp,1)$, define
$$
\nu = \min_{p \in [\pr,1]} \left\{ \lambda(p)[N\VNpr(j(p))-(N-1)\VIpp(j(p))-s]-rc(p) \right\} > 0,
$$
and choose $\varepsilon > 0$ such that $\pr + \varepsilon < p^\ddagger$ and
$$
(N \lambda(\pr + \varepsilon) + r)\left[\VNpr(\pr + \varepsilon) - s\right] < \nu/3.
$$

In the remainder of the proof, we write $p^K_J$ for the posterior belief starting from $p$ when $K$ players use the risky arm and $J$ lump-sums arrive within the length of time $\Delta$.

For $p \in (\pr,\pr+\varepsilon]$ and $\Delta \in (0,\Delta_0)$,
\begin{eqnarray*}
\lefteqn{(1-\delta) m(p) + \delta \E_N \wr(p)} \\
& \geq & (1-\delta) m(p) + \delta \E_N \VNpr(p) \\
& = & r\Delta\, m(p)
+ (1-r\Delta) \left\{
N\lambda(p)\Delta\, \VNpr(p^N_1)
+ (1-N\lambda(p)\Delta)\, \VNpr(p^N_0)
\right\}
+O(\Delta^2) \\
& = & \VNpr(p^N_0)
+ \left\{
r m(p)
+ N \lambda(p) \VNpr(p^N_1)
- (N \lambda(p) + r) \VNpr(p^N_0)
\right\} \Delta
+ O(\Delta^2),
\end{eqnarray*}
while
\begin{eqnarray*}
\lefteqn{(1-\delta) s + \delta \E_{N-1} \wp(p)} \\
& = & r\Delta\, s
+ (1-r\Delta) \left\{
(N-1)\lambda(p)\Delta\, \wp(p^{N-1}_1)
+ [1-(N-1)\lambda(p)\Delta]\, \wp(p^{N-1}_0)
\right\}
+O(\Delta^2) \\
& = & \wp(p^{N-1}_0)
+ \left\{
rs
+ (N-1) \lambda(p) \wp(p^{N-1}_1)
- [(N-1) \lambda(p) + r] \wp(p^{N-1}_0)
\right\} \Delta + O(\Delta^2).
\end{eqnarray*}
As $\VNpr(p^N_0) \geq s = \wp(p^{N-1}_0)$,
the difference
$(1-\delta) m(p) + \delta \E_{N}\wr(p) - \left[ (1-\delta) s + \delta \E_{N-1}\wp(p) \right]$
is no smaller than $\Delta$ times
$$
\lambda(p) \left[ N \VNpr(p^N_1) - (N-1) \wp(p^{N-1}_1) - s \right]
- r c(p)
- (N \lambda(p) + r) \left[ \VNpr(p^N_0) - s \right],
$$
plus terms of order $\Delta^2$ and higher.

Let $\xi=\frac{\nu}{6(N-1)\lambda_1}$.
By Lemma \ref{lem:convergence-punish-Delta} as well as Lipschitz continuity of $\VNpr$ and $\VIpp$, there exists $\Delta_1 \in (0,\Delta_0)$ such that
$\|\wp - \VIpp\|$,
$\max_{\pr \leq p \leq p^\ddagger} |\VNpr(p^N_1) - \VNpr(j(p))|$
and
$\max_{\pr \leq p \leq p^\ddagger} |\VIpp(p^{N-1}_1) - \VIpp(j(p))|$
are all smaller than $\xi$ when $\Delta < \Delta_1$.
For such $\Delta$ and $p \in (\pr,p^\ddagger]$, we thus have
$\VNpr(p^N_1) > \VNpr(j(p)) - \xi$
and
$\wp(p^{N-1}_1) < \VIpp(j(p)) + 2 \xi$,
so that the expression displayed above is larger than
$\nu - 2 (N-1) \lambda(p) \xi - \nu/3 > \nu/3$.
This implies existence of a $\Delta_{(\pr,\pr+\varepsilon]} \in (0,\Delta_1)$ as in the statement of the lemma.
}

\proofof{Proposition}{prop:SSE-Poisson}{
Given $\pr$ as in \eref{eq:thresholds-Poisson}, take $p^\sharp$ as in Lemma \ref{lem:k=1;pr-pr+epsilon-Poisson} and fix $\pp > p^\sharp$.
Choose $\varepsilon > 0$ and $\Delta_{(\pr,\pr+\varepsilon]}$ as in Lemma \ref{lem:k=1;pr-pr+epsilon-Poisson},
and
$\Delta_{(p^\ddagger,\pp]}$,
$\Delta_{(\pr+\varepsilon,\pp]}$
and
$\Delta_{(\pp,1]}$
as in Lemmas
\ref{lem:k=0;pddagger-pp},
\ref{lem:k=1;pr+epsilon-pp}
and
\ref{lem:k=1;pp-1}.
The two-state automaton is an SSE for all
$$\Delta < \min\left\{
\Delta_{(p^\ddagger,\pp]},
\Delta_{(\pr,\pr+\varepsilon]},
\Delta_{(\pr+\varepsilon,\pp]},
\Delta_{(\pp,1]}
\right\}.
$$
So the statement of the proposition holds with $p^\flat=p^\ddagger$ and the chosen $p^\sharp$.
}

For the proof of Proposition \ref{prop:SSE-exponential}, we modify notation slightly, writing $\l$ for the probability that, conditional on $\theta=1$, a player has at least one success on his own risky arm in any given round, and $g$ for the corresponding expected payoff per unit of time.\footnote{\textit{I.e.}, $\l = 1- e^{-\lambda_1\Delta}$ and $g = m_1$.}

Consider an SSE played at a given prior $p$, with associated payoff $W$.
If $K \ge 1$ players unsuccessfully choose the risky arm, the belief jumps down to a posterior denoted $p_K$.
Note that an SSE allows the continuation play to depend on the identity of these players.
Taking the expectation over all possible combinations of $K$ players who experiment, however,
we can associate with each posterior $p_K$, $K \ge 1$, an expected continuation payoff $W_K$.
If $K=0$, so that no player experiments, the belief does not evolve, but there is no reason that the continuation strategies (and so the payoff) should remain the same.
We denote the corresponding payoff by $W_0$.
In addition, we write $\pi \in [0,1]$ for the probability with which each player experiments at $p$, and
$q_K$ for the probability that at least one player has a success, given $p$, when $K$ of them experiment.
The players' common payoff must then satisfy the following optimality equation:
\begin{eqnarray*}
W
& = &\max \left\{(1-\d)p_0 g+\d \sum_{K=0}^{N-1}\binom{N-1}{K}\pi^K(1-\pi)^{N-1-K}[q_{K+1} g+(1-q_{K+1})W_{K+1})]\right.,\\
&&  \hspace{-2em} \left.(1-\d)s+\d \sum_{K=1}^{N-1}\binom{N-1}{K}\pi^K(1-\pi)^{N-1-K}(q_{K} g+(1-q_{K})W_K)+\d(1-\pi)^{N-1}W_0)\right\}.
\end{eqnarray*}
The first term corresponds to the payoff from playing risky, the second from playing safe.

As it turns out, it is more convenient to work with
odds ratios
$$\omega=\frac{p}{1-p} \quad \mbox{and} \quad \omega_K=\frac{p_K}{1-p_K},$$
which we refer to as ``belief'' as well. Note that
\[
p_K=\frac{p\,(1-\omega)^{\!K}}{p\,(1-\omega)^{\!K}+1-p}
\]
implies that $\omega_K=(1-\l)^{\!K} \omega.$ Note also that
\[
1-q_K=p\,(1-\l)^{\!K}+1-p=(1-p)(1+\omega_K),\quad q_K=p-(1-p)\omega_K=(1-p)(\omega-\omega_K).
\]
We
define
$$
m=\frac{s}{g-s},
\quad
\upsilon=\frac{W-s}{(1-p)(g-s)},
\quad
\upsilon_K=\frac{W_K-s}{(1-p_K)(g-s)}.
$$
Note that $\upsilon \ge 0$ in any equilibrium, as $s$ is a lower bound on the value.
Simple computations now give
\begin{eqnarray*}
\upsilon &=&\max \left\{ \omega-(1-\d)m+\d \sum_{K=0}^{N-1}\binom{N-1}{K}\pi^K(1-\pi)^{N-1-K}(\upsilon_{K+1}-\omega_{K+1})\right.,\\
&&\hspace{2.75em}\left.\d \omega+\d \sum_{K=0}^{N-1}\binom{N-1}{K}\pi^K(1-\pi)^{N-1-K}(\upsilon_K-\omega_K)\right\}.
\end{eqnarray*}
It is also useful to introduce $w=\upsilon-\omega$ and $w_K=\upsilon_K-\omega_K$.
We then obtain
\begin{eqnarray}\label{webern}
w&=&\max \left\{-(1-\d)m+\d \sum_{K=0}^{N-1}\binom{N-1}{K}\pi^K(1-\pi)^{N-1-K}w_{K+1}\right.,\notag \\
&&\hspace{2.75em}\left.-(1-\d)\omega+\d\sum_{K=0}^{N-1}\binom{N-1}{K}\pi^K(1-\pi)^{N-1-K}w_K\right\}.
\end{eqnarray}
We define
\[
\omega^*=\frac{m}{1+\frac{\d}{1-\d}\l}\,.
\]
This is the odds ratio corresponding to the single-agent cutoff $p_1^\Delta$, \textit{i.e.}, $\omega^*=p_1^\Delta/(1-p_1^\Delta)$. Note that $p_1^\Delta>p_1^*$ for $\Delta>0$.

As stated in Section \ref{sec:construction-Poisson}, no PBE involves experimentation below $p_1^\Delta$ or, in terms of odds ratios, $\omega^*$.
For all beliefs $\omega < \omega^*$, therefore, any equilibrium has $w=-\omega$, or $\upsilon=0$, for each player.
\vskip8truept

\proofof{Proposition}{prop:SSE-exponential}{
Following terminology from repeated games, we say that we can \emph{enforce} action $\pi\in\{0,1\}$ at belief $\omega$ if we can construct an SSE for the prior belief $\omega$ in which players prefer to choose $\pi$ in the first round rather than deviate unilaterally.

Our first step is to derive sufficient conditions for enforcement of $\pi\in\{0,1\}$. The conditions to enforce these actions are intertwined, and must be derived simultaneously.

\emph{Enforcing $\pi=0$ at $\omega$.} To enforce $\pi=0$ at $\omega$, it suffices that one round of using the safe arm followed by the best equilibrium payoff at $\omega$ exceeds the payoff from one round of using the risky arm followed by the resulting continuation payoff at belief $\omega_1$ (as only the deviating player will have experimented). See below for the precise condition.

\emph{Enforcing $\pi=1$ at $\omega$.} If a player deviates to $\pi=0$, we jump to $w_{N-1}$ rather than $w_N$ in case all experiments fail.
Assume that at $\omega_{N-1}$ we can enforce $\pi=0$. As explained above,
this implies that at $\omega_{N-1}$, a player's continuation payoff can be pushed down to what he would get by unilaterally deviating to experimentation, which is at most  $-(1-\d)m+\d w_N$ where $w_N$ is the highest possible continuation payoff at belief $\omega_N$.
To enforce $\pi=1$ at $\omega$, it then suffices that
\[
w=-(1-\d)m+\d w_N \ge -(1-\d)\omega+\d(-(1-\d)m+\d w_N),
\]
with the same continuation payoff $w_N$ on the left-hand side of the inequality.
The inequality simplifies to
\[
\d w_N \ge (1-\d)m-\omega;
\]
by the formula for $w$, this is equivalent to $w \ge -\omega$, \textit{i.e.}, $\upsilon \ge 0$.
Given that
\[
\upsilon
=\omega-(1-\d)m+\d(\upsilon_N-\omega_N)
=(1-\d(1-\l)^N) \omega-(1-\d)m+\d \upsilon_N,
\]
to show that $\upsilon \ge 0$, it thus suffices that
\[
\omega \ge \frac{m}{1+\frac{\d}{1-\d}(1-(1-\l)^N)}=\tilde{\omega},
\]
and that $\upsilon_N \ge 0$, which is necessarily the case if $\upsilon_N$ is an equilibrium payoff.
Note that $(1-\l)^N\tilde{\omega} \le \omega^*$, so that $\omega_N \ge \omega^*$ implies $\omega \ge \tilde{\omega}$.
In summary, to enforce $\pi=1$ at $\omega$, it suffices that $\omega_N \ge \omega^*$ and $\pi=0$ be enforceable at $\omega_{N-1}$.

\emph{Enforcing $\pi=0$ at $\omega$ (continued).}
Suppose we can enforce it at $\omega_1,\omega_2,\ldots,\omega_{N-1}$,  and that $\omega_N \ge \omega^*$. Note that $\pi=1$ is then enforceable at $\omega$  from our previous argument, given our hypothesis that $\pi=0$ is enforceable at $\omega_{N-1}$.
It then suffices that
\[
-(1-\d)\omega+\d (-(1-\delta)m+\d w_N) \ge -(1-\d^N)m+\d^N w_N,
\]
where again it suffices that this holds for the highest value of $w_N$. To understand this expression, consider a player who deviates by experimenting. Then the following period the belief is down one step, and if $\pi=0$ is enforceable at $\omega_1$, it means that his continuation payoff there can be chosen to be no larger than what he can secure at that point by deviating and experimenting again, etc. The right-hand side is then obtained as the payoff from $N$ consecutive unilateral deviations to experimentation (in fact, we have picked an upper bound, as the continuation payoff after this string of deviations need not be the maximum $w_N$). The left-hand side is the payoff from  playing safe one period before setting $\pi=1$ and getting the maximum payoff $w_N$, a continuation strategy that is sequentially rational given that $\pi=1$ is enforceable at $\omega$ by our hypothesis that $\pi=0$ is enforceable at $\omega_{N-1}$.

Plugging in the definition of $\upsilon_N$, this inequality simplifies to
\[
(\d^2-\d^N)\upsilon_N \ge (\d^2-\d^N)(\omega_N-m)+(1-\d)(\omega-m),
\]
which is always satisfied for beliefs $\omega\leq m$, \textit{i.e.}, below the myopic cutoff $\omega^m$ (which coincides with the normalized payoff $m$).

To summarize, if $\pi=0$ can be enforced at the $N-1$ consecutive beliefs $\omega_1,\ldots,\omega_{N-1}$, with $\omega_N \ge \omega^*$  and $\omega\le \omega^m$,
then both $\pi=0$ and $\pi=1$ can be enforced at $\omega$.
By induction, this implies that if we can find an interval of beliefs $[\omega_N,\omega)$ with $\omega_N \ge \omega^*$ for which $\pi=0$ can be enforced, then $\pi=0,1$ can be enforced at all beliefs $\omega' \in (\omega,\omega^m)$.

Our second step is to establish that such an interval of beliefs exists. This second step involves itself three steps. First, we derive some ``simple'' equilibrium, which is a symmetric Markov equilibrium. Second, we  show that we can enforce $\pi=1$ on sufficiently (finitely) many consecutive values of beliefs building on this equilibrium; third, we show that this can be used to enforce $\pi=0$ as well.

It will be useful to distinguish beliefs according to whether they belong to the interval $[\omega^*,(1+\lambda_1 \Delta)\omega^*),[(1+\lambda_1 \Delta)\omega^*,(1+2\lambda_1 \Delta)\omega^*),\ldots$
For $\tau \in \N$, let $I_{\tau+1}=[(1+\tau \lambda_1  \Delta)\omega^*,(1+(\tau+1)\lambda_1 \Delta)\omega^*)$. For fixed $\Delta$, every $\omega \ge \omega^*$ can be uniquely mapped into a pair  $(x,\tau) \in [0,1) \times \N$ such that $\omega=(1+\lambda_1(x+\tau)\Delta)\omega^*$, and we alternatively denote beliefs by such a pair. Note also that, for small enough $\Delta>0$, one unsuccessful experiment takes a belief that belongs to  the interval $I_{\tau+1}$ to (within $O(\Delta^2)$ of) the interval $I_{\tau}$. (Recall that $\l=\lambda_1 \Delta +O(\Delta^2)$.)

Let us start with deriving a symmetric Markov equilibrium. Hence, because it is Markovian, $\upsilon_0=\upsilon$ in our notation, that is, the continuation payoff when nobody experiments is equal to the payoff itself.

Rewriting the equations, using the risky arm gives the payoff\footnote{To pull out the terms involving the belief $\omega$ from the sum appearing in the definition of $\upsilon$, use the fact that $ \sum_{K=0}^{N-1}\binom{N-1}{K}\pi^K(1-\pi)^{N-1-K}(1-\Lambda)^K=(1-\pi \Lambda)^N/(1-\pi \Lambda)$.}
\[
\upsilon=\omega-(1-\d)m-\d (1-\l)(1-\pi \l)^{N-1}\omega+\d \sum_{K=0}^{N-1}\binom{N-1}{K}\pi^K(1-\pi)^{N-1-K}\upsilon_{K+1},\]
while using the safe arm yields
\[
\upsilon=\d (1-(1-\pi \l)^{N-1})\omega+\d(1-\pi)^{N-1}\upsilon+\d\sum_{K=1}^{N-1}\binom{N-1}{K}\pi^K(1-\pi)^{N-1-K}\upsilon_K.
\]
In the Markov equilibrium we derive, players are indifferent between both actions, and so their payoffs are the same. Given any belief $\omega$ or corresponding pair $(\tau,x)$, we conjecture an equilibrium in which $\pi=a(\tau,x) \Delta^2+O(\Delta^3)$, $\upsilon=b(\tau,x) \Delta^2+O(\Delta^3)$, for some functions $a,b$ of the pair $(\tau,x)$ only. Using the fact that $\l=\lambda_1 \Delta +O(\Delta^2),1-\delta=r \Delta +O(\Delta^2)$, we replace this in the two payoff expressions, and take Taylor expansions to get, respectively,
\[
0=\left( rb(\tau,x)+\frac{\lambda_1 m}{\lambda_1+r}(N-1)a(\tau,x) \right)\Delta^3+O(\Delta^4),
\]
and
\[
0=\left[b(\tau,x)-r m \lambda_1(\tau+x)\right]\Delta^2+O(\Delta^3).
\]
We then solve for $a(\tau,x)$, $b(\tau,x)$, to get
\[
\pi_-=\frac{r(\lambda_1+r)(x+\tau)}{N-1}\Delta^2+O(\Delta^3),
\]
with corresponding value
\[
\upsilon_-=\lambda_1 m r(x+\tau)\Delta^2+O(\Delta^3).
\]
This being an induction on $K$, it must be verified that the expansion indeed holds at the lowest interval, $I_1$, and this verification is immediate.\footnote{
Note that this solution is actually continuous at the interval endpoints.
It is not the only solution to these equations; as mentioned in the text, there are intervals of beliefs for which multiple symmetric Markov equilibria exist in discrete time. It is easy to construct such equilibria in which $\pi=1$ and the initial belief is in (a subinterval of) $I_1$.}

We now turn to the second step and argue that we can find $N-1$
consecutive beliefs at which $\pi=1$ can be enforced. We then verify that incentives can be provided to do so, assuming that $\upsilon_-$ are the continuation values used by the players whether a player deviates or not from $\pi=1$. Assume that $N-1$ players choose $\pi=1$. Consider the remaining one. His incentive constraint to choose $\pi=1$ is
\begin{equation}\label{zelenka}
-(1-\d)m+\d \upsilon_N-\d (1-\l)^N \omega \ge -(1-\d)\omega-\d(1-\l)^{N-1}\omega+\d \upsilon_{N-1},
\end{equation}
where $\upsilon_N,\upsilon_{N-1}$ are given by $\upsilon_-$ at $\omega_N$, $\omega_{N-1}$. The interpretation of both sides is as before, the payoff from abiding with the candidate equilibrium action vs. the payoff from deviating.
Fixing $\omega$ and the corresponding pair $(\tau,x)$, and assuming that $\tau \ge N-1$,\footnote{Considering $\tau<N-1$ would lead to $\upsilon_N=0$, so that the explicit formula for $\upsilon_-$ would not apply at $\omega_N$. Computations are then easier, and the result would hold as well.} we insert our formula for $\upsilon_-$, as well as $\l=\lambda_1 \Delta+O(\Delta),1-\delta=r \Delta+O(\Delta)$. This gives
\[
\tau \ge (N-1)\left(2+\frac{\lambda_1}{\lambda_1+r}\right)-x.
\]
Hence, given any integer $N' \in \N$, $N'>3(N-1)$, there exists $\bar{\Delta}>0$ such that for every $\Delta\in (0,\bar{\Delta})$, $\pi=1$ is an equilibrium action at all beliefs $\omega=\omega^*(1+\tau\Delta)$, for $\tau =3(N-1),\ldots,N'$ (we pick the factor 3 because $\lambda_1/(\lambda_1+r) <1$).

Fix $N-1$ consecutive beliefs such that they all belong to intervals $I_\tau$ with $\tau \ge 3(N-1)$ (say, $\tau \le {4N}$), and fix $\Delta$ for which the previous result holds, \textit{i.e.}, $\pi=1$ can be enforced at all these beliefs. We now turn to the third step, showing how $\pi=0$ can be enforced as well for these beliefs.

Suppose that players choose $\pi=0$. As a continuation payoff, we can use the payoff from playing $\pi=1$ in the following round, as we have seen that this action can be enforced at such a belief. This gives
\[
\d \omega+\d(-(1-\d)m-\d(1-\l)^Nl+\d \upsilon_-(\omega_N)).
\]
(Note that the discounted continuation payoff is the left-hand side of \eqref{zelenka}.)
By deviating from $\pi=0$, a player gets at most
\[
\omega+\left(-(1-\d)m-\d(1-\l)\omega+\d\upsilon_-(\omega_1)\right).
\]
Again inserting our formula for $\upsilon_-$, this reduces to
\[
\frac{mr(N-1)\lambda_1}{\lambda_1+r}\Delta \ge 0.
\]
Hence we can also enforce $\pi=0$ at all these beliefs. We can thus apply our induction argument: there exists $\bar{\Delta}>0$ such that, for all $\Delta \in (0,\bar{\Delta})$, both $\pi=0,1$ can be enforced at all beliefs $\omega \in ( \omega^*(1+4N\Delta),\omega^m)$.

Note that we have not established that, for such a belief $\omega$, $\pi=1$ is enforced with a continuation in which $\pi=1$ is being played in the next round (at belief $\omega_N>\omega^*(1+4N\Delta)$). However, if $\pi=1$ can be enforced at belief $\omega$, it can be enforced when the continuation payoff at $\omega_N$ is highest possible; in turn, this means that, as $\pi=1$ can be enforced at $\omega_N$, this continuation payoff is at least as large as the payoff from playing $\pi=1$ at $\omega_N$ as well. By induction, this implies that the highest equilibrium payoff at $\omega$ is at least as large as the one obtained by playing $\pi=1$ at all intermediate beliefs in $(\omega^*(1+4N\Delta),\omega)$ (followed by, say, the worst equilibrium payoff once beliefs  below this range are reached).

Similarly, we have not argued that, at belief $\omega$, $\pi=0$ is enforced by a continuation equilibrium in which, if a player deviates and experiments unilaterally, his continuation payoff at $\omega_1$ is what he gets if he keeps on experimenting alone. However, because $\pi=0$ can be enforced at $\omega_1$, the lowest equilibrium payoff that can be used after a unilateral deviation at $\omega$ must be at least as low as what the player can get at $\omega_1$ from deviating unilaterally to risky again. By induction, this implies that the lowest equilibrium payoff at belief $\omega$ is at least as low as the one obtained if a player experiments alone for all beliefs in the range $(\omega^*(1+4N\Delta),\omega)$ (followed by, say, the highest equilibrium payoff once beliefs below this interval are reached).

Note that, as $\Delta \rightarrow 0$, these bounds converge (uniformly in $\Delta$) to the cooperative solution (restricted to no experimentation at and below $\omega=\omega^*$) and the single-agent payoff, respectively, which was to be shown. (This is immediate given that these values correspond to precisely the cooperative payoff (with $N$ or $1$ player) for a cutoff that is within a distance of order $\Delta$ of the cutoff $\omega^*$, with a continuation payoff at that cutoff which is itself within $\Delta$ times a constant of the safe payoff.)

This also immediately implies (as for the case $\lambda_0>0$) that for fixed $\omega>\omega^m$, both $\pi=0,1$ can be enforced at all beliefs in $[\omega^m,\omega]$ for all $\Delta<\bar{\Delta}$, for some $\bar{\Delta}>0$: the gain from a deviation is of order $\Delta$, yet the difference in continuation payoffs (selecting as a continuation payoff a value close to the maximum if no player unilaterally defects, and close to the minimum if one does) is bounded away from 0, even as $\Delta \rightarrow 0$.\footnote{This follows by contradiction. Suppose that for some $\Delta \in (0,\bar{\Delta})$, there is $\hat{\omega} \in [\omega^m,\omega]$ for which either $\pi=0$ or 1 cannot be enforced. Consider the infimum over such beliefs. Continuation payoffs can then be picked as desired, which is a contradiction as it shows that at this presumed infimum belief $\pi=0,1$ can in fact be enforced.} Hence, all conclusions extend: fix $\omega \in (\omega^*,\infty)$; for every $\varepsilon>0$, there exists $\bar{\Delta}>0$ such that for all $\Delta<\bar{\Delta}$, the best SSE payoff starting at belief $\omega$ is at least as much as the payoff from all players choosing $\pi=1$ at all beliefs in $(\omega^*+\varepsilon,\omega)$ (using $s$ as a lower bound on the continuation once the belief $\omega^*+\varepsilon$ is reached); and the worst SSE payoff starting at belief $\omega$ is no more than the payoff from a player whose opponents choose $\pi=1$ if, and only if, $\omega \in (\omega^*,\omega^*+\varepsilon)$, and 0 otherwise.

The first part of the proposition follows immediately, picking arbitrary $\pr \in (p_1^*,p^m)$ and $\pp \in (p^m,1)$. The second part follows from the fact that (i) $p_1^*<p_1^\Delta$, as noted, and (ii) for any $p \in [p_1^\Delta,\pr]$, player $i$'s payoff in any equilibrium is weakly lower than his best-reply payoff against $\kappa(p)=1$ for all $p \in [p_1^*,\pr]$, as easily follows from \eqref{webern}, the optimality equation for $w$.\footnote{Consider the possibly random sequence of beliefs visited in an equilibrium. At each belief, a flow loss of either $-(1-\delta)m$ or $-(1-\delta)\omega$ is incurred. Note that the first loss is independent of the number of other players' experimenting, while the second is necessarily lower when at each round all other players experiment.}
}

\proofof{Proposition}{prop:limit-Poisson}{
For $\lambda_0 > 0$, the proof is the same as that of Proposition \ref{prop:limit-Brownian}, except for the fact that it deals with $\VNphat$ rather than $V_N^*$ and relies on Propositions \ref{prop:thresh}--\ref{prop:SSE-Poisson} rather than Proposition \ref{prop:SSE-Brownian}.

For $\lambda_0 = 0$, the proof of Proposition \ref{prop:SSE-exponential} establishes that there exists a natural number $M$ such that, given $\pr$ as stated, we can take $\bar{\Delta}$ to be $(\pr-p_1^*)/M$.
Equivalently, $p_1^*+M\bar{\Delta}=\pr$.
Hence, Proposition \ref{prop:SSE-exponential} can be restated as saying that, for some $\bar{\Delta}>0$, and all $\Delta \in (0,\bar{\Delta})$, there exists $p_\Delta \in (p_1^*,p_1^*+M\Delta)$ such that the two conclusions of the proposition hold with $\pr=p_\Delta$.
Fixing the prior, let $\wr ,\wp $ denote the payoffs in the first and second SSE from the proposition, respectively.\footnote{
Hence, to be precise, these payoffs are only defined on those beliefs that can be reached given the prior and the equilibrium strategies.}
Given that $\pr \rightarrow p_1^*$ and  $\wr(p) \rightarrow s,\wp(p) \rightarrow s$ for all $p \in (p_1^*,p_\Delta)$ as $\Delta \to 0$, it follows that we can pick $\Delta^\dagger \in (0,\bar{\Delta})$ such that for all $\Delta \in (0,\Delta^\dagger)$,
$\WsupPBE \leq \VNphat + \varepsilon$,
$\wr \geq \VNpr-\varepsilon$,
$\|W_1^\Delta-V_1^*\| < \varepsilon$
and $\|\wp-\VIpp\| < \frac{\varepsilon}{2}$.
The obvious inequalities follow as in the proof of Proposition \ref{prop:limit-Brownian} with the subtraction of an additional $\varepsilon$ from the left-hand side of the first one; and the conclusion follows as before, using $2\varepsilon$ as an upper bound.
}

\AppendixOut                        

\setstretch{\refstretch}

\def\next{\noindent \hangindent = 3em \hangafter 1}

\def \aer  {American Economic Review}
\def \ema  {Econometrica}
\def \geb  {Games and Economic Behavior}
\def \jet  {Journal of Economic Theory}
\def \qje  {Quarterly Journal of Economics}
\def \rand {RAND Journal of Economics}
\def \res  {Review of Economic Studies}
\def \te   {Theoretical Economics}


\newcommand{\AJ}[6]{\next {#1}
(#2): ``#3," {\it #4}, {\bf #5}, #6.}

\newcommand{\ajo}[6]{\AJ{\sc #1}{#2}{#3}{#4}{#5}{#6}}

\newcommand{\ajt}[7]{\AJ{{\sc #1} and {\sc #2}}{#3}{#4}{#5}{#6}{#7}}

\newcommand{\AB}[6]{\next {#1}
(#2): ``#3," in {\it #4}, #5. #6.}

\newcommand{\abo}[7]{\AB{\sc #1}{#2}{#3}{#4}{#5}{#6}{#7}}

\newcommand{\abt}[8]{\AB{{\sc #1} and {\sc #2}}{#3}{#4}{#5}{#6}{#7}{#8}}

\newcommand{\BK}[4]{\next {#1}
(#2): {\it #3}. #4.}

\newcommand{\bko}[4]{\BK{\sc #1}{#2}{#3}{#4}}

\newcommand{\bkt}[5]{\BK{{\sc #1} and {\sc #2}}{#3}{#4}{#5}}

\newcommand{\WP}[4]{\next {#1}
(#2): ``#3," #4.}

\newcommand{\wpo}[4]{\WP{\sc #1}{#2}{#3}{#4}}

\newcommand{\wpt}[5]{\WP{{\sc #1} and {\sc #2}}{#3}{#4}{#5}}

\newpage

\section*{References}

\ajo{Abreu, D.}  {1986}
{Extremal Equilibria of Oligopolistic Supergames}
{\jet} {39} {195--225}

\ajo{Abreu, D., D.\  Pearce and E.\ Stacchetti} {1986}
{Optimal Cartel Equilibria with Imperfect Monitoring}
{\jet} {39} {251--269}

\ajo{Abreu, D., D.\  Pearce and E.\ Stacchetti} {1993}
{Renegotiation and Symmetry in Repeated Games}
{\jet}  {60} {217--240}

\ajt{Bergin, J.}{W.B.\ MacLeod}{1993}
{Continuous Time Repeated Games}
{International Economic Review}{34}{21--37}

\ajt{Besanko, D.}{J.\ Wu} {2013}
{The Impact of Market Structure and Learning on the Tradeoff between R\&D Competition and Cooperation}
{Journal of Industrial Economics} {61} {166--201}

\ajt{Besanko, D., Tong, J.}{J.\ Wu} {2018}
{Subsidizing Research Programs with `If' and `When' Uncertainty in the Face of Severe Informational Constraints}
{\rand} {49} {285--310}

\ajt{Biais, B., T.\ Mariotti, G.\ Plantin}{J.-C.\ Rochet} {2007}
{Dynamic Security Design: Convergence to Continuous Time and Asset Pricing Implications}
{\res} {74} {345--390}

\ajt{Bolton, P.}{C.\ Harris} {1999}
{Strategic Experimentation}
{\ema} {67} {349--374}

\abt{Bolton, P.}{C.\ Harris} {2000}
{Strategic Experimentation: the Undiscounted Case}
{Incentives, Organizations and Public Economics -- Papers in Honour of Sir James Mirrlees}
{P.J.\ Hammond and G.D.\ Myles (Eds.)}
{Oxford: Oxford University Press, pp.\ 53--68}

\ajt{Bonatti, A.}{J. H\"{o}rner} {2011}
{Collaborating}
{\aer} {101} {632–-663}

\ajt{Cohen, A.}{E.\ Solan} {2013}
{Bandit Problems with L\'{e}vy Payoff Processes}
{Mathematics of Operations Research} {38} {92--107}

\ajt{Cronshaw, M.B.}{D.G.\ Luenberger} {1994}
{Strongly Symmetric Subgame Perfect Equilibria in Infinitely Repeated Games with Perfect Monitoring and Discounting}
{\geb} {6} {220--237}

\ajt{Das, K., N.\ Klein}{K.\ Schmid} {2020}
{Strategic Experimentation with Asymmetric Players}
{Economic Theory} {69} {1147--1175}

\bkt{Dixit, A.K.}{R.S.\ Pindyck} {1994}
{Investment under Uncertainty}
{Princeton: Princeton University Press}

\wpo{Dong, M.}{2018}
{Strategic Experimentation with Asymmetric Information}
{Working paper, Pennsylvania State University}

\ajo{Dutta, P.K.} {1995}
{A Folk Theorem for Stochastic Games}
{\jet} {66} {1--32}

\ajo{Fudenberg, D. and D.K.\ Levine} {2009}
{Repeated Games with Frequent Signals}
{\qje} {124} {233--265}

\ajo{Fudenberg, D., D.K.\ Levine and S.\ Takahashi} {2007}
{Perfect Public Equilibrium when Players Are Patient}
{\geb} {61} {27--49}

\ajt{Heidhues, P., S.\ Rady}{P.\ Strack} {2015}
{Strategic Experimentation with Private Payoffs}
{\jet} {159} {531--551}

\ajt{Hoelzemann, J.}{N.\ Klein} {2021}
{Bandits in the Lab}
{Quantitative Economics} {12} {1021--1051}

\wpt{H\"{o}rner, J., N.\ Klein}{S.\ Rady}{2014}
{Strongly Symmetric Equilibria in Bandit Games}
{Cowles Foundation Discussion Paper No.\ 1956}

\ajo{H\"{o}rner, J., T.\ Sugaya, S.\ Takahashi and N.\ Vieille} {2011}
{Recursive Methods in Discounted Stochastic Games: An Algorithm for $\delta \rightarrow 1$ and a Folk Theorem}
{\ema} {79} {1277--1318}

\ajt{H\"{o}rner, J.}{L.\ Samuelson} {2013}
{Incentives for Experimenting Agents}
{\rand} {44} {632--663}

\bkt{Johnson, N.L., S.\ Kotz}{N.\ Balakrishnan} {1994}
{Continuous Univariate Distributions: Volume 1 {\rm (second edition)}}
{New York: Wiley}

\ajt{Keller, G.}{S.\ Rady} {2010}
{Strategic Experimentation with Poisson Bandits}
{\te} {5} {275--311}

\ajt{Keller, G.}{S.\ Rady} {2015}
{Breakdowns}
{\te} {10} {175--202}

\ajt{Keller, G.}{S.\ Rady} {2020}
{Undiscounted Bandit Games}
{\geb} {124} {43--61}

\ajt{Keller, G., S.\ Rady}{M.\ Cripps} {2005}
{Strategic Experimentation with Exponential Bandits}
{\ema} {73} {39--68}

\ajt{Klein, N.}{S.\ Rady} {2011}
{Negatively Correlated Bandits}
{\res} {78} {693--732}

\ajt{Marlats, C.}{L.\ M\'{e}nager} {2021}
{Strategic Observation with Exponential Bandits}
{\jet} {193} {105232}

\bkt{Mertens, J.F., Sorin, S.}{S.\ Zamir}{2015}
{Repeated Games {\rm (Econometric Society Monographs, Vol.~55)}}{Cambridge: Cambridge University Press}

\ajo{M\"{u}ller, H.M.} {2000}
{Asymptotic Efficiency in Dynamic Principal-Agent Problems}
{\jet} {39} {251--269}

\bkt{Peskir, G.}{A.\ Shiryaev} {2006}
{Optimal Stopping and Free-Boundary Problems}
{Basel: Birkh\"{a}user Verlag}

\ajt{Rosenberg, D., E.\ Solan}{N.\ Vieille} {2007}
{Social Learning in One-Arm Bandit Problems}
{\ema} {75} {1591–-1611}


\ajt{Sadzik, T.}{E.\ Stacchetti} {2015}
{Agency Models with Frequent Actions}
{\ema} {83} {193--237}

\ajt{Simon, L.K.}{M.B.\ Stinchcombe}{1995}
{Equilibrium Refinement for Infinite Normal-Form Games}
{\ema}{63}{1421--1443}

\ajo{Thomas, C.D.}{2021}
{Strategic Experimentation with Congestion}
{American Economic Journal: Microeconomics} {13} {1--82}

\end{document}